\documentclass{jfm}
\usepackage{comment}
\usepackage{graphicx}
\usepackage{epstopdf,epsfig}
\usepackage{newtxtext}
\usepackage{newtxmath}
\usepackage{natbib}
\usepackage{hyperref}
\usepackage{subcaption}
\usepackage{bm}
\usepackage{overpic}
\usepackage{tikz}
\usepackage{import}
\usepackage{setspace}
\hypersetup{
    colorlinks = true,
    urlcolor   = blue,
    citecolor  = black,
}

\newcommand{\RomanNumeralCaps}[1]
\nolinenumbers
\AppendGraphicsExtensions{.tiff}

\newcommand{\axesIlesSlice}{%
            \def\rad{0.1} 
            \draw[thick,-stealth] (0.5,0.5) -- (0.5,1) node[above,inner sep = 0.2] {$x$} ; 
            \draw[thick,-stealth] (0.5,0.5) -- (1.0,0.5) node[right, inner sep = 0.1] {$y$}; 
            \node[anchor=south west] at (0.5,0.5) {$z$};
            \draw[thick] (0.5,0.5) circle (\rad);
            \draw[thick,rotate around={45:(0.5,0.5)}] (0.5-\rad,0.5) -- (0.5+\rad,0.5) ;
            \draw[thick,rotate around={-45:(0.5,0.5)}] (0.5-\rad,0.5) -- (0.5+\rad,0.5) ;
}
        
\newcommand{\axesIlesIsosurface}{%
            \def\xc{5.5}
            \def\yc{-0.5}
            \draw[thick,-stealth] (\xc,\yc) -- (\xc,\yc+0.45) node[above,inner sep = 0.2] {$x$} ; 
            \draw[thick,-stealth] (\xc,\yc) -- (\xc-0.26,\yc-0.22) node[above,inner sep = 0.2,anchor=north east] {$y$} ; 
            \draw[thick,-stealth] (\xc,\yc) -- (\xc+0.45,\yc-0.13) node[above,inner sep = 0.2,anchor=west] {$z$} ; 
        }
        
\newcommand{\axesIlesIsosurfaceCompression}{%
            \def\xc{2.8}
            \def\yc{-0.7}
            \draw[thick,-stealth] (\xc,\yc) -- (\xc,\yc+0.45) node[above,inner sep = 0.2] {$x$} ; 
            \draw[thick,-stealth] (\xc,\yc) -- (\xc-0.26,\yc-0.22) node[above,inner sep = 0.2,anchor=north east] {$y$} ; 
            \draw[thick,-stealth] (\xc,\yc) -- (\xc+0.45,\yc-0.13) node[above,inner sep = 0.2,anchor=west] {$z$} ; 
        }

\newcommand{\FigureAndLabel}[4]{%
            \begin{subfigure}[h]{#3\textwidth}
                \centering
                \begin{tikzpicture}
                    \node[anchor=north west] (image) at (0,0) { \includegraphics[width=#4\textwidth,trim=0.2cm 0.3cm 0.2cm 0.2cm, clip]{#1}};
                    \node[anchor=north west] at (0,-0) {(\textit{#2})};
                \end{tikzpicture}
            \end{subfigure}
        }

\title{Impact of transverse strain on linear, transitional and self-similar turbulent mixing layers}

\author{B. Pascoe\aff{1}
  \corresp{\email{bradley.pascoe@sydney.edu.au}},
  M. Groom\aff{1,2}, 
  D.L Youngs\aff{1,3}
 \and B. Thornber\aff{1,4}}

\affiliation{\aff{1}School of Aerospace, Mechanical and Mechatronic Engineering, University of Sydney, Sydney, NSW 2006, Australia\\
\aff{2}Geophysical Fluids Team, CSIRO Environment, Eveleigh NSW 2015, Australia\\
\aff{3}AWE, Aldermaston, Reading, RG7 4PR, United Kingdom\\
\aff{4}School of Mechanical and Aerospace Engineering, Queen’s University Belfast, Belfast, BT9 5AH, Northern Ireland, United Kingdom
}

\begin{document}
\maketitle
\setstretch{1.5}
\begin{abstract}
The growth of interfacial instabilities such as the Rayleigh--Taylor (RTI) and Richtmyer--Meshkov instability (RMI) are modified when developing in convergent geometries. Whilst these modifications are usually quantified by the compression rate and convergence rate of the mixing layer, an alternative framework is proposed, describing the evolution of the mixing layer through the effects of the mean strain rates experienced by the mixing layer. An investigation into the effect of the transverse strain rate on the mixing layer development is conducted through application of transverse strain rates in planar geometry. A model for the linear regime in planar geometry with transverse strain rate is derived, with equivalent solutions to convergent geometry, and validated with two-dimensional simulations demonstrating the amplification of the instability growth under transverse compression. The effect of the transverse strain rate on the transitional-to-turbulent mixing layer is investigated with implicit large eddy simulation based on the multi-mode quarter-scale $\theta$-group case by Thornber \textit{et al.} (\textit{Phys. Fluids}, vol. 29, 2017, 105107). The mixing layer's growth exhibits the opposite trend to the linear regime model, with reduced growth under transverse compression. The effect of shear-production under transverse compression causes the mixing layer to become more mixed and the turbulent kinetic energy is increasingly dominated by the transverse directions, deviating from the unstrained self-similar state. The mixing layer width is able to be predicted by adjusting the buoyancy-drag model by Youngs \& Thornber (\textit{Physica D}, vol. 410, 2020, 132517) to utilise a drag length scale that scales with the transverse expansion.

\end{abstract}


\section{Introduction}

The development of a turbulent mixing layer from the Rayleigh--Taylor instability \citep{Rayleigh_1882_InvestigationCharacterEquilibrium,Taylor_1950_InstabilityLiquidSurfaces} and the Richtmyer--Meshkov instability \citep{Richtmyer_1960_TaylorInstabilityShock,Meshkov_1969_InstabilityInterfaceTwo} can be observed in a variety of contexts ranging from inertial confinement fusion (ICF) \citep{Lindl_2004_PhysicsBasisIgnition,Lindl_2014_ReviewNationalIgnition}, supernova explosions \citep{Arnett_2000_RoleMixingAstrophysics}, and supersonic combustion \citep{Yang_2014_RichtmyerMeshkovInstabilityInduced}. The Rayleigh--Taylor instability (RTI) occurs when a light fluid is accelerated into a heavy fluid with a perturbed interface between the two. The misalignment between the density and pressure gradients causes a continual baroclinic deposition of vorticity, amplifying the instability. The Richtmyer--Meshkov instability (RMI) is similar, however the acceleration is taken in the impulsive limit, such as experienced from a shock-wave. With a short/transient deposition of vorticity, the instability is unstable whether accelerated heavy-to-light or light-to-heavy. Both instabilities can cause degradation in the performance of ICF, where fusion is attained by compressing a pellet of fuel in an implosion in-directly or directly driven by lasers. The implosion profile of the pellet utilises multiple shocks to compress the fuel, however this amplifies any imperfections in the fuel-shell interface. As the pellet implodes, the ablation front is RT unstable during the acceleration phase which can feed through to the fuel-shell interface. During the deceleration phase, the fuel-shell interface becomes unstable, further driving the growth of the interface perturbations and mixing cold pellet material into the hot core and degrading ICF performance. For more details on RTI and RMI, thorough reviews have been conducted in the works of \cite{Zhou_2017_ReviewA,Zhou_2017_ReviewB,Zhou_2021_JourneyThroughScales}.

For small amplitudes, typically taken to be $a\leq 0.1\lambda$, RMI and RTI are in the linear regime limit, with Fourier modes evolving independently. For RMI this corresponds to a linear growth rate, whilst RTI experiences exponential growth. As a mode continues to grow it will enter the non-linear regime, experiencing decreased growth and secondary instabilities such as the Kelvin--Helmholtz forming, causing the modes to roll-up and asymmetries in the mixing layer to form. The penetrating structures are labelled spikes for where the heavy fluid penetrates the lighter fluid, and bubbles where the light fluid is penetrating into the heavy. For sufficiently high Reynolds numbers, these structures will further breakdown, creating a self-similar turbulent mixing layer. The mixing layer width $h$ in the late time grows according to $h\propto At g t^2$ for RTI where $At=(\rho_2-\rho_1)/(\rho_2+\rho_1)$ is the Atwood number and $g$ is the acceleration, whilst the RMI mixing layer grows according to $h\propto t^\theta$, where $\theta$ is a sub-linear power law. For narrowband initial conditions ($k_{max}/k_{min} \le 2$) the observed power-law falls within the theoretical bounds of 1/4 \citep{Soulard_2022_PermanenceLargeEddies} and 1/3 \citep{Elbaz_2018_ModalModelMean}. Broadband initial conditions are observed to produce larger power-law growth rates in comparison \citep{Thornber_2010_InfluenceInitialConditions,Groom_2020_InfluenceInitialPerturbation}. The bubble and spike heights are considered to grow according to the same power laws when analysed over a long enough period of time \citep{Youngs_2020_BuoyancyDragModelling,Groom_2023_NumericalSimulationIdealised}.

The instabilities of ICF and supernova do not occur in planar geometry, causing the behaviour of RMI and RTI to differ from the canonical, planar cases. Convergent geometry, used to describe both cylindrical and spherical geometry, is well described in the linear regime limit by Bell--Plesset effects \citep{Penney_1942_ChangingFormNearly,Bell_1951_TaylorInstabilityCylinders,Plesset_1954_StabilityFluidFlows}. Whilst the work of \cite{Bell_1951_TaylorInstabilityCylinders} was limited to looking at the compressible and incompressible cases where one fluid was of negligible density, \cite{Plesset_1954_StabilityFluidFlows} instead considered the incompressible limit between two fluids for any density ratio. The combined modelling of the two approaches is common, and the modified growth rate can be nicely expressed as a differential equation for the amplitude that depends on the fluid compression rate, radius, and radius convergence rate \citep{Epstein_2004_BellPlessetEffects}. Bell--Plesset models have been validated against experiments for single-mode convergent RMI \citep{Vandenboomgaerde_2018_NonlinearGrowthConverging} and divergent RMI \citep{Li_2020_DivergentRichtmyerMeshkov,Zhang_2023_DivergentRichtmyerMeshkov}. Models have been adjusted to account for re-shock in single-mode simulations \citep{Flaig_2018_SinglemodePerturbationGrowth}, and have also been able to predict multi-mode initial conditions prior to mode saturation \citep{ElRafei_2019_ThreedimensionalSimulationsTurbulent}. Weakly non-linear models which account for higher-order harmonics for incompressible RTI have been derived for cylindrical geometry \citep{Wang_2015_WeaklyNonlinearBellPlesset} and spherical geometry \citep{Zhang_2017_WeaklyNonlinearIncompressible}. The experiments of \cite{Luo_2019_NonlinearBehaviourConvergent} for convergent, cylindrical RMI able to be predicted by the model of \cite{Zhang_2017_WeaklyNonlinearIncompressible} until a non-dimensionalised time of 1. 

Whilst the linear regime is well studied, the late-time behaviour of the RMI-induced mixing layer is more complex in convergent geometry compared to the single shock RMI case in planar geometry. The primary reason for this is that in planar geometry, the mixing layer may evolve unaffected by further waves. Implosion profiles however can experience multiple wave interactions due to multiple inward shocks, as well as re-shock and reflected re-shocks due to the shock waves passing through the centre and reflecting off the interface. Re-shocks amplify the turbulent kinetic energy in the mixing layer for both planar and convergent geometry, accelerating the transition to turbulence. Due to the difficulty of capturing all of the complicated flow physics for implosions, numerical studies are an important tool to be utilised. Modelling convergent geometry is not without issues however, as using a Cartesian mesh for a cylindrical or spherical problem can affect the solution obtained. In the inviscid limit, the cross-code comparison of \cite{Joggerst_2014_CrosscodeComparisonsMixing} found the instability growth rates to converge whilst the small scale mixing was dependent upon the numerical scheme, and \cite{Flaig_2018_SinglemodePerturbationGrowth} observed greater difficulty in converging a 2D spherical implosion using a Cartesian mesh compared to the codes using a cylindrical mesh. The influence of these effects can be mitigated by improving the mesh resolution, thus increasing the computational power required \citep{Woodward_2013_SimulatingTurbulentMixing}, or looking only at initial perturbation spectra with large $a/\lambda$. Investigations into the turbulence statistics of RMI implosions have been performed with LES on a Cartesian mesh \citep{Lombardini_2014_TurbulentMixing1,Lombardini_2014_TurbulentMixing2}, ILES on a spherical mesh \citep{ElRafei_2024_TurbulenceStatisticsTransport}, and direct numerical simulation on a Cartesian mesh \citep{Li_2021_StatisticalCharacteristicsTurbulent}.

Whilst planar geometry RMI will not typically experience the compression rates or convergence rates associated with convergent geometry, it is possible to reproduce these effects. \cite{Epstein_2004_BellPlessetEffects} re-derived the Bell--Plesset model for cylindrical and spherical geometry, as well as a model for planar geometry with a compression rate. The compression rate used was equivalent to a mean strain rate in the direction normal to the interface, hence labelled an axial strain rate. The axial strain rate acts to stretch or compress the mixing layer depending upon the sign of the strain rate. \cite{Li_2019_RoleRarefactionCompression,Li_2021_GrowthMechanismInterfacial} observed the influence of this term due to the strain rates that manifest across the mixing layer due to transient waves passing through. \cite{Ge_2020_LatetimeTurbulentMixing,Ge_2022_EvaluatingStretchingCompression} analysed the mixing layer's growth in cylindrical geometry, decomposing the growth into two main components: the compression/stretching effect from the axial strain rate, and the turbulent growth from fluctuating velocity. In the previous work of \cite{Pascoe2024}, the development of RMI from the linear to self-similar regime was analysed. The model of \cite{Epstein_2004_BellPlessetEffects} was able to describe the linear regime, but was inaccurate as the mixing layer transitioned to turbulence. The shear production from the strain rate opposed the strain rate effects, increasing the mixing layer growth under compression and decreasing growth under expansion.

A strain rate formulation can provide an alternate understanding of the modifications of the instability growth in convergent geometry, splitting the contributions into axial and transverse strain rates as opposed to compression and convergence rates. The effect of axial strain rate will stretch or compress the mixing layer, however the role of transverse strain rate is not as clear. The transverse strain rate corresponds to the convergence rate, but it is also a component of the compression rate; an incompressible model requires the axial strain rate and transverse strain rates to summate to zero. It is therefore difficult to isolate the transverse strain rate in convergent geometry. In planar geometry, there is no effective radius to consider for the convergence rate, as compared to the compression rate which has an analogue in the axial strain rate. In order to better understand the transverse strain rate contribution, a method of modelling convergence is performed in planar geometry, using transverse strain rates to replicate the effects of convergence.

In \S \ref{sec:problem_formulation} the definition of the convergence rate for convergent modelling is analysed and shown to be the same as a transverse strain rate for planar geometry, and the methods used to apply the transverse strain rates and simulate the flow are outlined. In \S \ref{sec:2DSingleMode} a linear regime model for planar geometry with axial and transverse strain rates is derived. The results are compared to the model for convergent geometry as well as to simulations of the linear regime under transverse strain are conducted. The analysis of a RMI-induced multi-mode narrowband mixing layer under transverse strain rates is performed in \S \ref{sec:SSML} using an implicit large eddy simulation (ILES). A summary of the findings are provided in \S\ref{sec:conclusion}, describing the performance of the different models used for predicting the effects of a transverse strain rate on RMI.

\section{Problem formulation}
\label{sec:problem_formulation}

\subsection{Transverse strain rate}

In convergent geometry, the linear regime solutions are functions of the convergence rate of the interface radius, $\gamma_R = \dot{R}/R$, and the fluid compression rate, $\gamma_\rho = \dot{\rho}/{\rho}$. The equation for conservation of mass shows that the compression rate is equal to the negative sum of the strain rates:
\begin{align}
    \frac{1}{\rho} \frac{D\rho}{Dt} = -div(\bf{u}),
\end{align}
where the divergence of the velocity field is the trace of the strain tensor, which takes on different forms depending upon the coordinate system used. The convergence rate can be re-written in terms of the local flow field, with the mean radial velocity corresponding to the mean interface velocity:
\begin{align}
    \gamma_R = \frac{\bar{u}_r(r,t)}{r} \bigg{|}_{r=R}.
\end{align}
In spherical geometry, the polar strain rate and azimuthal strain rate are given by
\begin{subeqnarray}
    S_{\theta \theta} &=& \frac{u_r} {r} + \frac{1}{r} \frac{\partial u_\theta}{\partial \theta},\\
    S_{\varphi \varphi} &=& \frac{u_r}{r} + \frac{1}{r} \left( \frac{\partial u_\varphi}{\partial \varphi} + u_\theta \cos(\theta)\right).
\end{subeqnarray}
In the case of the spherically symmetric flow, where mean flow has no polar or azimuthal component, the convergence rate is equal to the mean transverse strain rate at the interface.

The transverse strain rate can also be considered not in terms of the variation of the interface radius, but instead the variation of the wavelength. Angular perturbations in spherical or cylindrical geometry will have an effective wavelength which scales with the radius of the interface. With this proportionality, $R\propto \lambda$, the convergence rate is equivalent to
\begin{align}
    \gamma_R = \frac{\dot{\lambda}}{\lambda} \label{eqn:convergence_wavelength}.
\end{align}
In planar geometry there is no mean interface radius, only an arbitrary interface position. The application of a transverse strain rate is possible, stretching or compressing the domain in the transverse direction. For the wavelength aligned with the transverse direction, the wavelength will vary according to
\begin{align}
    \lambda = \lambda_0 \exp \left[ \int_{t_0}^t \bar{S}_{22}(t') dt'\right] \label{eqn:wavelength},
\end{align}
for a mean transverse strain rate $\bar{S}_{22}$, and with initial wavelength $\lambda_0$ at time $t_0$ that is aligned with $\bar{S}_{22}$. Evaluating equation (\ref{eqn:convergence_wavelength}), shows that $\gamma_R = \bar{S}_{22} = \bar{S}_{\theta \theta}$. It is useful to define the transverse expansion factor which is used throughout the paper:
\begin{align}
    \Lambda(t) = \exp\left[\int^t_0 \bar{S}_{22}(t') dt'\right]. \label{eqn:expansion_parameter}
\end{align}
The transverse expansion factor represents the change in the transverse lengthscale, as shown in equation (\ref{eqn:wavelength}) for the wavelength. 

Therefore, the transverse strain rate in convergent geometry is equal to the convergence rate and also contributes to the compression rate. The application of a transverse strain rate in planar geometry is possible, allowing for an investigation into the effects of convergence through planar simulations. The validity of this approach is further illustrated in \S\ref{sec:linear_model} where the linear regime model in planar geometry is expanded to include a transverse strain rate in the background flow.

\subsection{Strain rate profile}
Two different strain rate profiles are utilised for the investigation of transverse strain rate on the mixing layer development. Focusing on the application of transverse strain rates, the strain rates used will refer to the transverse strain rate applied in the $y$-direction for 2D flows ($\bar{S}=\bar{S}_{22}$), and in $y$- and $z$-direction for 3D flows ($\bar{S}=\bar{S}_{22}=\bar{S}_{33}$).
\subsubsection{Constant velocity}
\label{sec:CV}

The first strain rate profile arises from a domain growing or shrinking with a constant boundary velocity, denoted $V_0$. For this profile any unpertubed packet of fluid will maintain its original velocity throughout the strain profile. For a domain with an initial length of $L_0$, the domain length as a function time varies linearly by
\begin{align}
    L(t) = L_{0} + V_0 R(t-t_0),
\end{align}
where $R(\phi)=\max(0,\phi)$ is the ramp function, $t_0$ is the initial time at which strain is applied and $t-t_0$ indicates the time since strain is initially applied. Initialised with a linear velocity profile, the mean strain rate is initially given by $\bar{S}_{0}= V_0/L_0$. The mean strain rate will change as the length of the domain changes. The time-varying strain rate may be expressed as a function of the initial strain rate:
\begin{subeqnarray}
    \bar{S}(t) & = &\frac{V_0}{L(t)} \label{eqn:CV_S1},\\
    & = & \frac{\bar{S}_{0}}{1+\bar{S}_{0} R(t-t_0)} .\label{eqn:CV_S2}
\end{subeqnarray}
The expansion factor for the constant velocity case is simply given by
\begin{align}
    \Lambda(t) = 1 + \bar{S}_0 R(t-t_0).
\end{align}


\subsubsection{Constant strain rate}
The second strain rate profile used is designed for a constant strain rate. For the system with strain applied at time $t_0$, the strain rate is defined by
\begin{align}
    \bar{S}(t) = \bar{S} H(t-t_0),
\end{align}
where $H(\phi)$ is the Heaviside step function, equal to unity for $\phi\ge 0$ and zero otherwise. The domain experiences exponential growth,
\begin{align}
    \Lambda(t) = \exp\left[\bar{S} R(t-t_0) \right],
\end{align}
which requires some acceleration to drive flow. To accelerate the flow without a pressure differential, thereby isolating the strain rate effects from RT effects, a potential forcing is required \citep{Yu_2007_ExtensionCompressibleIdealgas}. Analysis of the background flow requires the potential term to be
\begin{align}
    g_i= \bar{S}_{ii}^2 x_i,
\end{align}
where $i$ indicates a single direction and not summation.

\subsection{Non-dimensionalisation}
The strain rate has units of inverse time, lending itself to the strain time-scale of $1/\bar{S}$. This time-scale can be used to evaluate the when turbulence is in the rapid distortion limit, requiring the strain time-scale to be much shorter than the turbulence timescale $(k/\epsilon)$. The cases presented in this paper are RMI-induced, for which the typical non-dimensionalised time is calculated according to
\begin{align}
    \tau = \frac{t \dot{h}}{\lambda},
\end{align}
for the impulsive RMI linear growth rate of the mixing layer $\dot{h}$, and an initial perturbation wavelength $\lambda$. The ratio of $\lambda/\dot{h}$ approximates the initial eddy turnover time at the start-up of the instability. For mixing layers induced by alternate means, a different expression may be used for the dominant eddy timescale. It is worth noting that the turbulence in an RMI-induced mixing layer is anisotropic and decaying, and so this non-dimensionalisation is not representative of eddy turnover time throughout the simulation. The same non-dimensionalisation can be applied to the strain rate, in essence comparing the initial eddy turnover time to the strain time-scale:
\begin{align}
    \hat{S} = \frac{\bar{S}\lambda}{\dot{W}}.
\end{align}
As discussed in \cite{Pascoe2024}, typical values for implosions or explosions will vary with time but tend around the order of unity. The timescales are therefore of a similar order, and whilst not necessarily in the rapid-distortion regime, the influence of the strain rate on the mixing layer should not be neglected.
\subsection{Governing equations}

The number fraction model of \cite{Thornber_2018_FiveequationModelSimulation} represents an extension of the non-conservative five-equation model of \cite{Allaire_2002_FiveEquationModelSimulation} and \cite{Massoni_2002_PropositionMethodesModeles} to include the effects of viscosity and diffusion: 

\begin{align}
    \frac{\partial \rho}{\partial t} + \frac{\partial}{\partial x_j} \left(\rho u_j\right)  &= 0,\\
    \frac{\partial \rho u_i}{\partial t} + \frac{\partial}{\partial x_j} \left( \rho u_i u_j + p\delta_{ij}\right)  &=  \frac{\partial \sigma_{ij}}{\partial x_j} + \rho g_i,\\
    \frac{\partial \rho E}{\partial t} + \frac{\partial}{\partial x_j} \left( \left(\rho E + p\right)u_j \right)  &=  \frac{\partial }{\partial x_j} \left( \sigma_{ij} u_i + q_j + {q_d}_j \right) + \rho g_i u_i,\\
    \frac{\partial \rho Y_a}{\partial t} + \frac{\partial}{\partial x_j} \left( \rho Y_a u_j \right)  &=  \frac{\partial}{\partial x_j} \left( D_{12} \rho \frac{\partial Y_a}{\partial x_j} \right),\\
    \frac{\partial f_a}{\partial t} + u_j \frac{\partial f_a}{\partial x_j} = \frac{\partial}{\partial x_j} \left(D_{12} \frac{\partial f_a}{\partial x_j}\right) &- \mathcal{M}D_{12} \frac{\partial f_1}{\partial x_j}\frac{\partial f_a}{\partial x_j} + D_{12} \frac{\partial f_a}{\partial x_j} \frac{\partial N}{\partial x_j} \frac{1}{N}.
    \label{eqn:5eqn_model}
\end{align}
The model makes use of the total number density, $N=p/k_b T$, and the value $\mathcal{M}=(W_1-W_2)/(W_1 f_1 + W_2 f_2)$. The viscous stress tensor, heat flux, and enthalpy flux are given by
\begin{subeqnarray}
    \sigma_{ij} & = & \bar{\mu} \left(\frac{\partial u_i}{\partial x_j} + \frac{\partial u_j}{\partial x_k}  -\frac{2}{3} \frac{\partial u_k}{\partial x_k} \delta_{ij}\right),\\
    q_j & = & \bar{\kappa} \frac{\partial T}{\partial x_j},\\
    {q_d}_j & = & \rho D_{12} \frac{\partial Y_a h_a}{\partial x_j}.
\end{subeqnarray}
The fluids simulated are treated as an ideal gas, with a caloric equation of state, thermal equation of state, and enthalpy relation for each species given by
\begin{subeqnarray}
    e &= \frac{p}{\rho (\gamma -1)},\\
    p &= \rho \frac{\mathcal{R}}{W} T,\\
    h &= c_{p} T.
\end{subeqnarray}
The thermal conductivity of each species is calculated using kinetic theory,
\begin{align}
    \kappa = \mu \left(\frac{5 \mathcal{R}}{4 W} + c_{p}\right),
\end{align}
in terms of the molecular weight of each species, $W$, and the specific heat capacity at constant pressure, $c_{p}$. The mixture quantities for viscosity, $\bar{\mu}$, and thermal conductivity, $\bar{\kappa}$ are calculated from the species' values using Wilke's rule. The binary diffusion coefficient, $D_{12}$, is calculated using the Lewis number which is assumed to be equal for both species:
\begin{align}
    D_{12} = \frac{\bar{\kappa}}{Le \rho \bar{c}_p}.
\end{align}
For multi-species closure within a cell, an isobaric approximation is used:
\begin{align}
    \frac{1}{\gamma -1} = \frac{f_a}{\gamma_a-1}.
\end{align}

\subsection{Numerical methods}

The equations are solved using \textsf{FLAMENCO}, a finite-volume algorithm that is nominally fifth-order in space and second-order in time. The inviscid fluxes are evaluated using the method of characteristics, solving the Riemann problem at the interface using the HLLC Riemann solver \citep{Toro_1994_RestorationContactSurface}. The values at the interface are reconstructed using a scheme that is up to fifth-order in one-dimension in smooth flow regions \citep{Kim_2005_AccurateEfficientMonotonic}, and the reconstructed values are corrected to ensure the correct dissipation scaling at low-Mach-numbers \citep{Thornber_2008_EntropyGenerationDissipation,Thornber_2008_ImprovedReconstructionMethod}. The viscous and diffusion terms are calculated using centred second-order finite differences. The time-stepping is performed with a second-order total variation diminishing Runge-Kutta method \citep{Spiteri_2002_NewClassOptimal}.

With the inclusion of the mean velocity gradients in the simulation, the mesh is deformed with the mean velocity gradients by using an arbitrary Lagrange-Eulerian moving mesh scheme \citep{Thomas_1979_GeometricConservationLaw,Farhat_2001_DiscreteGeometricConservation,Luo_2004_ComputationMultimaterialFlows}. By deforming the mesh with the specified strain rate profile, the simulation enforces the desired mean velocity gradient. The boundary conditions in the direction of the applied strain rates are moving, reflecting free-slip walls, such that the ghost cell quantities are symmetric with respect to the boundary and the strain rate profile is maintained through the ghost cells.

\section{Two-dimensional linear regime}
\label{sec:2DSingleMode}

\newcommand{\axesTwoDim}{%
            \def\xc{0.5}
            \def\yc{0.5}
            \def\len{0.5}
            \draw[thick,-stealth] (\xc,\yc) -- (\xc,\yc+\len) node[above,inner sep=0] {$x$}; 
            \draw[thick,-stealth] (\xc,\yc) -- (\xc+\len,\yc) node[right,inner sep=0] {$y$}; 
        }

\subsection{Linear potential flow}
\subsubsection{Convergent models}
The differential equation provided by \cite{Epstein_2004_BellPlessetEffects} for the amplitude growth rate in spherical geometry is
\begin{align}
    \left(-\gamma_\rho - \gamma_R + \frac{d}{dt}\right) \frac{d}{dt}\left(a_l \rho R^2\right) &= \gamma_0^2 \left(a_l \rho R^2\right).
\end{align}
where $a_l$ is the spatial amplitude of the spherical harmonic perturbation of degree $l$, $Y_l^m (\theta,\varphi)$. The driving term, $\gamma_0^2$ is given by 
\begin{align}
  \gamma_0^2 = \frac{l(l+1)}{R} \frac{\rho^+ - \rho^-}{l\rho^+ + (l+1)\rho^-} g_p \label{eqn:spherical_driving},
\end{align}
where $\rho^+$ and $\rho^-$ are densities above and below the interface respectively, and $g_p$ is the linearised pressure acceleration at the interface, $g_p = -(1/\rho)[\partial p/\partial x]_{r=R}$. 

To obtain a strain rate formulation, the compression rate and convergence rate are converted to the strain rates equivalents,
\begin{subeqnarray}
    \gamma_\rho &=& -\bar{S}_{11} - 2\bar{S}_{22},\\
    \gamma_R &=& \bar{S}_{22} ,
\end{subeqnarray}
where $\bar{S}_{11}$ is the mean radial/axial strain rate and $\bar{S}_{22}$ is the symmetric transverse strain rate. This assumes a spherically symmetric mean velocity profile, with a singular transverse strain rate, $\bar{S}_{22} = \bar{S}_{\theta\theta} = \bar{S}_{\varphi\varphi}$. Expanding the derivatives and simplifying gives the solution
\begin{align}
    \ddot{a}_l + \dot{a}_l \left(\bar{S}_{22} - \bar{S}_{11}\right) + a_l \left(-\bar{S}_{11} \bar{S}_{22} - \dot{\bar{S}}_{11}\right) = \gamma_0^2 a_l \label{eqn:spherical_strain}.
\end{align}

The same process can applied to the solution for cylindrical geometry given by \cite{Epstein_2004_BellPlessetEffects} with
\begin{align}
    \left(-\gamma_\rho + \frac{d}{dt}\right) \frac{d}{dt}\left(a_l \rho R\right) &= \gamma_0^2 \left(a_l \rho R\right).
\end{align}
The cylindrical model neglects activity in the longitudinal direction, such that the perturbations are only a function of polar coordinate, $\cos\left(l\theta\right)$. The compression rate is given by $\gamma_\rho = -\bar{S}_{11}-\bar{S}_{22}$, which is different to the spherical model which has an extra $\bar{S}_{22}$ component. The resulting equation as a function of the strain rates is identical to equation (\ref{eqn:spherical_strain}), with the adjustment for the driving term,
\begin{align}
    \gamma_0^2 = \frac{l}{R} \frac{\rho^+-\rho^-}{\rho^+ + \rho^-} g_p \label{eqn:cylindrical_driving}.
\end{align}
This similarity shows the strain rate formulation provides a more standardised and universal method to describe the effects of the convergent geometry on the amplitude growth for instabilities.

\subsubsection{Planar model}
\label{sec:linear_model}
To reproduce the differential equation for the growth rate in convergent geometry, both axial and transverse strain rates need to be applied in the planar geometry. The background fluid velocities, denoted with an overbar, are then given by
\begin{align}
    \bar{u}_1 (x,t) &= \dot{x}_0(t) + \bar{S}_{11} (x-x_0(t)),\\
    \bar{u}_2 (y,t) &= \bar{S}_{22} y,
\end{align}
where $x_0$ is the mean interface position, $\dot{x}_0$ is the mean interface velocity, and the transverse velocity is stationary at $y=0$. The velocity potential, $\bar{u}_i = \partial \Phi/\partial x_i$, for the background flow is then
\begin{align}
   \Phi(x,y,t) = \Phi_0 (t) + \dot{x}_0(x-x_0) + \frac{1}{2} \bar{S}_{11} (x-x_0(t))^2 + \frac{1}{2} \bar{S}_{22} y^2.
\end{align}
The potential field is linearised about the mean interface,
\begin{align}
    U(x,y,t) = U_0 + g_{U_x} (x-x_0) + g_{U_y} y,
\end{align}
where $g_{U_x} = [\partial U/\partial x]_{x=x_0}$ and $g_{U_y} = [\partial U/\partial y]_{y=0}$. The pressure field is likewise linearised, only allowing for a pressure gradient in the $x$-direction,
\begin{align}
    p(x,t) = p_0 - \rho g_p (x-x_0).
\end{align}
where $g_p=-(1/\rho)[\partial p/\partial x]_{x=x_0}$. The acceleration of the interface is the sum of the potential and pressure acceleration,
\begin{align}
    \ddot{x}_0 = g_{U_x} + g_p.
\end{align}

To take into account the transverse expansion or compression, the perturbed interface uses a time-varying wave-number,
\begin{align}
    x_{int}(y,t) = x_0(t) + a_k(t) \cos(k(t) y) \label{eqn:interface}
\end{align}
where $k(t) = 2\pi/\lambda(t)$. The amplitude $a_k(t)$ corresponds to a specific wavenumber $k(t)$. Under a transverse strain rate of $S_{22}$, the expansion factor will scale the wavelength, and therefore the wave-number evolves as
\begin{align}
    k(t) = k_0 \exp\left[-\int_{t_0}^t S_{22}(t') dt'\right],
\end{align}
where $k_0=2\pi/\lambda_0$ is the reference wave-number at some time $t_0$. For simplicity, the dependence of $k$, $a$, $x_{int}$, and the strain rates will not be written further. The corresponding incompressible velocity potential is given by
\begin{align}
    \phi^+_k &= b^+_k \cos(k y) \exp \left[-kx\right],\\
    \phi^-_k &= b^-_k \cos(k y) \exp \left[kx\right],
\end{align}
where the $+$ superscript denotes the fluid above the interface ($x>x_0$), and $-$ superscript denotes for below the interface ($x<x_0$). Both terms decay to zero as $x$ heads towards the relative infinity. The total velocity potential is then given by
\begin{align}
    \phi^\pm = \Phi + \phi^\pm_k \label{eqn:full_potential}.
\end{align}
The $b_k^\pm$ terms can be removed by equating the interface velocity (total time derivative of equation (\ref{eqn:interface})) with the fluid's $x$-velocity at the interface ($x$-derivative of equation (\ref{eqn:full_potential})):
\begin{align}
    \frac{d x_{int}}{dt} &= \frac{\partial \phi^\pm}{\partial x} \bigg{|}_{x=x_{int}},\\
    b^\pm_k &= \pm \frac{a_k\bar{S}_{11} - \dot{a}_k}{k} \exp \left[\pm k x_{int}\right],\\
    \phi^\pm_k &= \pm \frac{a_k\bar{S}_{11}-\dot{a}_k}{k} \cos(ky) \exp \left[\mp k(x-x_{int})\right].
\end{align}
The Bernoulli equation for unsteady, irrotational flow is given by
\begin{align}
    \frac{\partial \phi}{\partial t} + \frac{1}{2} u_i u_i + U + \frac{p}{\rho} = 0.
\end{align}
The equation is evaluated at the perturbed interface, equating the pressure on each side of the interface, $p^+ = p^-$. The cosine harmonic of the solution provides the equation,
\begin{align}
    \ddot{a}_k + \dot{a}_k \left(\bar{S}_{22} -\bar{S}_{11}\right) + a_k \left(-\bar{S}_{11}\bar{S}_{22} - \dot{\bar{S}}_{11}\right) = a_k k g_p At,
\end{align}
where the Atwood number is given by $At = (\rho^+-\rho^-)/(\rho^++\rho^-)$. The differential equation is identical to the solution obtained from the spherical model in equation (\ref{eqn:spherical_strain}), with the exception of the different driving term on the right-hand side. The wave-number in the driving term is slightly different between the models due to the geometry. The cylindrical wave-number in the model is $l/R$ which can be derived by taking the wavelength as $2\pi R/l$. For the spherical harmonic, the effective wavelength can be calculated using Jeans' relation, giving $\lambda = 2\pi R/\sqrt{l(l+1)}$ \citep{Jeans_1997_PropagationEarthquakeWaves,Wieczorek_2018_SHToolsToolsWorking}. The cylindrical model uses the standard Atwood number definition, whilst the spherical model uses the definition, $At = \sqrt{l(l+1)} (\rho^+-\rho^-)/(l\rho^+ + (l+1)\rho^-)$. For large mode numbers $l$, the spherical Atwood number will approach the value from the standard definition.

The planar model was derived for a two-dimensional flow, however it produces the same differential equation as the three-dimensional spherical model. A third dimension could be added to the planar model, such that the interface is a function of $y$ and $z$, however the same solution is obtained for a uniform transverse strain rate in both directions, such that all wavelengths are affected uniformly. 

For the case with only axial strain rate, $\bar{S}_{22}=0$, the differential equation collapses down to
\begin{align}
    \ddot{a}_k - \frac{d}{dt}\left(a_k \bar{S}_{11}\right) = a_k k g_p At.
\end{align}
This model was investigated in the work of \cite{Pascoe2024} for RMI, which showed the model was accurate within the linear regime. For only a transverse strain rate, $\bar{S}_{11}=0$, the equation becomes
\begin{align}
    \ddot{a}_k + \dot{a}_k \bar{S}_{22} = a_k k g_p At.
\end{align}
For a single-mode RMI where the shock provides the acceleration $g_p = \Delta u \delta(t)$, the amplitude growth rate is
\begin{align}
    \dot{a}(t) = U_0 \exp\left[-\int_0^t \bar{S}_{22}(t') dt'\right],
\end{align}
where $U_0$ is the impulsive velocity as specified by \cite{Richtmyer_1960_TaylorInstabilityShock},
\begin{align}
    U_0 = a_0 k_0 \Delta u At.
\end{align}
This solution shows an amplification of the growth rate $\dot{a}$ as the system compresses, or a reduction as the system expands. The solution can also be written as
\begin{align}
    \dot{a}(t) = a_0 k(t) \Delta u At \label{eqn:linear_sol},
\end{align}
 showing the linear growth rate scales as if the impulse was applied at the current wave-number, as opposed to the initial wave-number.

\subsection{Initial conditions}
The cases were simulated in a two-dimensional domain with a 3:1 density ratio for the two fluids. The single-mode instability is initialised with a velocity perturbation instead of an amplitude perturbation and shock interaction. The velocity perturbation is designed to produce an initial linear growth rate, as described in \cite{Thornber_2010_InfluenceInitialConditions} and \cite{Pascoe2024}. Starting from a flat interface, the velocity perturbation is designed to grow at a velocity of $U_0=1$m/s for the initial wavelength of $\lambda_0=0.2$m. The fluids properties for the initial conditions are prescribed in table \ref{tab:linear_fluid_properties}. As the initial amplitude of the instability starts from zero, there is no directly equivalent shock-induced RMI, but a comparison can be made if a small, finite amplitude is assumed. For example, a shock strength of Ma=1.8439 impacting the ideal gases with initial Atwood of 0.52 at pressure $p=36kPa$ will give approximately the same parameters for an initial linearity of $ak=0.015$. Smaller shock strengths can be used, requiring larger initial amplitudes to compensate. As the viscous and diffusive five-equation model (see equation (\ref{eqn:5eqn_model})) is in use, the initial interface is diffuse, defined by an error function for the volume-fraction profile,
\begin{align}
    f_1 = \frac{1}{2} \left( 1 - \text{erf}\left( \frac{\sqrt{\pi} (x_1-x_0)}{h} \right)\right),
\end{align}
where $h$ is the initial diffusion width set to $\lambda/64$. The initial Reynolds number of the system is given by 
\begin{align}
    \Rey = \bar{\rho} U_0 \lambda_0/\bar{\mu} = 2048,
\end{align}
where the initial wavelength, average density, average viscosity, and the prescribed linear amplitude growth rate have been used. This Reynolds number is sufficiently high that the amplitude growth rate is not significantly affected by the viscosity, as seen in \cite{Walchli_2017_ReynoldsNumberEffects} for unstrained RMI.

\begin{table}
  \begin{center}
\def~{\hphantom{0}}
  \begin{tabular}{lccc}
      Property                      & Fluid 1       & Fluid 2       & Units \\[3pt]
      Density, $\rho$              & 3             & 1             & kg\,m$^{-3}$\\
      Pressure, $p$                 & 100           & 100           & kPa\\
      Molecular Mass, $W$           & 90            & 30            & g\,mol$^{-1}$\\
      Viscosity, $\mu$              & 1.953125$\times 10^{-4}$   & 1.953125$\times 10^{-4}$   & Pa\,s \\
      Specific Heat Ratio, $\gamma$ & 5/3           & 5/3           \\
      Lewis Number, $Le$            & 1             & 1             \\
      \end{tabular}
  \caption{Fluid properties for the linear regime cases.}
  \label{tab:linear_fluid_properties}
  \end{center}
\end{table}

A total of eight strain cases are conducted, four cases for each strain profile, split between expansion and compressive strain rates, as listed in table \ref{tab:linear_cases}. The high-magnitude expansion strain rate cases expand by a factor of four, which requires the initial mesh to four times denser in the $y$-direction. These cells for the expansion case are initially skewed, but become closer to cubic as the simulations progress. The compression cases in contrast start with a isotropic mesh, but the cells become skewed as the domain compresses. 

\begin{table}
  \begin{center}
\def~{\hphantom{0}}
  \begin{tabular}{lrrcccc}
	  &  &  & Simulation  & Initial domain  \\
	  Strain Profile        & $\bar{S}_0$ (s$^{-1}$) & $\hat{S}_0~$ &  time (s)   &  size (m$^2$)   & Grid resolution & $\Lambda_f$ \\[3pt]
      Unstrained            & ~~0.0~            & ~0.0~  & 0.02   & $1.0\times0.2$                & $640\times128$ & 1.00\\
      Constant velocity     & $-37.50$            & -7.50  & 0.02   & $1.0\times0.2$                & $640\times128$ & 0.25\\
      Constant velocity     & $-18.75$            & -3.75  & 0.02   & $1.0\times0.2$                & $640\times512$ & 0.62\\
      Constant velocity     & ~75.0~            & 15.0~  & 0.02   & $1.0\times0.2$                & $640\times512$ & 2.50\\
      Constant velocity     & 150.0~            & 30.0~  & 0.02   & $1.0\times0.2$                & $640\times128$ & 4.00\\
      Constant strain rate  & $-70.0$~            & -14.0~ & 0.02   & $1.0\times0.2$                & $640\times128$ & 0.25\\
      Constant strain rate  & $-35.0$~            & -7.0~  & 0.02   & $1.0\times0.2$                & $640\times128$ & 0.50\\
      Constant strain rate  & ~35.0~            & ~7.0~  & 0.02   & $1.0\times0.2$                & $640\times512$ & 2.01\\
      Constant strain rate  & ~70.0~            & 14.0~  & 0.02   & $1.0\times0.2$                & $640\times512$ & 4.06\\
  \end{tabular}
  \caption{The strain rates, total simulation time, domain size, grid resolution, and final expansion factor for each of the linear regime cases.}
  \label{tab:linear_cases}
  \end{center}
\end{table}

\subsection{Results}
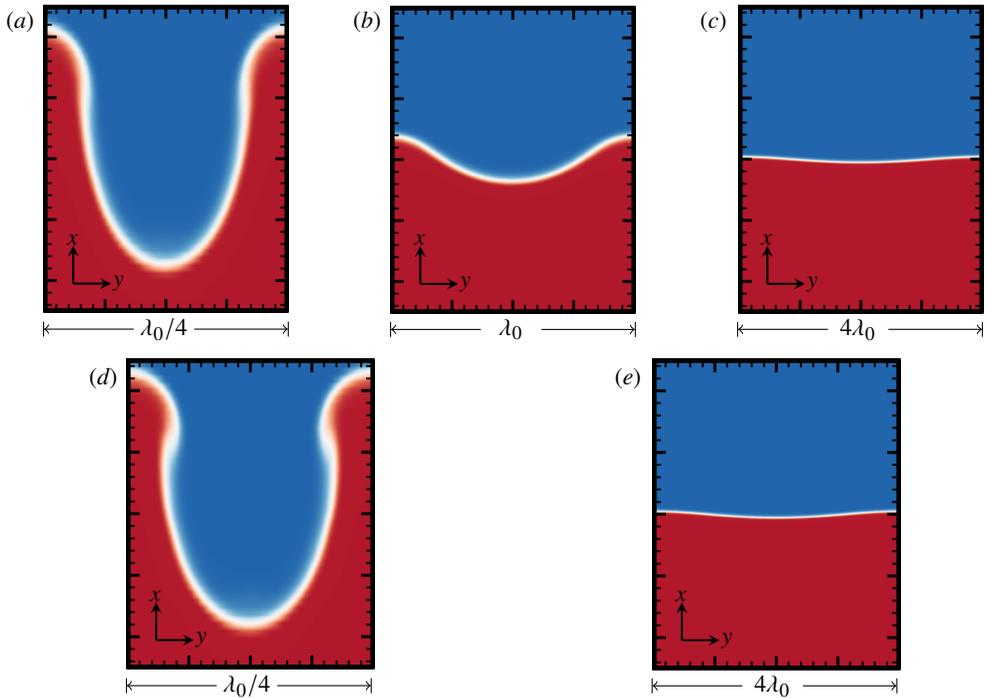
\begin{figure}
    \centering
    \begin{subfigure}[h]{0.32\textwidth}
        \centering
        \begin{tikzpicture}
            \node[anchor=south west] at (0,0) {\includegraphics[width=0.75\textwidth]{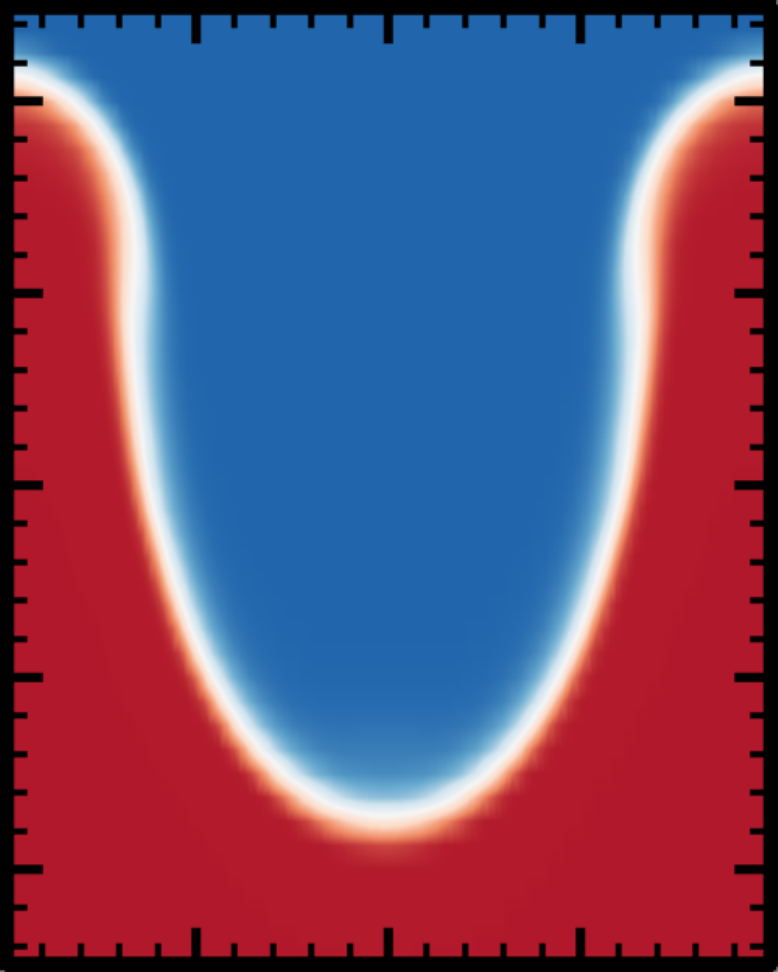}};
            \axesTwoDim
            \node[anchor=north west] at (-0.5,4.2) {(\textit{a})};
            \node at (1.7,-0.1) {$\lambda_0/4$};
            \draw [->|] (2.1,-0.1) -- (3.35,-0.1);
            \draw [->|] (1.3,-0.1) -- (0.1,-0.1);
        \end{tikzpicture}
    \end{subfigure}
    \hfill
    \begin{subfigure}[h]{0.32\textwidth}
        \centering
        \begin{tikzpicture}
            \node[anchor=south west] at (0,0) {\includegraphics[width=0.75\textwidth]{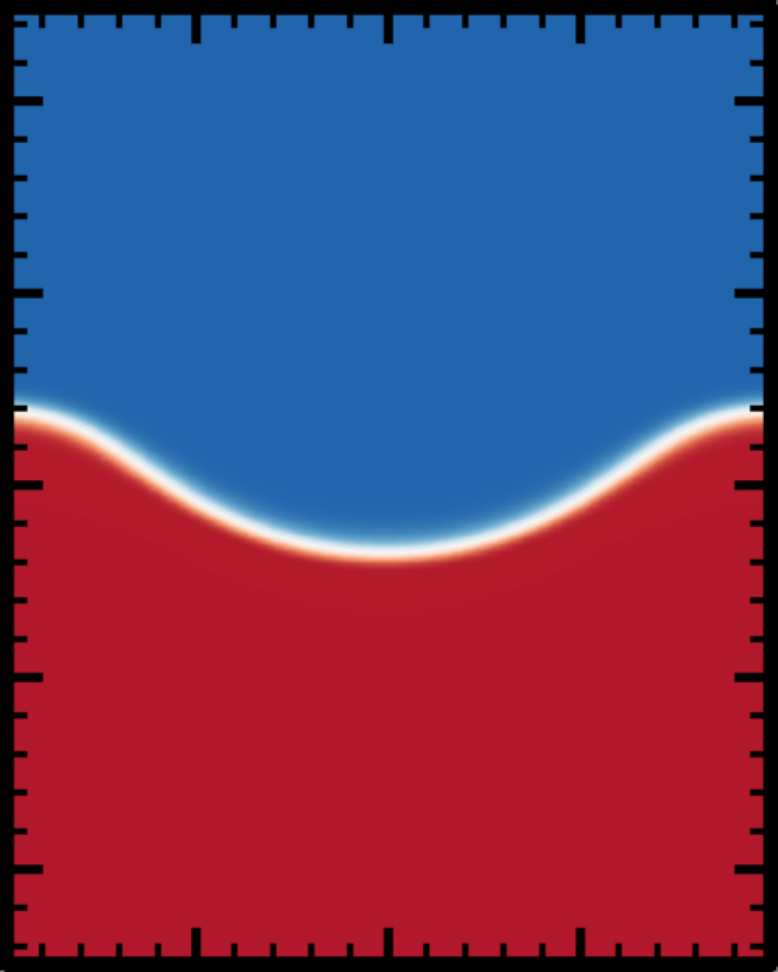}};
            \axesTwoDim
            \node at (1.7,-0.1) {$\lambda_0$};
            \draw [->|] (2.1,-0.1) -- (3.35,-0.1);
            \draw [->|] (1.3,-0.1) -- (0.1,-0.1);
            \node[anchor=north west] at (-0.5,4.2) {(\textit{b})};
        \end{tikzpicture}
    \end{subfigure}
    \hfill
    \begin{subfigure}[h]{0.32\textwidth}
        \centering
        \begin{tikzpicture}
            \node[anchor=south west] at (0,0) {\includegraphics[width=0.75\textwidth]{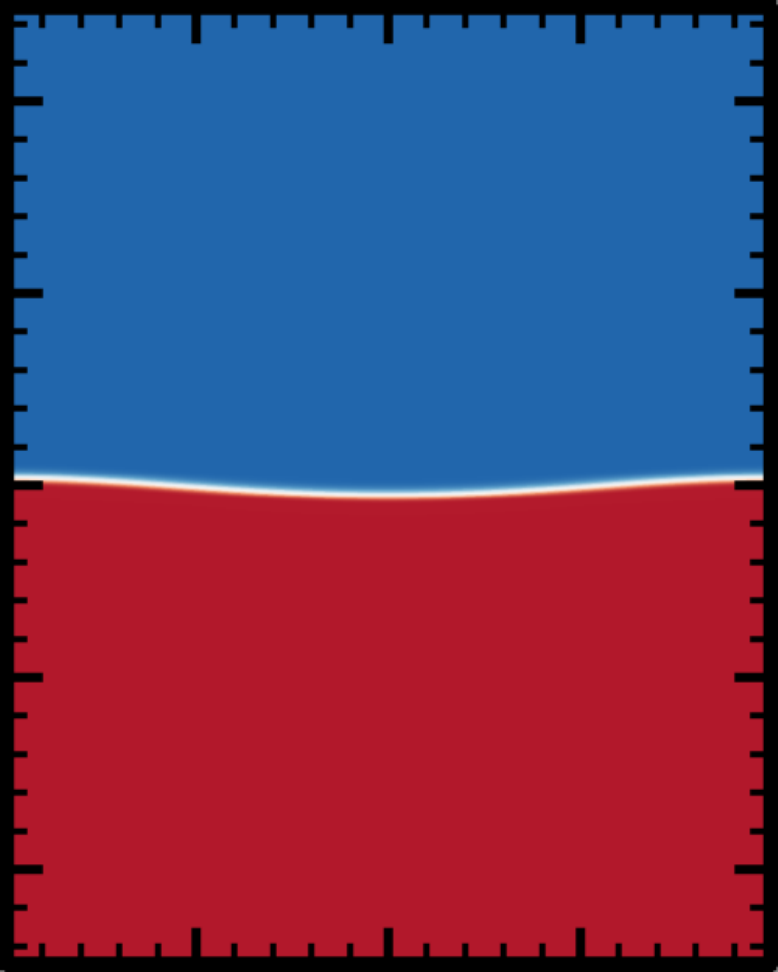}};
            \axesTwoDim
            \node at (1.7,-0.1) {$4\lambda_0$};
            \draw [->|] (2.1,-0.1) -- (3.35,-0.1);
            \draw [->|] (1.3,-0.1) -- (0.1,-0.1);
            \node[anchor=north west] at (-0.5,4.2) {(\textit{c})};
        \end{tikzpicture}
    \end{subfigure}
    \hfill
    \begin{subfigure}[h]{0.485\textwidth}
        \centering
        \begin{tikzpicture}
            \node[anchor=south west] at (0,0) {\includegraphics[width=0.5\textwidth]{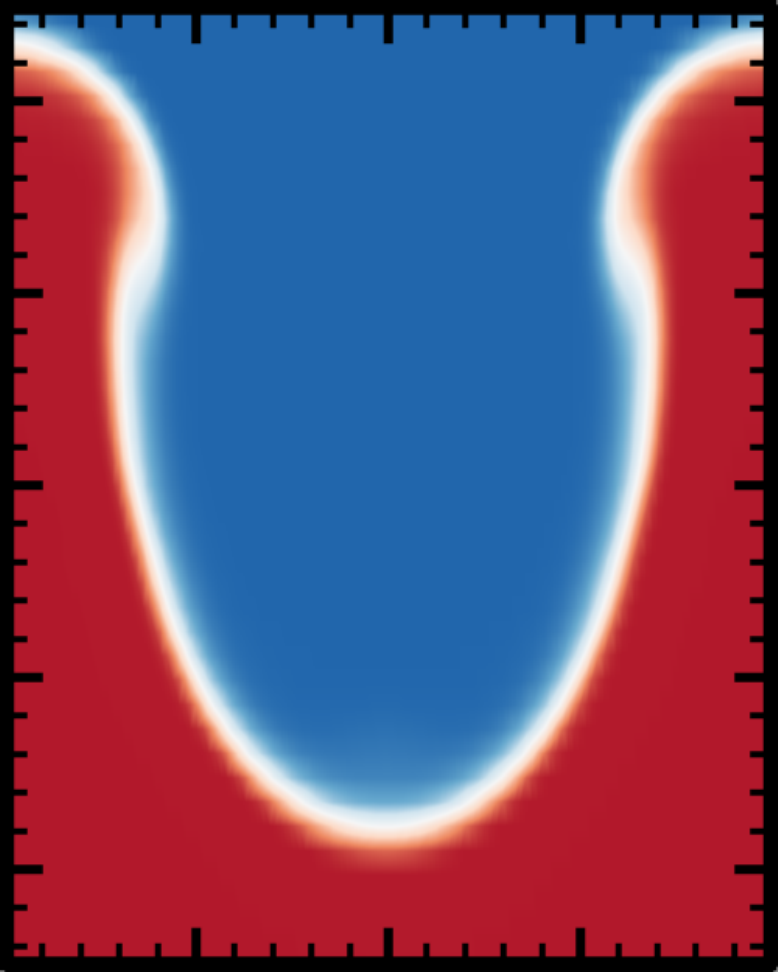}};
            \axesTwoDim
            \node at (1.7,-0.1) {$\lambda_0/4$};
            \draw [->|] (2.1,-0.1) -- (3.35,-0.1);
            \draw [->|] (1.3,-0.1) -- (0.1,-0.1);
            \node[anchor=north west] at (-0.5,4.2) {(\textit{d})};
        \end{tikzpicture}
    \end{subfigure}
    \hfill
    \begin{subfigure}[h]{0.485\textwidth}
        \centering
        \begin{tikzpicture}
            \node[anchor=south west] at (0,0) {\includegraphics[width=0.5\textwidth]{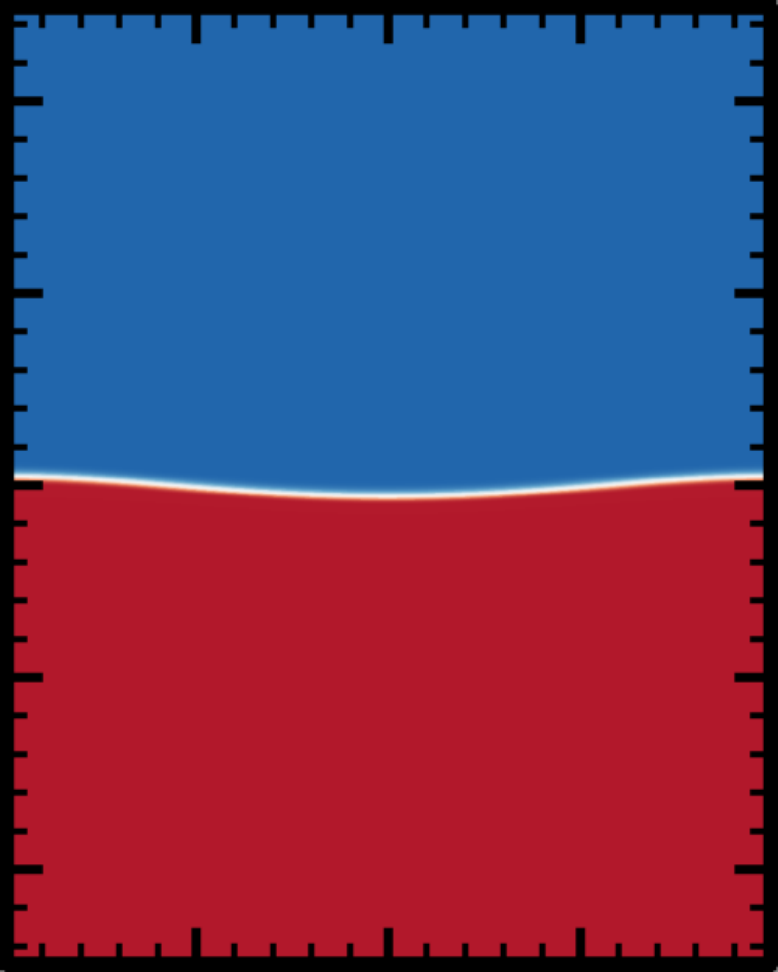}};
            \axesTwoDim
            \node at (1.7,-0.1) {$4\lambda_0$};
            \draw [->|] (2.1,-0.1) -- (3.35,-0.1);
            \draw [->|] (1.3,-0.1) -- (0.1,-0.1);
            \node[anchor=north west] at (-0.5,4.2) {(\textit{e})};
        \end{tikzpicture}
    \end{subfigure}
    \caption{Interface at $\tau=0.1$ for the 2-D single-mode simulations. Heavy fluid ($f_1=1$) is red, light fluid ($f_1=0$) is blue. Major ticks indicate a distance of $\lambda(t)/4$, with the final wavelength marked below the plot. (\textit{a}) Constant velocity, $\hat{S}_0 = -7.5$; (\textit{b}) Unstrained case; (\textit{c}) Constant velocity, $\hat{S}_0 = 30$; (\textit{d}) Constant strain rate, $\hat{S} = -14$;  (\textit{e}) Constant strain rate, $\hat{S} = 14$.}
    \label{fig:Linear_ContourPlots}
\end{figure}

Visualisations of the volume fraction contour for the unstrained case and the high-magnitude strain rate cases for each profile are shown in figure \ref{fig:Linear_ContourPlots}. Each plot is scaled to the final wavelength of the simulation, where the compression cases have a wavelength that is 16 times smaller than the expansion cases. Relative to the final wavelengths, the compression cases have a much larger value of $a/\lambda(t)$, displaying the formation of penetrating bubbles and spikes that are beginning to roll-up. The unstrained case is around the limit of the linear regime ($a\le 0.1 \lambda$), and the interface appears to be well described by a cosine function. For comparison, the expansion cases show a very small $a/\lambda(t)$ value, suggesting it is still within the linear regime.

\begin{figure}
    \centering
    \begin{subfigure}[h]{0.485\textwidth}
        \centering
        \begin{tikzpicture}
            \node[anchor=south west] at (0,0){\includegraphics[width=\textwidth]{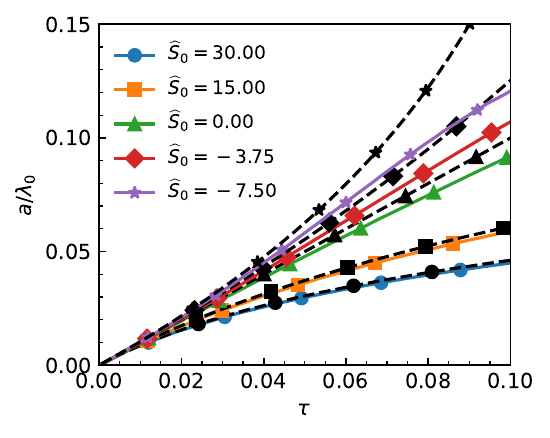}};
            \node[anchor=north west] at (0.,5.3) {(\textit{a})};
        \end{tikzpicture}
    \end{subfigure}
    \begin{subfigure}[h]{0.485\textwidth}
        \centering
        \begin{tikzpicture}
            \node[anchor=south west] at (0,0){\includegraphics[width=\textwidth]{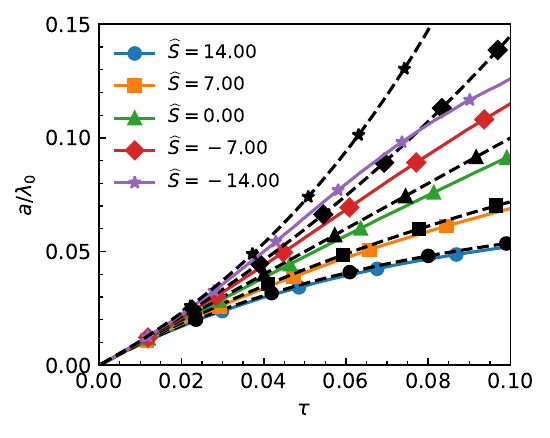}};
            \node[anchor=north west] at (0,5.3) {(\textit{b})};
        \end{tikzpicture}
    \end{subfigure}
    \caption{Amplitude of the single mode linear regime, non-dimensionalised for the initial wavelength,for (\textit{a}) constant velocity and (\textit{b}) constant strain rate. Solid lines indicate numerical results, dashed lines indicate the linearised potential model.}
    \label{fig:Linear_Amplitudes}
\end{figure}

\begin{figure}
    \centering
    \begin{subfigure}[h]{0.485\textwidth}
        \centering
        \begin{tikzpicture}
            \node[anchor=south west] at (0,0){\includegraphics[width=\textwidth]{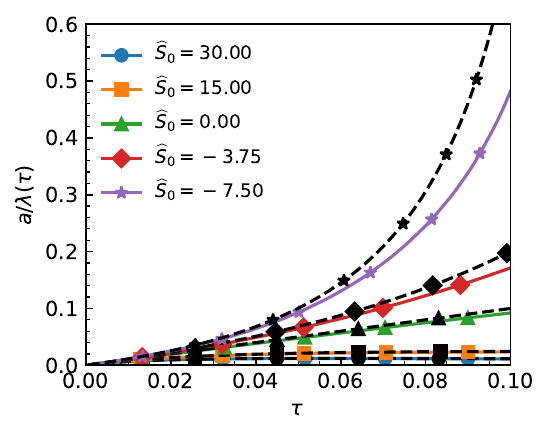}};
            \node[anchor=north west] at (0.,5.3) {(\textit{a})};
        \end{tikzpicture}
    \end{subfigure}
    \begin{subfigure}[h]{0.485\textwidth}
        \centering
        \begin{tikzpicture}
            \node[anchor=south west] at (0,0){\includegraphics[width=\textwidth]{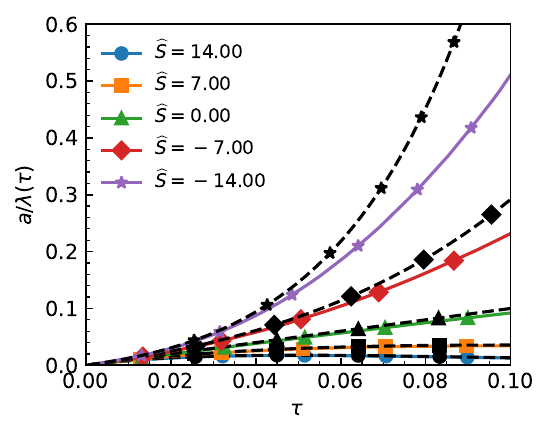}};
            \node[anchor=north west] at (0,5.3) {(\textit{b})};
        \end{tikzpicture}
    \end{subfigure}
    \caption{Amplitude of the single mode linear regime, non-dimensionalised by the time-varying wavelength for (\textit{a}) constant velocity and (\textit{b}) constant strain rate. Solid lines indicate numerical results, dashed lines indicate the linearised potential model.}
    \label{fig:Linear_Amplitudes_t}
\end{figure}

\begin{figure}
    \centering
    \begin{subfigure}[b]{0.485\textwidth}
        \centering
        \begin{tikzpicture}
            \node[anchor=south west] at (0,0){\includegraphics[width=\textwidth]{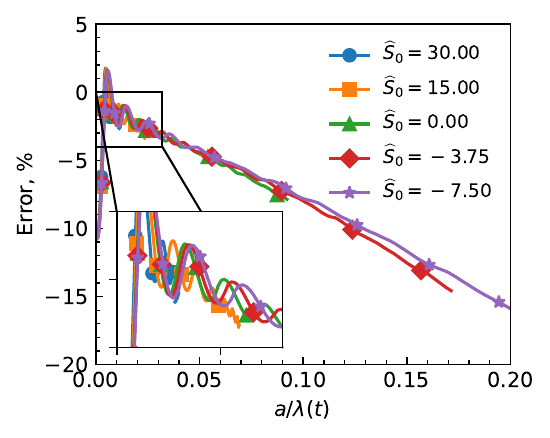}};
            \node[anchor=north west] at (0,5.3) {(\textit{a})};
        \end{tikzpicture}
    \end{subfigure}
    \hfill
    \begin{subfigure}[b]{0.485\textwidth}
        \centering
        \begin{tikzpicture}
            \node[anchor=south west] at (0,0){\includegraphics[width=\textwidth]{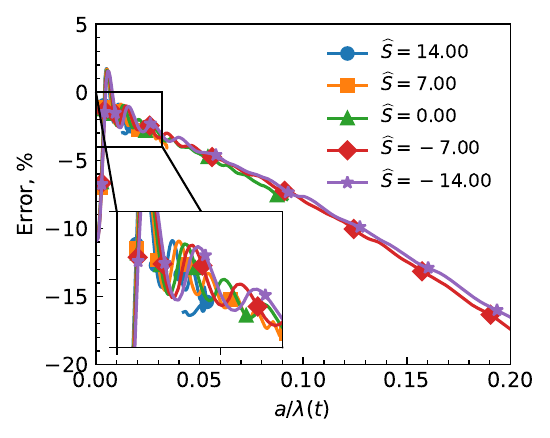}};
            \node[anchor=north west] at (0,5.3) {(\textit{b})};
        \end{tikzpicture}
    \end{subfigure}
    \caption{Error in the amplitude for the linear regime under (\textit{a}) constant velocity and (\textit{b}) constant strain rate.}
    \label{fig:Linear_Error_Plots}
\end{figure}

As the interface is yet to roll-up and remains smoothly connected, the mean interface position is taken along the  $f_1=0.5$ volume fraction isocontour line. The amplitude of the interface is taken to be half of the distance between the maximum and minimum of the isocontour, representing the peak and trough of the perturbation, given by
\begin{equation}
    a = 0.5\left(\max(x_{f_1=0.5}) - \min(x_{f_1=0.5}) \right).
\end{equation}
A second-order interpolation scheme is used to locate the minimum and maximum isocontour positions from the simulation's cell average values. The amplitudes non-dimensionalised by the initial wavelength are plotted in figure \ref{fig:Linear_Amplitudes}, along with the theoretical model given in equation (\ref{eqn:linear_sol}). The simulation results show that the expansion cases grow the slowest and the compression cases grow the fastest. The model is able to accurately predict the growth rate of the expansion cases, with a small final error for the unstrained case which can be attributed to saturation as the mode becomes non-linear. The compression cases have a larger error, with the high magnitude compression (negative) strain rate cases the least accurate at the final simulation time. These cases have the largest amplitudes, such that they can be expected to be saturating by the final simulation time, as observed in the volume fraction contour plots of figure \ref{fig:Linear_ContourPlots}. It is worthwhile to instead look at the performance of the model when plotting as a function of the amplitude non-dimensionalised by the time-varying wavelength, as done in figure \ref{fig:Linear_Amplitudes_t}. The same trends can be observed in these plots, with the expansion cases growing the slowest and the compression cases growing the fastest. For the highest magnitude expansion cases, the growth of $a/\lambda(t)$ goes negative, a result of the wavelength growing faster than the amplitude. This suggests that for highly strained expansion cases the perturbation will remain in the linear regime indefinitely. The performance of the model confirms that the perturbation becomes non-linear depending upon the time-varying wavelength and not the initial wavelength. Figure \ref{fig:Linear_Error_Plots} reinforces this by plotting the error between the model and simulation as a function of $a/\lambda(t)$. After some initial noise from the interpolation approximating the peak/trough of the initially flat interface, the error profile collapses to a rather straight line for the unstrained and compression cases. The expansion cases fall within the general trend, however the highly expanded cases are not monotonically increasing for $a/\lambda(t)$. The constant strain rate case of $\hat{S}=14$ shows a final trajectory towards the origin, decreasing in error magnitude and $a/\lambda(t)$.

\section{Self-similar mixing layer}
\label{sec:SSML}

The RMI-induced mixing layer becomes self-similar at late-time, exhibiting asymptotic values for quantities such as the mixedness and anisotropy of the turbulent kinetic energy. One prominent example of the late-time analysis of the RMI-induced mixing layer is the $\theta$-group collaboration \citep{ThetaGroup}, which performed a cross-code validation of the development of the mixing layer, using eight different codes to perform large eddy simulations. The quarter-scale case from the $\theta$-group is utilised to investigate how the application of transverse strain rates affect the development of the mixing layer towards the self-similar state. Several models are proposed in order to capture the effects of the transverse strain rate and to help further understand how strain affects the mixing layer behaviour.

\subsection{Initial conditions}

ILES cases are conducted using the quarter-scale narrowband case from the $\theta$-group collaboration \citep{ThetaGroup}. Relying on the dissipation of the numerical scheme to mimic cascade of energy and dissipation, the simulations are conducted in \textsf{FLAMENCO} using the inviscid five-equation model presented in equation (\ref{eqn:5eqn_model}). With the omission of the viscous, diffusive and conductive fluxes, the simulations represent the high-Reynolds number limit. The ILES simulations conducted use the same initialisation but have a slightly different domain set-up. The initial domain size is the same, with size $x\times y \times z =\mathcal{L}_x\times \mathcal{L}\times \mathcal{L}= 2.8\pi \times 2\pi \times 2\pi$m$^3$, however the boundary conditions are different. Whilst the original quarter-scale case used periodic boundary conditions in $y$- and $z$-directions, for the application of transverse strain rates with moving mesh, it is necessary to use symmetry plane boundary conditions in \textsf{FLAMENCO}. An alternative approach to this could be to use the modelling approach of \cite{Rogallo_1981_NumericalExperimentsHomogeneous} to transform the domain and solve for the fluctuations. The $x$-direction boundary conditions remain as outflows. The fluids are initially set-up in a heavy-to-light order, with the mean interface position at $x=3.5$\,m, and the shock initialised below $x=3$\,m. The unshocked densities of the heavy and light fluids are 3\,kg\,m$^{-3}$ and 1\,kg\,m$^{-3}$ respectively. Both fluids have $\gamma=5/3$, and the initial shock strength of Mach 1.8439 achieves a four-fold increase in the shocked heavy fluid. The interface between the heavy and light fluid is perturbed with a narrowband spectrum, using a constant power spectrum from $\lambda_{\text{min}} = \mathcal{L}/32$ to $\lambda_{\text{max}}=\mathcal{L}/16$. The amplitude and phase of each mode is randomly generated from a Gaussian distribution. The random numbers and initial spectrum are reproducible however, as a specific seed is used with the Mersenne Twister algorithm. These amplitudes are scaled to ensure the amplitude of the final spectrum is equal to $0.1\lambda_\text{min}$. The interface also uses a diffuse thickness, given by
\begin{align}
    f_1 = \frac{1}{2} \left( 1 - \text{erf}\left( \frac{\sqrt{\pi}\left[x-S(y,z)\right]}{\delta}\right)\right),
\end{align}
where diffuse thickness of $\delta = \mathcal{L}/128$ is used, and $S(y,z)$ is the perturbation spectrum:
\begin{align}
    S(y,z) = 3.5 + \sum_{k_y,k_z} a_0 \cos\left(k_y y + k_z z + \phi\right).
\end{align}
To offset the velocity difference imparted by the shock, the velocity in the domain is given an offset of $\Delta u = -291.575$\,m\,s$^{-1}$, allowing the interface to remain close to stationary after shock transition. The lengthscales are non-dimensionalised by the mean wavelength, $\bar{\lambda} = \sqrt{12/7}\lambda_\text{min}$, and the velocities are non-dimensionalised by the initial growth rate of the integral width, $\dot{W}_0  = 12.649$\,m\,s$^{-1}$, where the integral width is defined by
\begin{align}
    W = \int_0^{\mathcal{L}_x} \bar{f}_1 \bar{f}_2 dx.
\end{align}
Further details on how these properties are calculated can be found in \cite{Thornber_2010_InfluenceInitialConditions,ThetaGroup}. Other variables are non-dimensionalised using a combination of $\bar{\lambda}$, $\dot{W}_0$, the mean post-shock density for the unstrained case $\bar{\rho}^+ = 3.51$\,kg\,m$^{-3}$, and the cross-sectional area $4\pi^2$.

The transverse strain rates are applied at $\tau=1$, when the mixing layer is beginning to transition. The strain rates are applied by adding the mean velocity gradient to the flow profile and moving the mesh with the prescribed strain rate profile. Eight strain cases are conducted as listed in table \ref{tab:ILES_cases}. The are four strain cases for each strain rate profile, which can be further subdivided into two expansion cases and two compression cases. The simulations are conducted until the domain changes in size by around a factor of two. In order to resolve the expansion cases, the solution is interpolated on a mesh with twice as many cells in the $y$ and $z$ directions to ensure the simulations resolve the same minimum scale at the late time as the unstrained case. As shown in Appendix \ref{sec:appendix}, the cases are converged for the integral properties such as the integral width and molecular mixedness, ensuring the strain rate effects on these properties are independent of the mesh resolution. Whilst the simulations are conducted to less extreme expansion factors compared to ICF problems, the range of strain rates investigated are representative of practical application and capture a noticeable change in the development of the mixing layer.

\begin{table}
  \begin{center}
\def~{\hphantom{0}}
  \begin{tabular}{lrrcccc}
        & & & Simulation & Initial domain  \\
      Strain Profile        & $\bar{S}_0$ (s$^{-1}$) & $\hat{S}_0~~$ & time (s) & size (m$^3$) & Grid resolution & $\Lambda_F$ \\[3pt]
      Unstrained            & ~0.0~~ & ~0.0~~ & 0.711 & $2.8\pi\times2\pi\times2\pi$ & $720\times~512^2$            & 1.00\\
      Constant velocity     & $-2.50$~ & $-0.051$ & 0.203 & $2.8\pi\times2\pi\times2\pi$ & $720\times~512^2$        & 0.54\\
      Constant velocity     & $-0.625$ & $-0.013$ & 0.711 & $2.8\pi\times2\pi\times2\pi$ & $720\times~512^2$        & 0.57\\
      Constant velocity     & 1.25~ & ~0.025 & 0.711 & $2.8\pi\times2\pi\times2\pi$ & $720\times1024^2$             & 1.86\\
      Constant velocity     & 5.0~~ & ~0.102 & 0.203 & $2.8\pi\times2\pi\times2\pi$ & $720\times1024^2$             & 1.91\\
      Constant strain rate  & $-4.0$~~ & $-0.081$ & 0.203 & $2.8\pi\times2\pi\times2\pi$ & $720\times~512^2$        & 0.48\\
      Constant strain rate  & $-1.0$~~ & $-0.020$ & 0.711 & $2.8\pi\times2\pi\times2\pi$ & $720\times~512^2$        & 0.50\\
      Constant strain rate  & 1.0~~ & ~0.020 & 0.711 & $2.8\pi\times2\pi\times2\pi$ & $720\times1024^2$             & 2.00\\
      Constant strain rate  & 4.0~~ & ~0.081 & 0.203 & $2.8\pi\times2\pi\times2\pi$ & $720\times1024^2$             & 2.08
  \end{tabular}
  \caption{The strain cases, total simulation time, domain size, grid resolution, and final expansion factor for each of the ILES cases.}
  \label{tab:ILES_cases}
  \end{center}
\end{table}

\subsection{Results}

\subsubsection{Visualisations}

Slices of volume fraction contours for the constant velocity cases are shown in figure \ref{fig:ILES_contours} for the middle $x$-$y$ plane of the simulations. The left columns shows the time $\tau = 9.84$ for all cases, whilst the right column shows the results for the low-magnitude strain rate cases at the later time of $\tau=34.35$. Due to the varying domain width with time, the scale of each plot is varied to fit the slice within the column. The compression cases appear to be dominated several large structures, whilst the expansion cases maintain more distinct small structures in the mixing layer which have not been broken down by turbulent mixing. 

\newcommand{\axesLength}[1]{%
            \node at (3.35,-0.05) {#1$\pi$};
            \draw [->|] (3.7,-0.05) -- (6.6,-0.05);
            \draw [->|] (3.0,-0.05) -- (0.10,-0.05);
        }
\begin{figure}
    \centering
    \begin{subfigure}{\textwidth}
         \begin{tikzpicture}
             \node[anchor=south west] (image) at (0,0) {\includegraphics[width=0.48\textwidth]{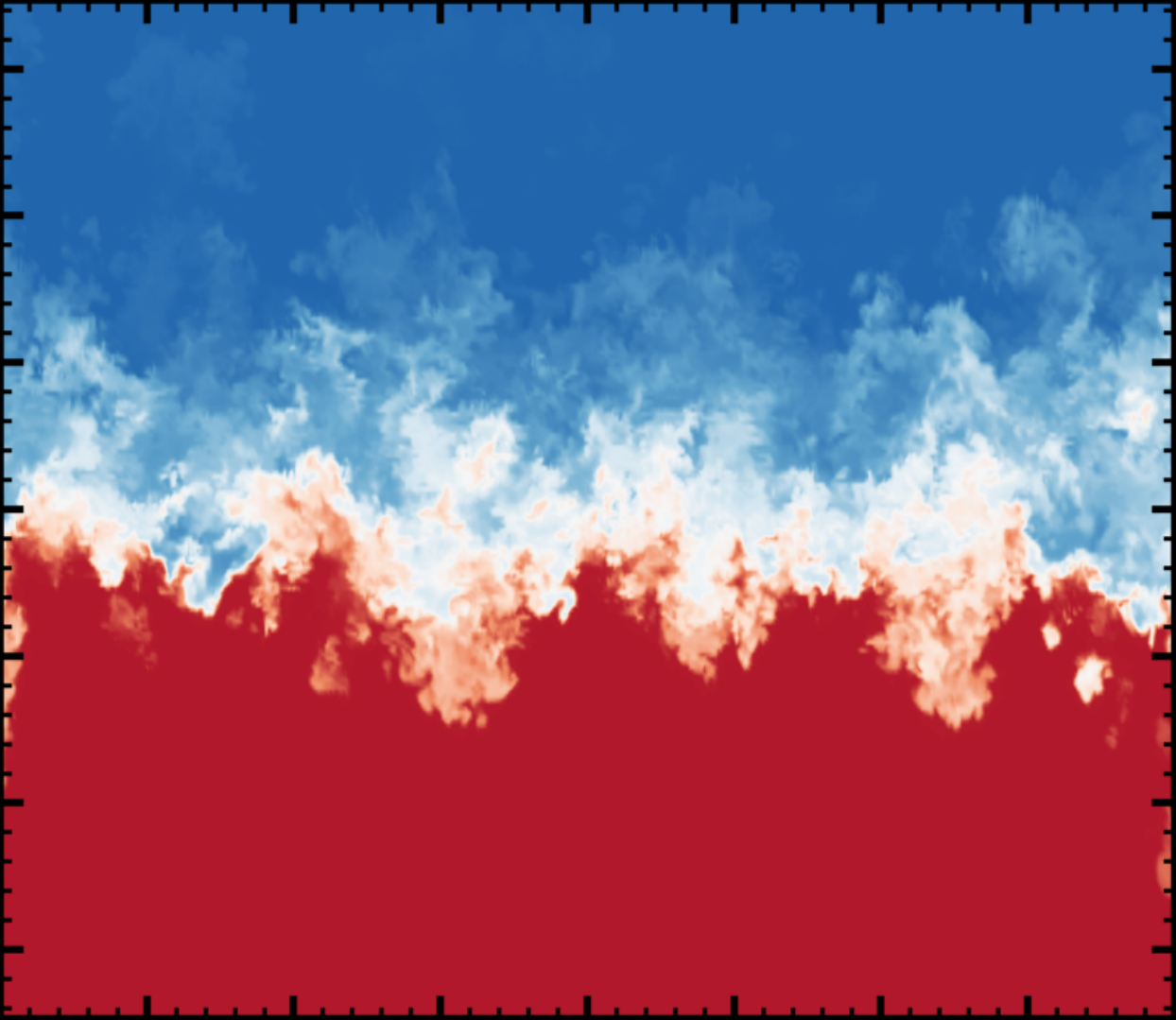}};
             \node[anchor=north west] at (-0.5,5.8) {(\textit{a})};
             \axesIlesSlice
             \axesLength{1.1} 
         \end{tikzpicture}
    \end{subfigure}
    \begin{subfigure}{0.48\textwidth}
         \begin{tikzpicture}
             \node[anchor=south west] (image) at (0,0) {\includegraphics[width=\textwidth]{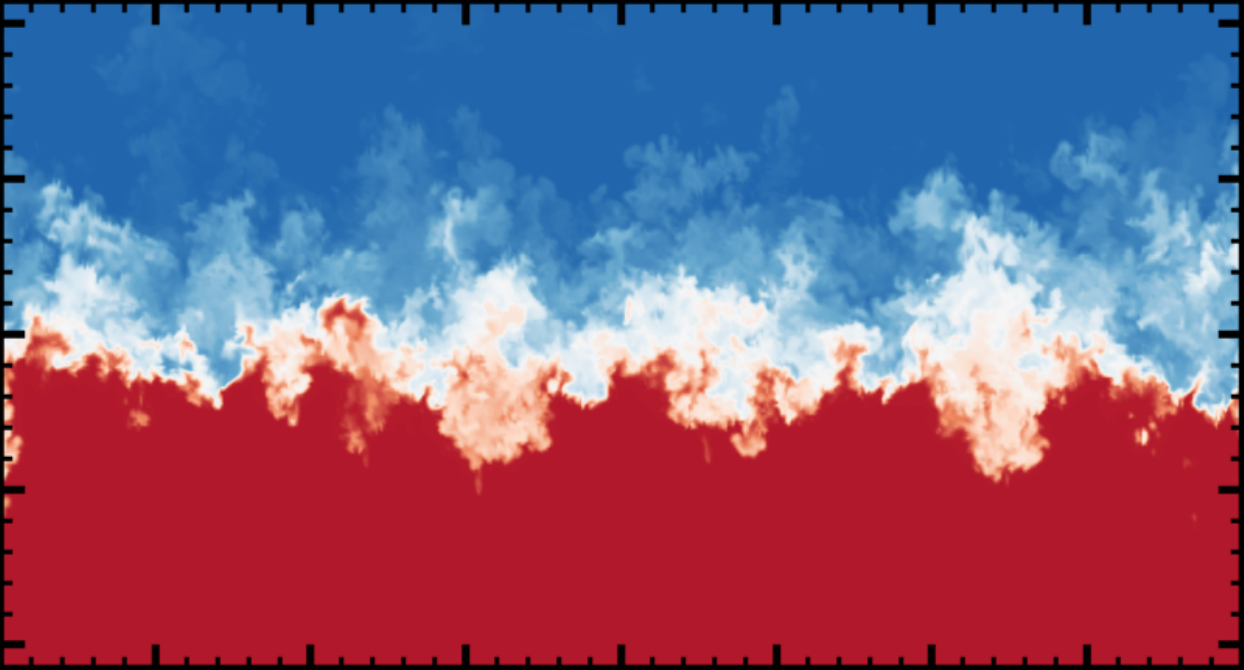}};
             \node[anchor=north west] at (-0.5,3.7) {(\textit{b})};
             \axesLength{1.8} 
         \end{tikzpicture}
    \end{subfigure}
    \hfill
    \begin{subfigure}{0.48\textwidth}
         \begin{tikzpicture}
             \node[anchor=south west] (image) at (0,0) {\includegraphics[width=\textwidth]{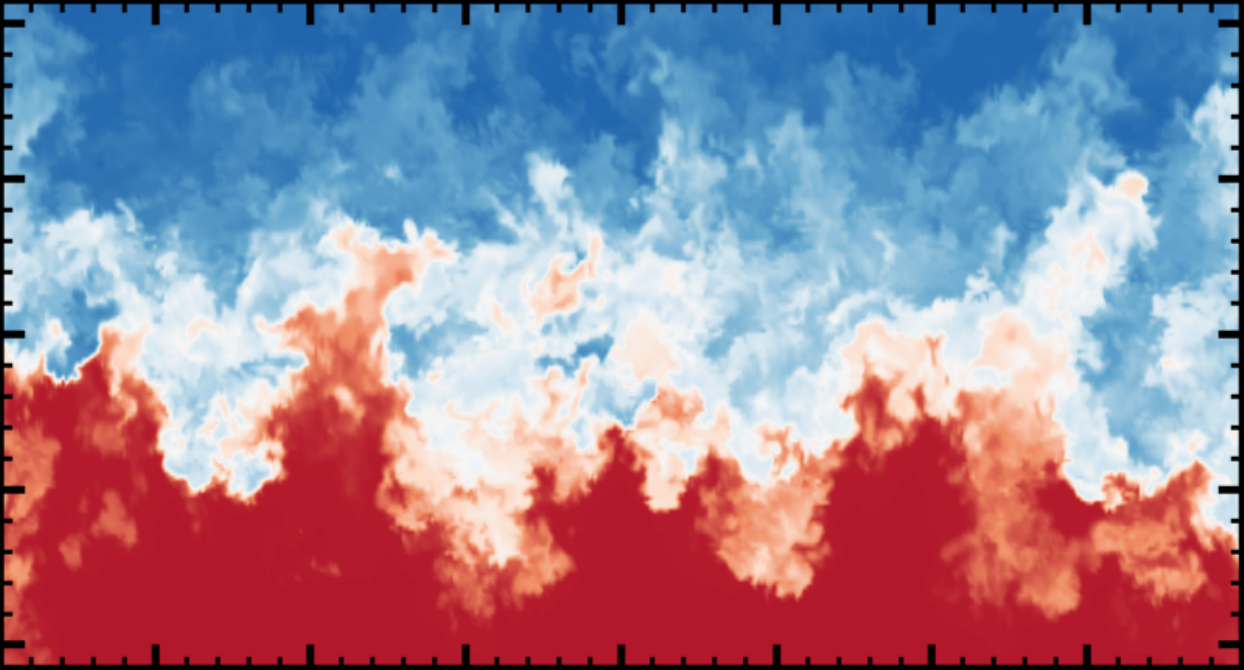}};
             \node[anchor=north west] at (-0.5,3.7) {(\textit{c})};
             \axesLength{1.2} 
         \end{tikzpicture}
    \end{subfigure}
    \begin{subfigure}{0.48\textwidth}
         \begin{tikzpicture}
             \node[anchor=south west] (image) at (0,0) {\includegraphics[width=\textwidth]{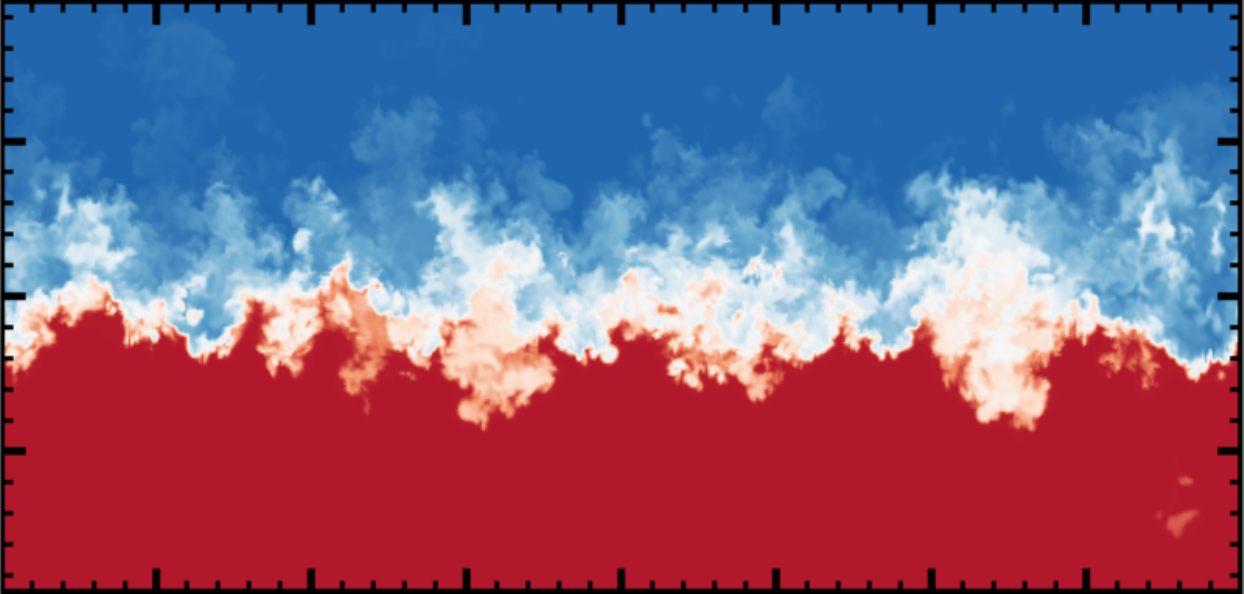}};
             \node[anchor=north west] at (-0.5,3.5) {(\textit{d})};
             \axesLength{2} 
         \end{tikzpicture}
    \end{subfigure}
    \hfill
    \begin{subfigure}{0.48\textwidth}
         \begin{tikzpicture}
             \node[anchor=south west] (image) at (0,0) {\includegraphics[width=\textwidth]{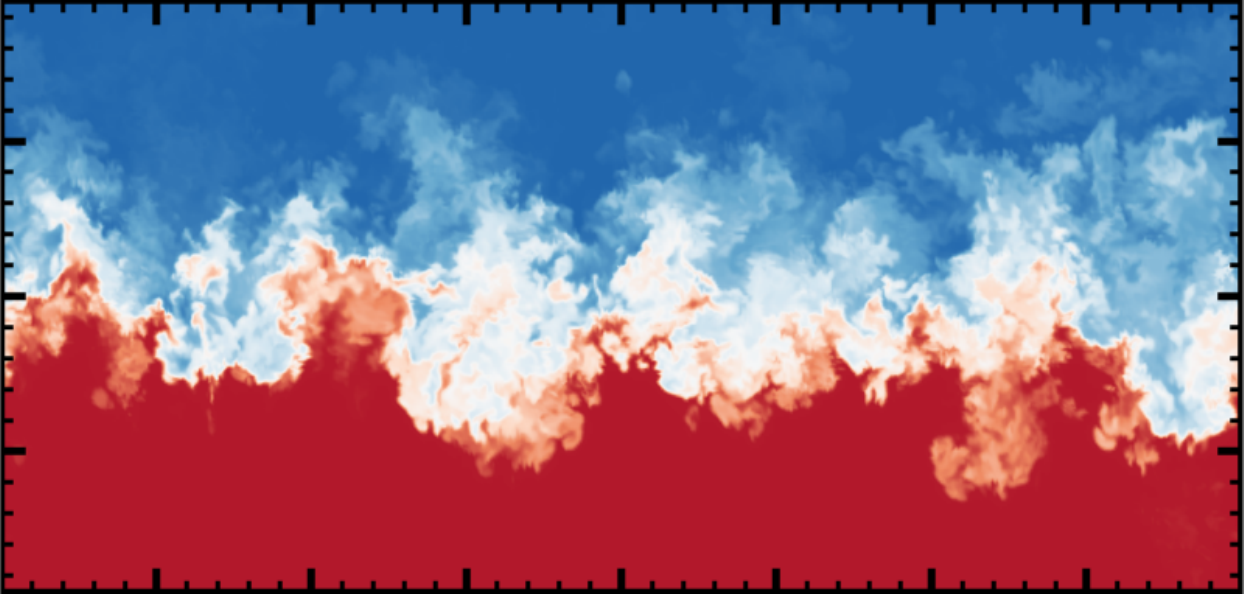}};
             \node[anchor=north west] at (-0.5,3.3) {(\textit{e})};
             \axesLength{2} 
         \end{tikzpicture}
    \end{subfigure}
    \begin{subfigure}{0.48\textwidth}
         \begin{tikzpicture}
             \node[anchor=south west] (image) at (0,0) {\includegraphics[width=\textwidth]{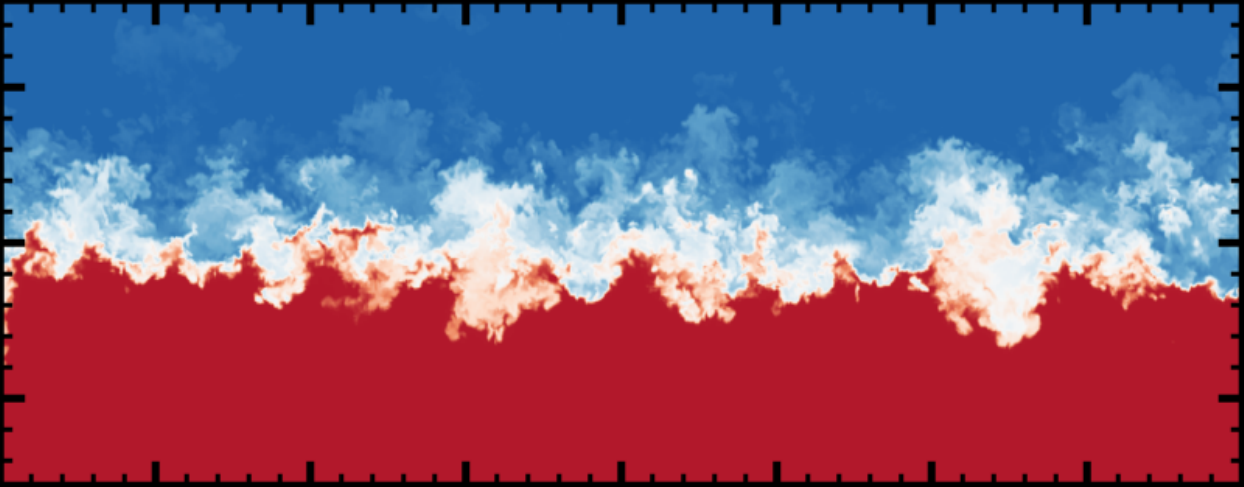}};
             \node[anchor=north west] at (-0.5,2.8) {(\textit{f})};
             \axesLength{2.4} 
         \end{tikzpicture}
    \end{subfigure}
    \hfill
    \begin{subfigure}{0.48\textwidth}
         \begin{tikzpicture}
             \node[anchor=south west] (image) at (0,0) {\includegraphics[width=\textwidth]{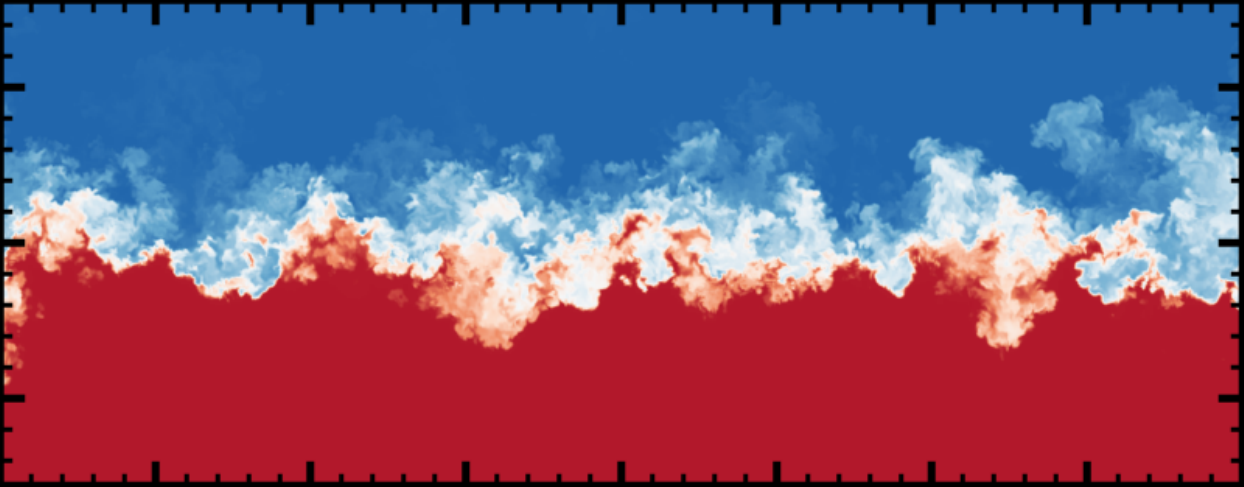}};
             \node[anchor=north west] at (-0.5,2.8) {(\textit{g})};
             \axesLength{3.7} 
         \end{tikzpicture}
    \end{subfigure}
    \begin{subfigure}{\textwidth}
         \begin{tikzpicture}
             \node[anchor=south west] (image) at (0,0) {\includegraphics[width=0.48\textwidth]{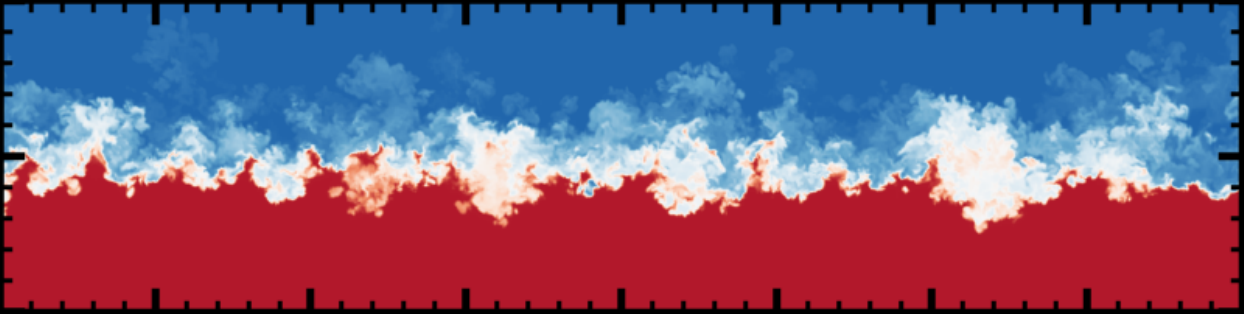}};
             \node[anchor=north west] at (-0.5,1.9) {(\textit{h})};
             \axesLength{3.8} 
         \end{tikzpicture}
    \end{subfigure}
    \caption{Contours of the volume fraction for the constant velocity ILES cases at the centre $x$-$y$ plane ($z=\mathcal{L}(t)/2$): (\textit{a,b,d,f,h}) $\tau = 9.843$; (\textit{c,e,g}) $\tau = 34.451$;  (\textit{a}) $\hat{S}_0=-0.051$; (\textit{b,c}) $\hat{S}_0=-0.013$; (\textit{d,e}) $\hat{S}_0 = 0$; (\textit{f,g}) $\hat{S}_0 = 0.025$; (\textit{h}) $\hat{S}_0 = 0.102$. Heavy fluid ($f_1=1$) is red, light fluid ($f_1=0$) is blue. Major ticks on the axes correspond to $\mathcal{L}(t)/8$;}
    \label{fig:ILES_contours}
\end{figure}
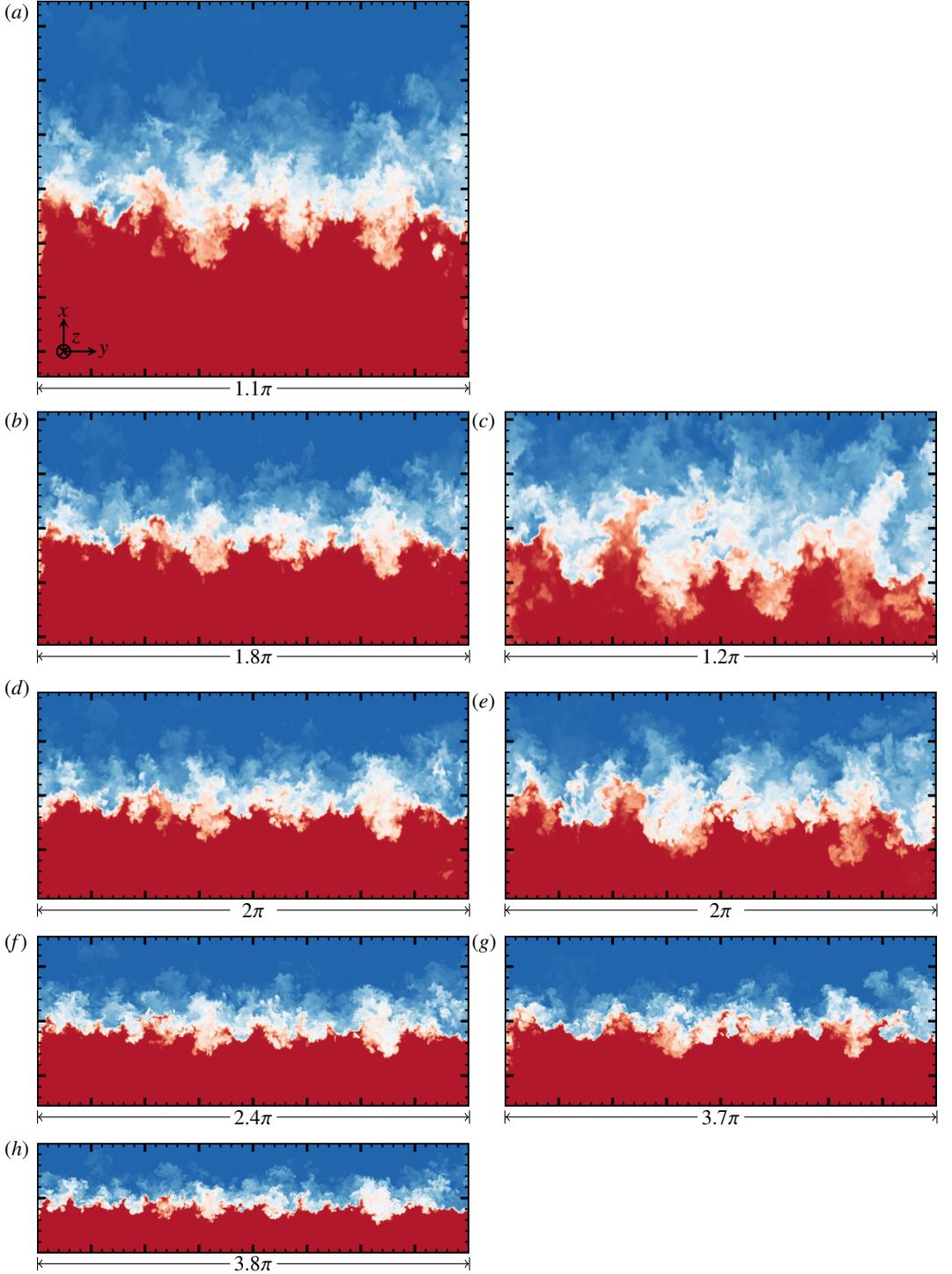

\begin{figure}
    \centering
    \begin{subfigure}{0.40\textwidth}
        \begin{tikzpicture}
            \node[anchor=north west] (image) at (0,0) {\includegraphics[width=\textwidth,trim=0.5cm 0.5cm 0.5cm 0.5cm, clip]{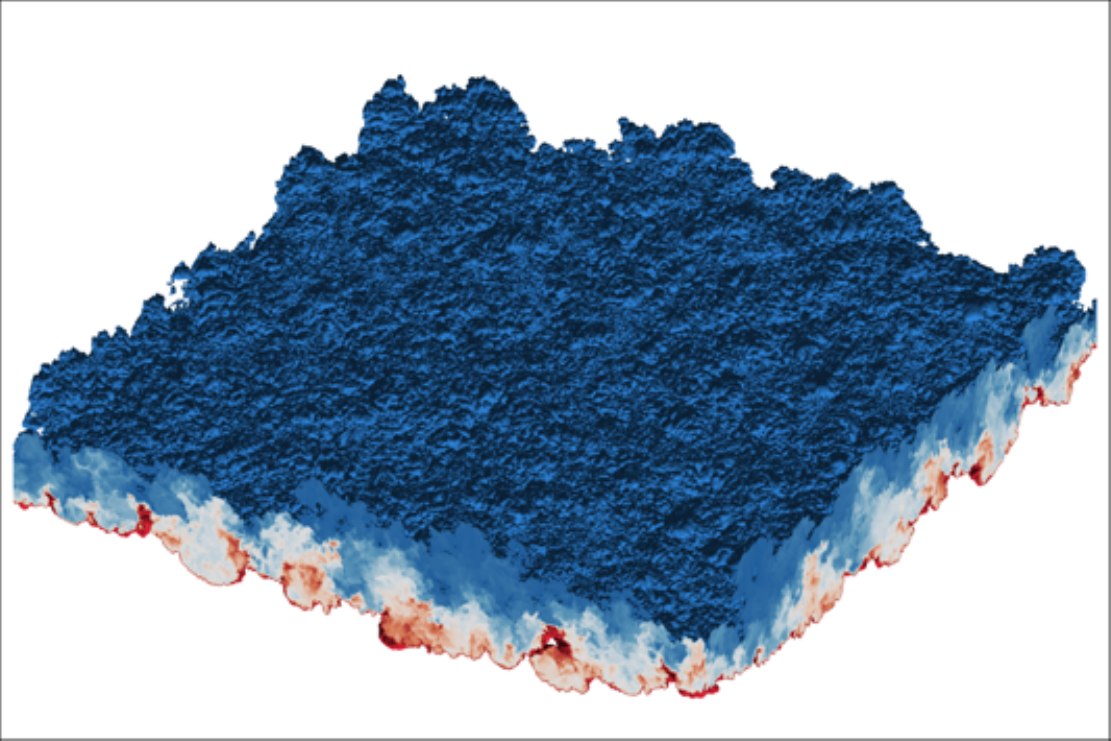}};
            \node[anchor=north west] at (0.2,-0.2) {(\textit{a})};
            \axesIlesIsosurface
        \end{tikzpicture}
    \end{subfigure}
    \hfill
    \begin{subfigure}{0.4\textwidth}
        \centering
        \begin{tikzpicture}
            \node[anchor=north west] (image) at (0,0) {\includegraphics[width=\textwidth,trim=0.5cm 0.5cm 0.5cm 0.5cm, clip]{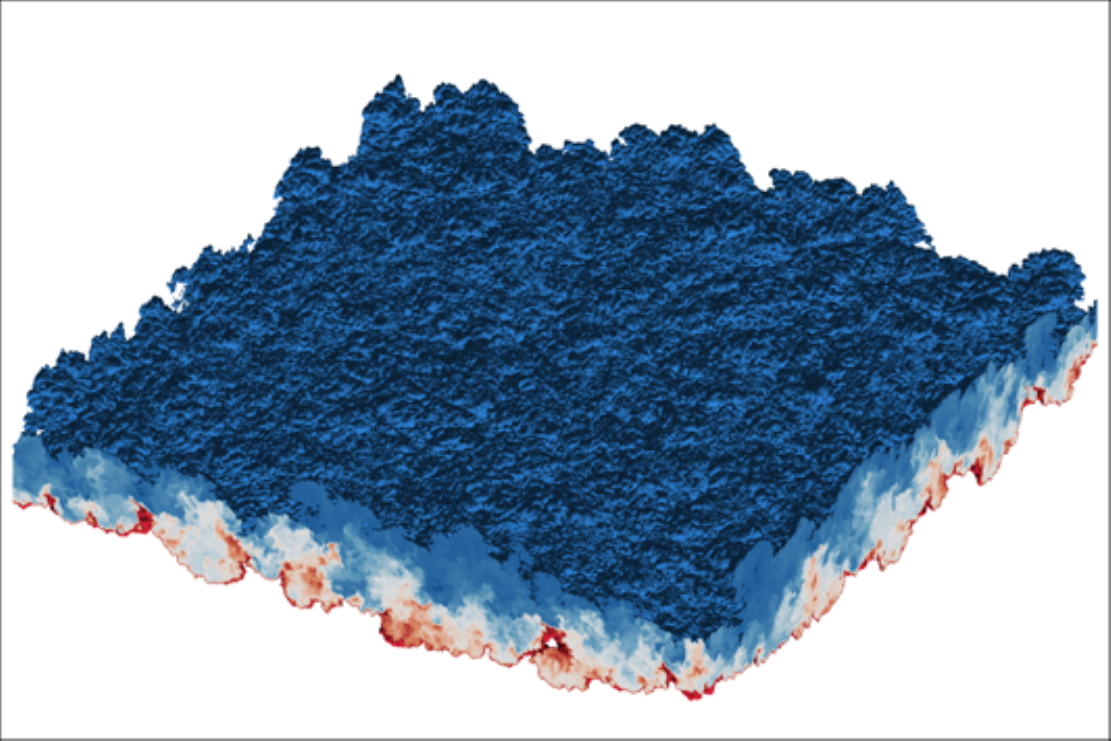}};
            \node[anchor=north west] at (0.2,-0.2) {(\textit{b})};
            \axesIlesIsosurface
        \end{tikzpicture}
    \end{subfigure}
    \hfill
    \begin{subfigure}{0.4\textwidth}
        \centering
        \begin{tikzpicture}
            \node[anchor=north west] (image) at (0,0) {\includegraphics[width=\textwidth,trim=0.5cm 0.1cm 0.5cm 0.5cm, clip]{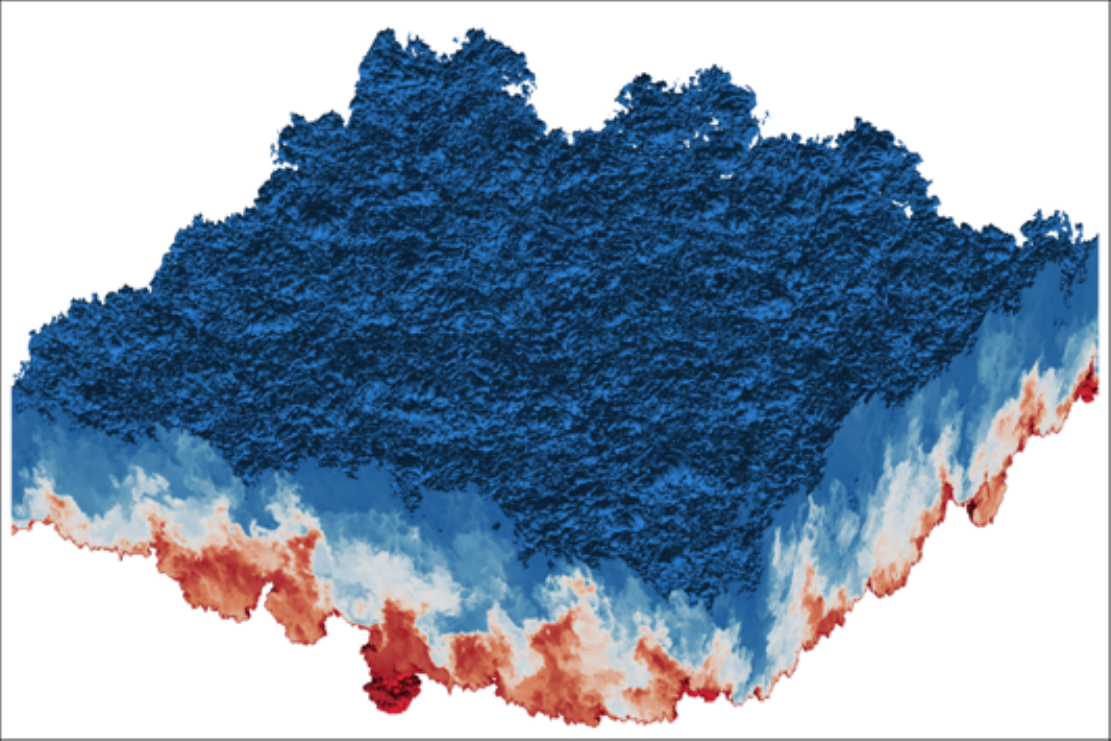}};
            \node[anchor=north west] at (0.2,-0.2) {(\textit{c})};
            \axesIlesIsosurface
        \end{tikzpicture}
    \end{subfigure}
    \hfill
    \begin{subfigure}{0.4\textwidth}
        \centering
        \begin{tikzpicture}
            \node[anchor=north west] (image) at (0,0) {\includegraphics[width=\textwidth,trim=0.5cm 0.1cm 0.5cm 0.5cm, clip]{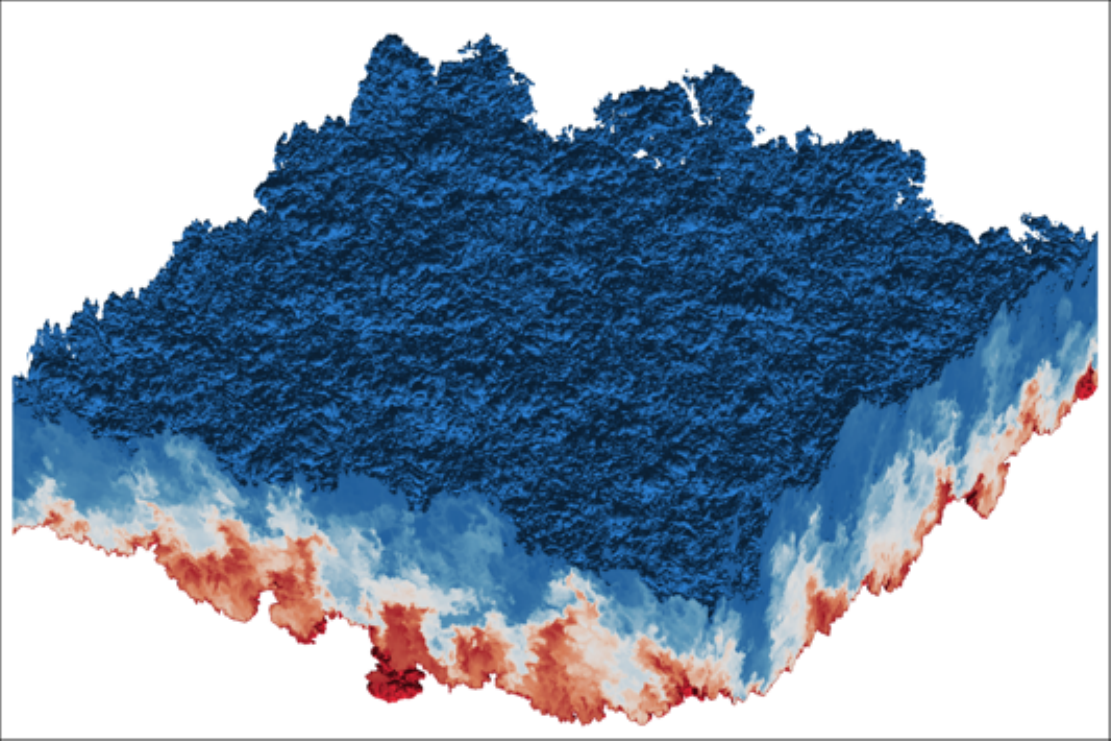}};
            \node[anchor=north west] at (0.2,-0.2) {(\textit{d})};
            \axesIlesIsosurface
        \end{tikzpicture}
    \end{subfigure}
    \caption{Contour of volume fraction $f_1$ for the expansion mixing layers at $\Lambda \approx 1.82$, bounded by $f_1=0.99$ (red) and $f_1=0.01$ (blue). (\textit{a}) $\hat{S}_0=0.102$, $\tau = 9.05$, (\textit{b}) $\hat{S}=0.081$, $\tau = 8.37$, (\textit{c}) $\hat{S}_0=0.025$, $\tau = 33.5$, (\textit{d}) $\hat{S}_0=0.020$, $\tau = 30.5$}
    \label{fig:Isosurface_Expansion}
\end{figure}

\begin{figure}
    \centering
    \begin{subfigure}{0.24\textwidth}
        \begin{tikzpicture}
            \node[anchor=north west] (image) at (0,-0.4) {\includegraphics[width=\textwidth,trim=0.5cm 0.1cm 0.5cm 0.5cm, clip]{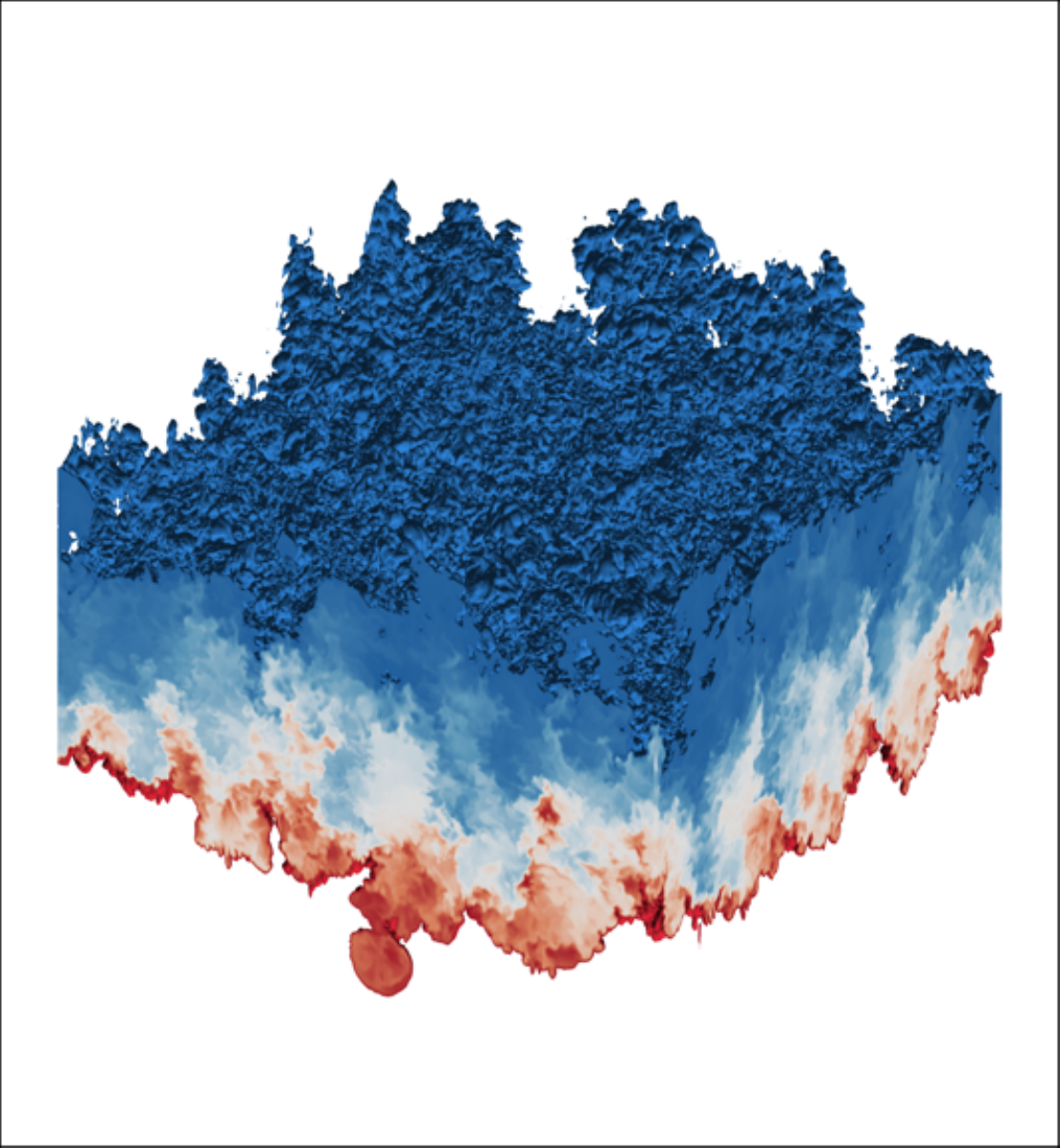}};
            \node[anchor=north west] at (0.2,-0.2) {(\textit{a})};
            \axesIlesIsosurfaceCompression
        \end{tikzpicture}
    \end{subfigure}
    \hfill
    \begin{subfigure}{0.24\textwidth}
        \centering
        \begin{tikzpicture}
            \node[anchor=north west] (image) at (0,-0.4) {\includegraphics[width=\textwidth,trim=0.5cm 0.1cm 0.5cm 0.5cm, clip]{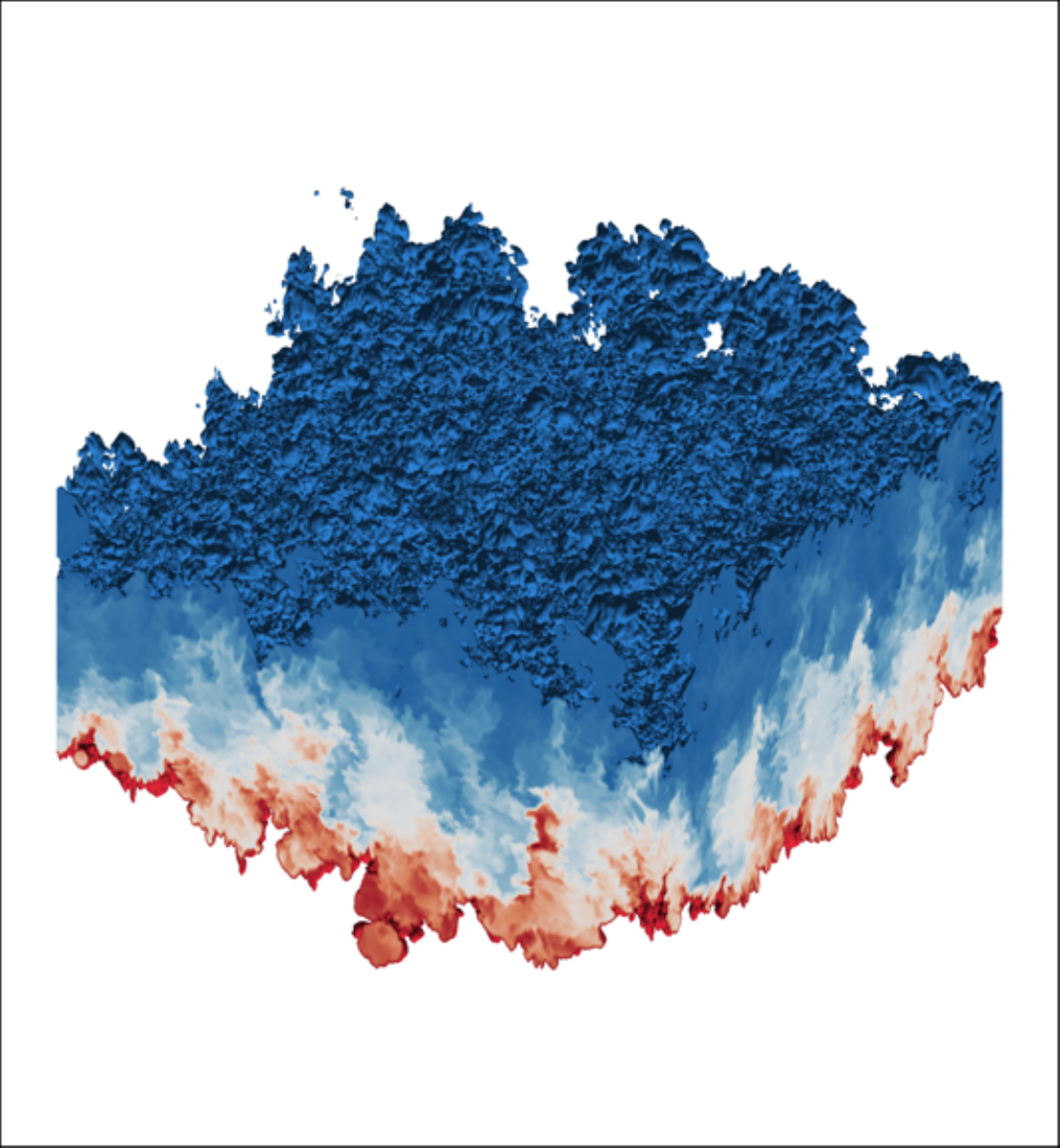}};
            \node[anchor=north west] at (0.2,-0.2) {(\textit{b})};
        \end{tikzpicture}
    \end{subfigure}
    \hfill
    \begin{subfigure}{0.24\textwidth}
        \centering
        \begin{tikzpicture}
            \node[anchor=north west] (image) at (0,-0.4) {\includegraphics[width=\textwidth,trim=0.5cm 0.1cm 0.5cm 0.5cm, clip]{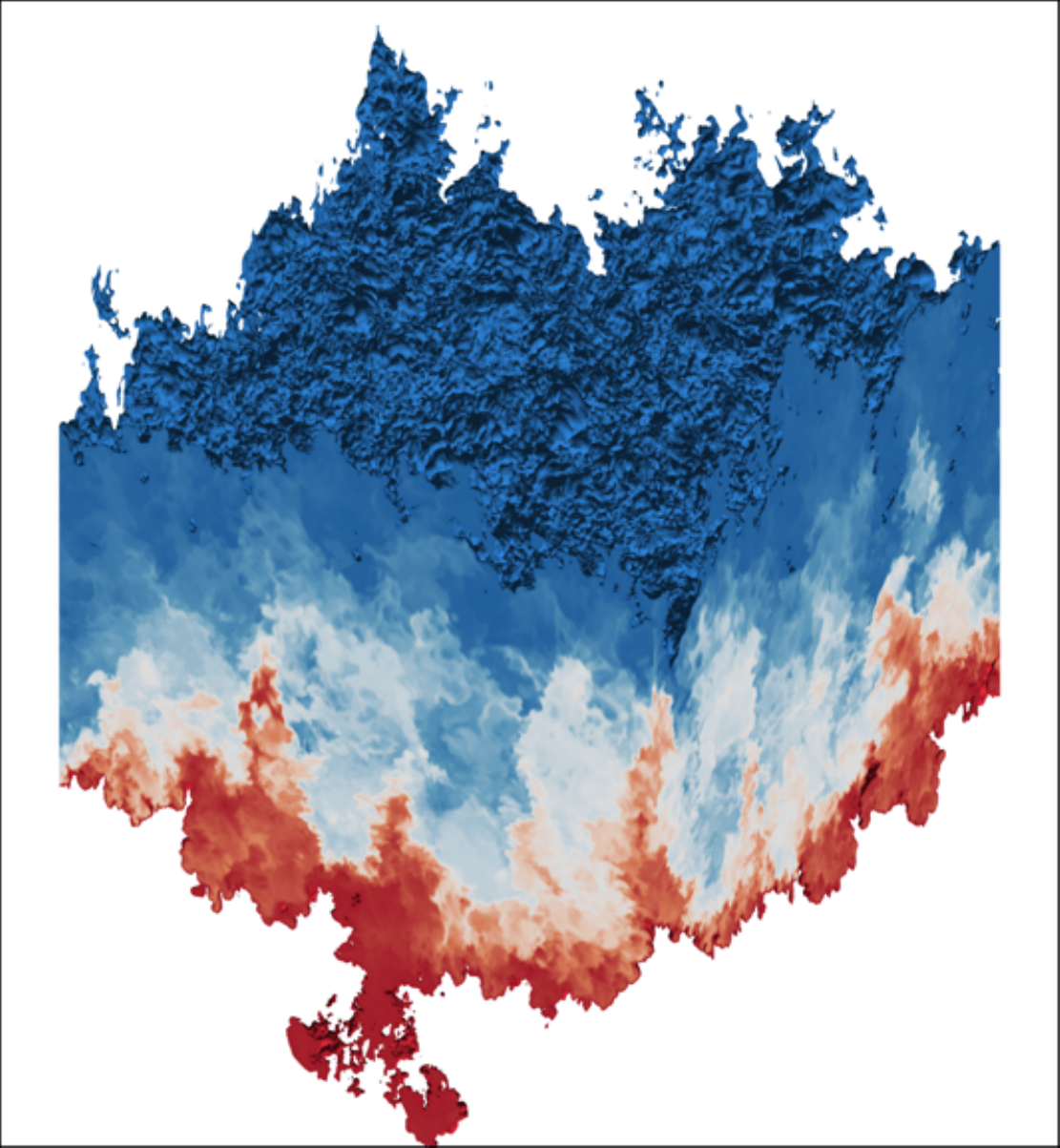}};
            \node[anchor=north west] at (0.2,-0.2) {(\textit{c})};
        \end{tikzpicture}
    \end{subfigure}
    \hfill
    \begin{subfigure}{0.24\textwidth}
        \centering
        \begin{tikzpicture}
            \node[anchor=north west] (image) at (0,-0.4) {\includegraphics[width=\textwidth,trim=0.5cm 0.1cm 0.5cm 0.5cm, clip]{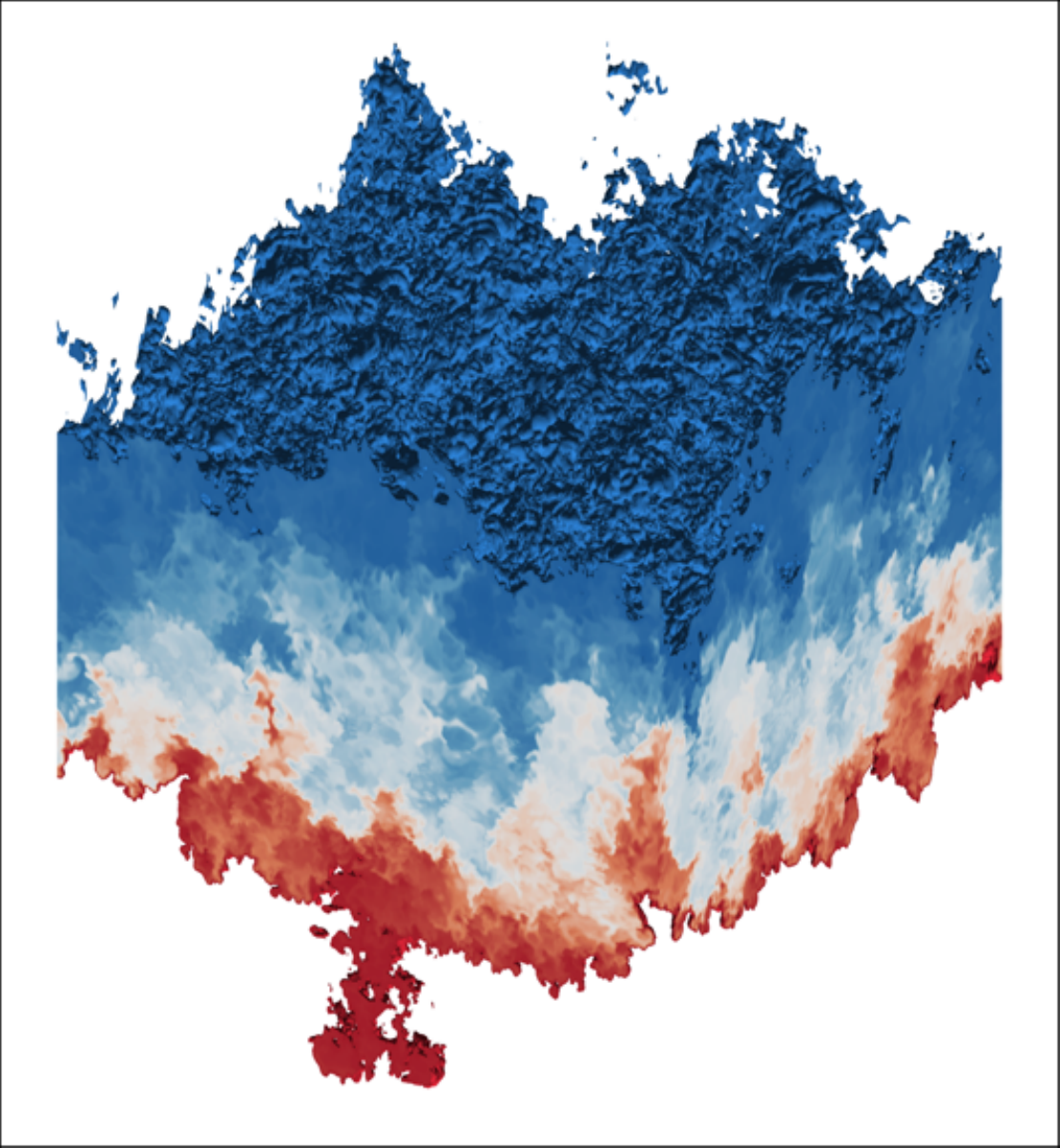}};
            \node[anchor=north west] at (0.2,-0.2) {(\textit{d})};
        \end{tikzpicture}
    \end{subfigure}
    \caption{Contour of volume fraction $f_1$ for the expansion mixing layers at $\Lambda \approx 0.57$, bounded by $f_1=0.99$ (red) and $f_1=0.01$ (blue). (\textit{a}) $\hat{S}_0=-0.051$, $\tau = 9.45$, (\textit{b}) $\hat{S}=-0.081$, $\tau = 7.88$, (\textit{c}) $\hat{S}_0=-0.013$, $\tau = 34.9$, (\textit{d}) $\hat{S}_0=-0.020$, $\tau = 28.5$}
    \label{fig:Isosurface_Compression}
\end{figure}

Isosurfaces of the volume fraction at $f_1=0.01$ and $f_1=0.99$ are shown in figure \ref{fig:Isosurface_Expansion} for the expansion cases at $\Lambda \approx 1.82$, and in figure \ref{fig:Isosurface_Compression} for the compression cases at $\Lambda\approx0.57$. The scale used in the expansion plots is different to the scale used for the compression plots, so a visual comparison of the mixing layer thickness between expansion and compression cases is not an accurate representation. For the lower magnitude strain rate cases, the specified expansion factor is achieved at a later non-dimensional time. This is evident from the increased mixing layer thickness observable for the (c) and (d) subplots compared to (a) and (b). These later time plots also show the penetration of a vortex ring into the heavy fluid. For the expansion cases, the structure remains intact, but for the compression cases the structure has bifurcated. This may be the result of vortex tilting and the vortex ring being pinched by the compressive strain. Vortex pinching has been investigated by \cite{Marshall_1994_EvolutionBreakupVortex} for planar strain configurations, where expansive and compressive strain rates are applied in perpendicular directions in the plane of the vortex ring. For sufficiently high strain rates this can cause the vortex ring to deform into an elliptical shape and curve into the third dimension, eventually pinching at the centre and producing two smaller vortex rings.

\subsubsection{Width and mix measures}

\begin{figure}
    \FigureAndLabel{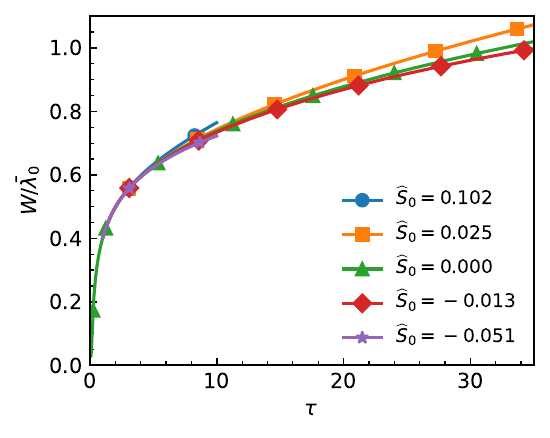}{a}{0.48}{1}
    \hfill
    \FigureAndLabel{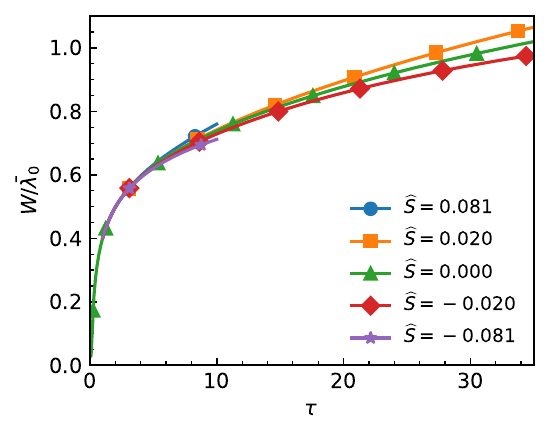}{b}{0.48}{1}
    \caption{Integral width for (\textit{a}) constant velocity, and (\textit{b}) constant strain rate.}
    \label{fig:ILES_IW}
\end{figure}

The integral width of the mixing layer normalised by the initial mean wavelength of the interface is plotted in figure \ref{fig:ILES_IW}. The cases with applied expansion transverse strain rates show a slight increase in the integral width, whilst the compression cases show a decrease. This behaviour is the opposite of what was observed in the linear regime where transverse compression increased the growth and was well captured by the Bell-Plesset model. \cite{ElRafei_2019_ThreedimensionalSimulationsTurbulent} also noted that once the modes begin to saturate, Bell-Plesset models fail to accurately predict the growth of the mixing layer for the spherical implosion. This observed trend suggests that the influence of the transverse strain rate on the transitional and turbulent mixing layer is fundamentally different from the linear regime and can not be modelled by the same approach. These results are for a narrowband perturbation spectrum, whilst a broadband initialisation will likely show different behaviour. With a broadband spectrum, the largest modes would not saturate until a later time and would remain governed by the linear regime. Whilst the high wave-number modes will transition to a turbulent state, the growth rate of the larger modes will be amplified, causing an increased mixing layer growth rate. It is also important to note that the integral widths of transverse strain rate cases are much closer to the unstrained simulation than was observed for cases with axial strain rates in \cite{Pascoe2024}, where the strain rate causes the mixing layer to stretch/compress directly from the background velocity difference. This is a similar observation to the one made by \cite{Lombardini_2014_TurbulentMixing1} which noted the compression effects were larger than the convergence effects for the analysed implosion profile.

\begin{figure}
    \FigureAndLabel{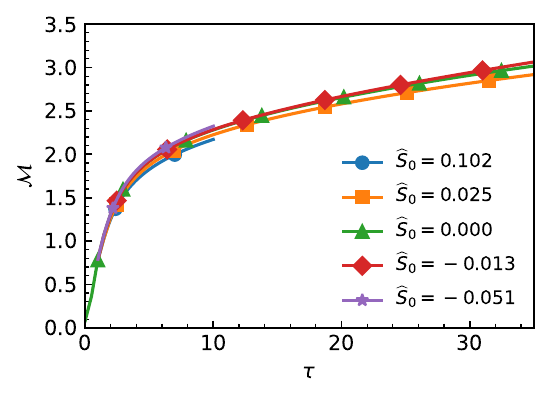}{a}{0.48}{1}
    \hfill
    \FigureAndLabel{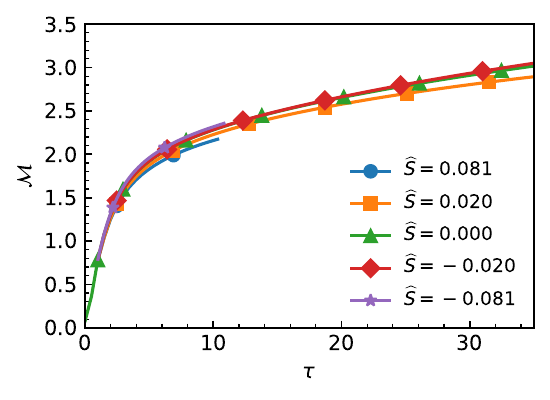}{b}{0.48}{1}
    \caption{Mixed mass for (\textit{a}) constant velocity and (\textit{b}) constant strain rate.}
    \label{fig:ILES_mixed_mass}
\end{figure}

The mixed mass is an alternate measure of the mixing layer, and measures how much of one fluid has mixed with another. An attractive feature of the mixed mass is the ability to derive the evolution equation for the mixed mass from the mass fraction transport equation \citep{Zhou_2016_AsymptoticBehaviorMixed}:
\begin{align}
    \mathcal{M} = \int 4 \rho Y_1 Y_2 dV
\end{align}
The profiles of the mixed mass are plotted in figure \ref{fig:ILES_mixed_mass}, and show a different trend compared to the integral width. For the mixed mass, the compression cases achieve slightly higher growth and the expansion cases achieve less growth. As the name mixed mass suggests, it is not purely a measure of the width of the mixing layer but also has dependence upon mixedness of the mixing layer, whereas the integral width depends on the mean volume fraction profile. Therefore, depending upon the choice of measurement metric for the mixing layer development, the influence of the transverse strain rate will change signs. The increased mixedness of the compression cases suggests a more mixed or homogeneous composition in the mixing layer as compared to the unstrained, and likewise a less mixed layer for the expansion cases. The compression cases show a smaller deviation from the unstrained case compared to the expansion cases. As the change in mixedness is opposing the change in mixing layer width for the mixed mass results, it is not surprising that the deviation is larger for the expansion cases. The expansion cases are becoming less mixed which is a simpler task compared to making the mixing layer even more mixed for the compression cases. The mixedness of the mixing layer is further explored using the molecular mixing fraction below.

\begin{figure}
    \FigureAndLabel{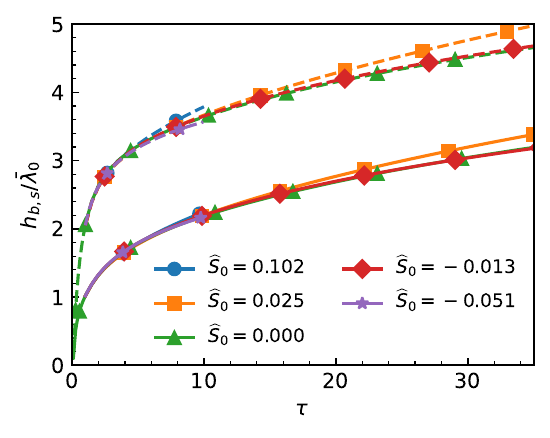}{a}{0.48}{1}
    \hfill
    \FigureAndLabel{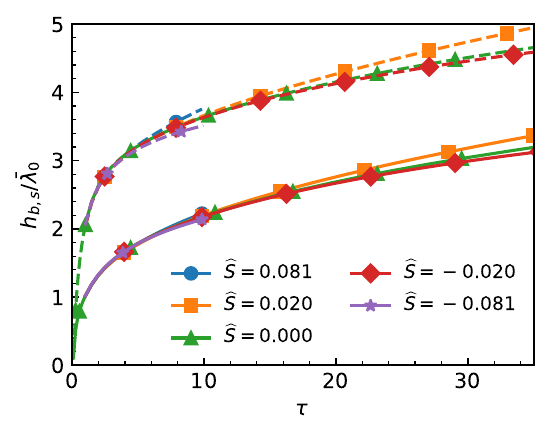}{b}{0.48}{1}
    \caption{Bubble and spike heights for (\textit{a}) constant velocity and (\textit{b}) constant strain rate. Solid lines indicate bubble height, dashed lines indicate spike height.}
    \label{fig:ILES_bubblespike}
\end{figure}

The bubble and spike heights of the mixing layer are defined relative to the mixing layer centre $x_C$, taken to be the planar location where there is an equal volume of penetrating fluid on either side:
\begin{align}
    \int_{-\infty}^{x_C} \bar{f}_2 dx = \int_{x_C}^{\infty} \bar{f}_1 dx
\end{align}
To reduce the impact of statistical fluctuations, the bubble and spike heights used are based off the integral measure proposed by \cite{Youngs_2020_BuoyancyDragModelling}:
\begin{subeqnarray}
    \bar{h}_b^{(m)} &=& \left[\frac{(m+1)(m+2)}{2} \frac{\int^0_{-\infty} |x'|^m (1-\bar{f}_1) dx'}{\int^0_{-\infty} (1-\bar{f}_1) dx'}\right]^{1/m} ,\\
    \bar{h}_s^{(m)} &=& \left[\frac{(m+1)(m+2)}{2} \frac{\int_0^{\infty} |x'|^m \bar{f}_1 dx'}{\int_0^{\infty} \bar{f}_1 dx'}\right]^{1/m}.
\end{subeqnarray}
The integrals are taken with respect to the mixing layer centre, $x' = x-x_C$, and the integral terms of the denominator are equal to the volume of the penetrating fluid on either side of the mixing layer centre. These definitions provided assume that fluid 1 is located below fluid 2 in the $x$-direction and that $\rho_1>\rho_2$. The bubble and spike heights plotted in figure \ref{fig:ILES_bubblespike} use the heights $h = 1.1\bar{h}^{(2)}$ which was found to be well aligned with the heights measured by the $1\%$ and $99\%$ mean volume-fraction cut-off but are less sensitive to statistical fluctuations \citep{Youngs_2020_BuoyancyDragModelling}.

\begin{figure}
    \FigureAndLabel{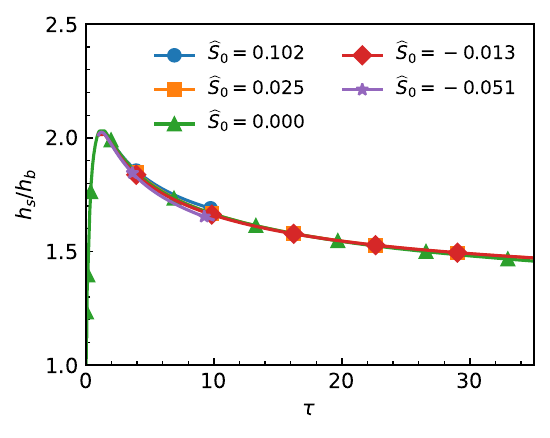}{a}{0.48}{1}
    \hfill
    \FigureAndLabel{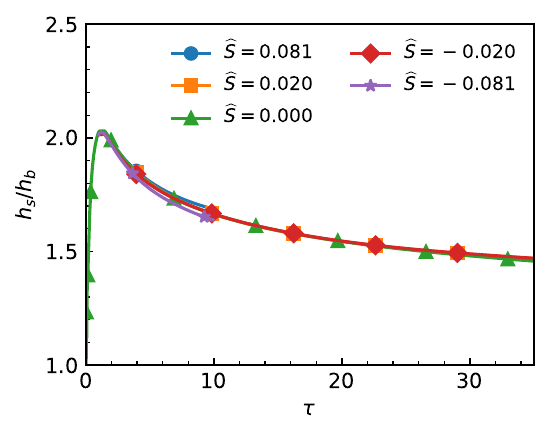}{b}{0.48}{1}
    \caption{Ratio of spike-to-bubble height for (\textit{a}) constant velocity and (\textit{b}) constant strain rate.}
    \label{fig:ILES_ratioBubblespike}
\end{figure}

The bubble and spike heights show the same behaviour as observed for the integral width, with the expansion cases growing slightly compared to the unstrained simulation. The compression cases tend to have smaller heights than the unstrained cases, however the influence of the transverse compression appears to be smaller, such that the results are closer to the unstrained simulation results. Despite this, the ratio of the spike height to bubble height shows a common trend with the unstrained case, as shown in figure \ref{fig:ILES_ratioBubblespike}. The effect of convergence doesn't appear to change the growth rate of the bubble and spike heights in a disproportionate manner, suggesting a common self-similar ratio between the heights may be obtained as seen for unstrained simulations. 

\begin{figure}
    \FigureAndLabel{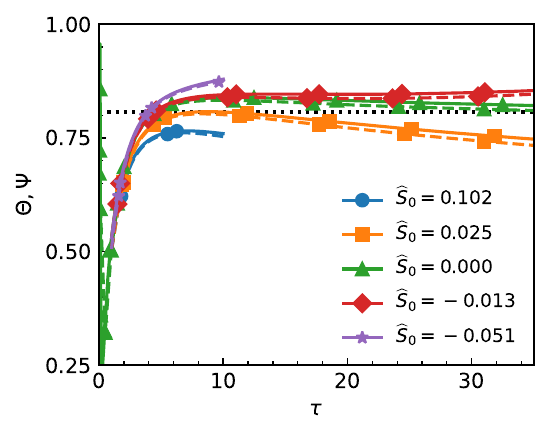}{a}{0.48}{1}
    \hfill
    \FigureAndLabel{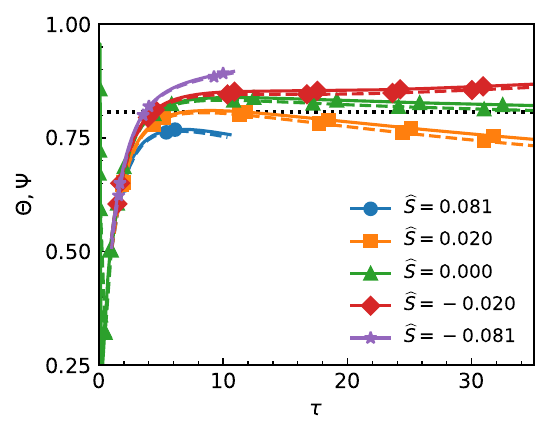}{b}{0.48}{1}
    \caption{Mixing measures for the (\textit{a}) constant velocity and (\textit{b}) constant strain rate. Solid lines indicate $\Theta$, dashed lines indicate $\Psi$, dotted line is \textsf{FLAMENCO}'s final $\Theta$ value at $\tau=246$ \citep{ThetaGroup}.}
    \label{fig:ILES_Theta}
\end{figure}

To measure the mixedness of the mixing layer, the molecular mixing fraction is used. The molecular mixing fraction measures how well the species in the mixing layer are mixed, as measured by the volume fraction. A value of $\Theta=0$ suggests complete segregation, whilst $\Theta=1$ suggests perfect homogeneity in the plane. The molecular mixing fraction is calculated by
\begin{align}
    \Theta (t) = \frac{\int \overline{f_1 f_2}dx}{\int \bar{f}_1 \bar{f}_2 dx}.
\end{align}
At late-time, a steady $\Theta$ indicates that a mixing layer has become self-similar. Included in figure \ref{fig:ILES_Theta}, is the final value of $\Theta$ from the quarter-scale case using \textsf{FLAMENCO} in the $\theta$-group collaboration at a time of $\tau=246$. The unstrained quarter-scale case simulation can be observed to be approaching the self-similar value marked by the black, dashed line. The strained cases do not appear to be approaching the same asymptote. Instead the compression cases show an increasing mixedness and the expansion cases are decreasing.

This is the same trend as observed when axial strain rates are applied to the mixing layer, such that a decrease in turbulent growth of the mixing layer corresponds to an increase in the mixedness. The difference between the transverse and axial strain rates is that the decreased turbulent growth occurs for compressive transverse strain rates, whilst the decreased turbulent growth occurs for expansive axial strain rates \citep{Pascoe2024}. The compressive transverse strain rates can be expected to enhance the turbulence in the transverse direction through shear-production, which will allow for greater mixing as shown in the results. This clearly affects the self-similarity of the simulation, potentially converging to a different self-similar state, or not converging at all. The normalised mixed mass is also plotted in figure \ref{fig:ILES_Theta}, using the definition
\begin{align}
    \Psi = \frac{\int \rho Y_1 Y_2 dV}{\int \bar{\rho}\bar{Y}_1\bar{Y}_2 dV}.
\end{align}
The results for $\Theta$ and $\Psi$ are well aligned, with the values of $\Psi$ attaining slightly smaller values compared to $\Theta$. This behaviour has previously been observed in several other studies \citep{Zhou_2016_AsymptoticBehaviorMixed,Zhou_2020_DependenceEnstrophyTransport,ElRafei_2020_NumericalStudyBuoyancy}.

\subsubsection{Self-similarity}
To further investigate the self-similarity of the simulation, the two contributions of the molecular mixing fraction may be analysed: the mean volume fraction profile $\bar{f}_1$, and the mean volume fraction product $\overline{f_1 f_2}$. These terms correspond to the denominator and numerator of the molecular mixing fraction, and for a self-similar mixing layer these profiles will collapse under non-dimensionalisation. 

\begin{figure}
    \FigureAndLabel{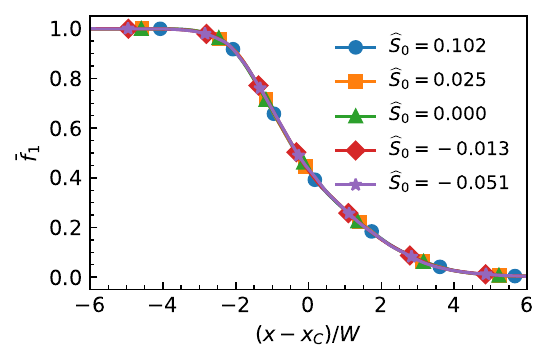}{a}{0.48}{1}
    \hfill
    \FigureAndLabel{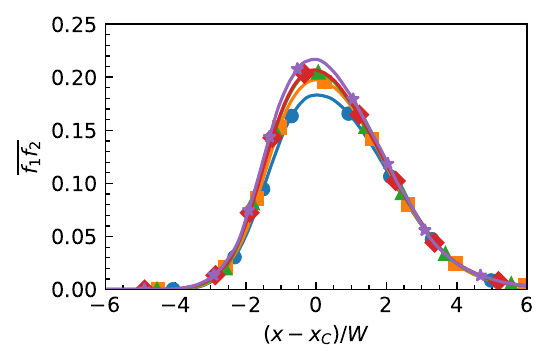}{b}{0.48}{1}
    \hfill
    \FigureAndLabel{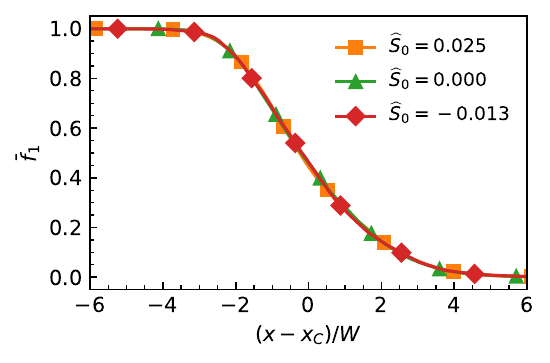}{c}{0.48}{1}
    \hfill
    \FigureAndLabel{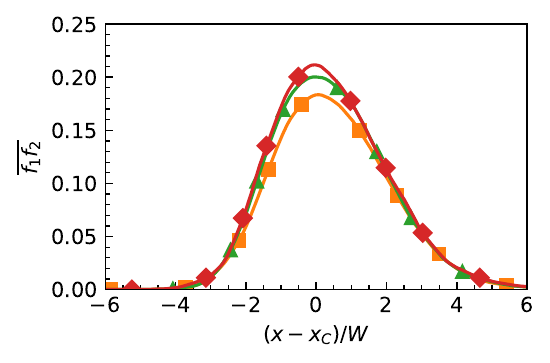}{d}{0.48}{1}
    \caption{Planar-averaged volume-fraction profiles for the constant velocity cases: (\textit{a,b}) $\tau=9.843$; (\textit{c,d}) $\tau=34.451$}
    \label{fig:SelfSimilar_CV}
\end{figure}

\begin{figure}
    \FigureAndLabel{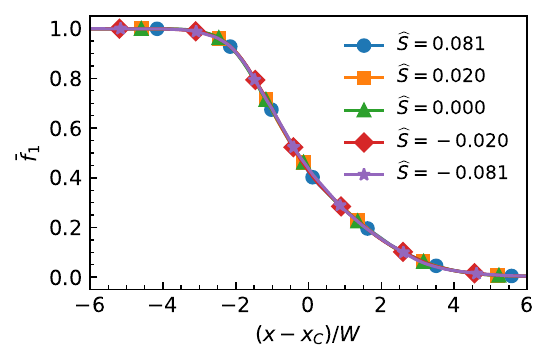}{a}{0.48}{1}
    \hfill
    \FigureAndLabel{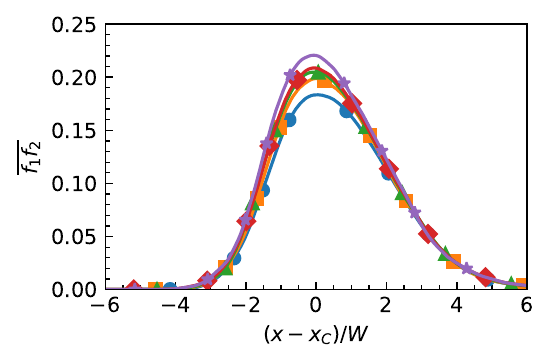}{b}{0.48}{1}
    \hfill
    \FigureAndLabel{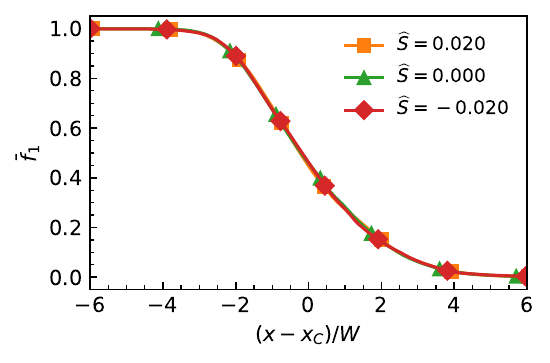}{c}{0.48}{1}
    \hfill
    \FigureAndLabel{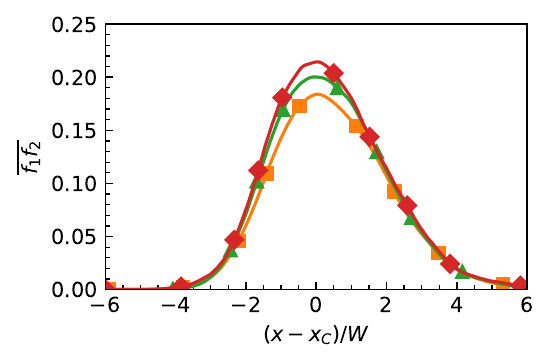}{d}{0.48}{1}
    \caption{Planar-averaged volume-fraction profiles for the constant strain rate cases: (\textit{a,b}) $\tau=9.843$; (\textit{c,d}) $\tau=34.451$}
    \label{fig:SelfSimilar_CS}
\end{figure}

The profiles of $\bar{f}_1$ and $\overline{f_1 f_2}$ are plotted in figure \ref{fig:SelfSimilar_CV} for the constant velocity cases, and in figure \ref{fig:SelfSimilar_CS} for the constant strain rate cases. The profiles are shown at two different times, corresponding to the end times of the high strain rate cases, $\tau = 9.843$, and the end time of the low strain rate cases, $\tau=34.451$. All cases are visible for the early time, whilst at the late time only information for the low strain rate cases are available.
The mean profile of $f_1$ is plotted in the sub-figures (a) and (c). The mean volume fraction profile does not appear to vary from the unstrained solution for both strain profiles and at both times shown. The evolution of the mean volume fraction profile for the quarter-scale $\theta$-group case almost collapsed to a single profile, however the profile slightly smoothed out with time around the inflection points at $(x-x_C)/W=-2$ and $2.5$. The agreement between the strain cases and the unstrained case is suggestive that the bubble and spike heights grow in the same proportion as in the unstrained case, reinforcing the results in figure \ref{fig:ILES_ratioBubblespike}.

The mean volume fraction product shown in subplots (b) and (d) of figures \ref{fig:SelfSimilar_CV} and \ref{fig:SelfSimilar_CS} show a larger deviation from the unstrained case. Whilst the unstrained quarter-scale $\theta$-group case was self-similar in the centre of the mixing layer for the mean volume fraction product, the strained cases do not match the profile, explaining the different $\Theta$ values obtained. The compression cases which possessed the largest $\Theta$ values have a larger peak value of $\overline{f_1 f_2}$, representing greater mixing at the centre of the mixing layer. This is due to the shear-production increasing the mixing, with the larger magnitude strain rates showing larger changes in the mean product. The low-magnitude strain rates can be observed to deviate further away from the unstrained profile as the simulation progresses, showing that the strained cases have not achieved a self-similar state.


\subsubsection{Turbulent kinetic energy}
\label{sec:iles_tke}

Under transverse strain rates, the shear-production contribution to the turbulent kinetic energy clearly affects the mixedness of the mixing layer. The total turbulent kinetic energy in the domain is calculated by
\begin{align}
    TKE = \iiint \frac{1}{2} \rho u_i'' u_i''\, \mathrm{d}x\,\mathrm{d}y\,\mathrm{d}z,
\end{align}
where the Favre fluctuations $u_i''$ are calculated from the flow field with the mean velocity gradients removed. To compare to the simulation results of the $TKE$, it is possible to define a simple model. As the quantity has been integrated over the domain, fluxes may be neglected and the model consists of shear production and dissipation,
\begin{align}
    \frac{\mathrm{d} TKE}{\mathrm{d}t} = \mathcal{P} - \varepsilon
\end{align}
It has also been assumed that there is negligible buoyancy production at this stage, and for simplicity it is assumed that the pressure-rate-of-strain tensor has no net effect on the turbulent kinetic energy, and only redistributes energy. Analysis of the turbulent kinetic energy budgets shows that the pressure-rate-of-strain has a net contribution to the turbulent kinetic energy for the unstrained case, however the significance of this term decreases with time \citep{Thornber_2019_TurbulentTransportMixing}. A compressible model for the pressure-dilation will include contributions for the total turbulent kinetic energy, whereas incompressible models do not. The compressible model of \cite{Sarkar_1992_PressureDilatationCorrelation} includes pressure-dilation contributions which are weighted by the turbulent Mach number, however the mean turbulent Mach number in the strained simulations is below 0.03, calculated using the planar-averaged turbulent kinetic energy. Using dimensional analysis, the dissipation rate is taken to be on the order of $\sim u^3/l$ \citep{Thornber_2010_InfluenceInitialConditions,Groom_2020_InfluenceInitialPerturbation}, 
\begin{align}
    \varepsilon = C_\epsilon \frac{TKE^{3/2}}{W\sqrt{M}} \label{eqn:base_dissipation}
\end{align}
where the integral width $W$ is interpolated from the simulation data and used as the lengthscale, and $M$ is the mass within the mixing layer, $M = 4\pi^2 \bar{\rho}_0 W$. An additional coefficient of $C_\epsilon = 1/140$ is included to calibrate the dissipation rate to match the unstrained simulation. To accurately calculate the shear production contributions it is necessary to also evolve the turbulent kinetic energy components for each direction, given by
\begin{eqnarray}
    TKX = \iiint \frac{1}{2} \rho u_1'' u_1'' \,\mathrm{d}x\,\mathrm{d}y\,\mathrm{d}z, \qquad&\qquad
    TKY = \iiint \frac{1}{2} \rho u_2'' u_2'' \,\mathrm{d}x\,\mathrm{d}y\,\mathrm{d}z.
\end{eqnarray}
Assuming homogeneity in the $y$- and $z$- directions allows for the simplification of $TKY=TKZ$, and by definition the total turbulent kinetic energy will be equal to $TKE=TKX+2TKY$. The model for the turbulent kinetic energy components is based upon Reynolds stress transport equations composing of production $(\mathcal{P})$, pressure-rate-of-strain tensor $(\mathcal{R})$, and isotropic dissipation where each direction uses a third of the total dissipation rate specified in equation (\ref{eqn:base_dissipation}),
\begin{align}
    \frac{d}{dt} \begin{pmatrix}
        TKX \\ TKY
    \end{pmatrix} = 
    \begin{pmatrix}
        0 \\ \mathcal{P}_{22}
    \end{pmatrix} + 
    \begin{pmatrix}
        \mathcal{R}_{11} \\ \mathcal{R}_{22}
    \end{pmatrix}
    - \frac{1}{3}\varepsilon.
\end{align}
The shear production only acts in the transverse direction due to the applied transverse strain rate and is calculated by $\mathcal{P}_{22}=2 (TKY) \bar{S}$. The total production for the $TKE$ is the sum of the production terms for $y$ and $z$, resulting in $\mathcal{P} = 2\mathcal{P}_{22}$.
The pressure-rate-of-strain tensor is calculated using the LRR-IP model by \cite{Launder_1975_ProgressDevelopmentReynoldsstress}, which is a combination of the return-to-isotropy model by \cite{Rotta_1951_StatistischeTheorieNichthomogener} and the isotropisation of production model by \cite{Naot_1970_InteractionsComponentsTurbulent}. The LRR-IP model is commonly used for Reynolds stress transport models  for compressible turbulent mixing \citep{Gregoire_2005_SecondorderTurbulenceModel,Schwarzkopf_2011_ApplicationSecondmomentClosure}. Adjusting the model to use the domain integrals instead of the Reynolds stress terms gives the expressions:
\begin{subeqnarray}
    \mathcal{R}_{11} &=& -C_R \frac{\varepsilon}{TKE} \left(TKX - \frac{1}{3} TKE \right) - C_2 \left( -\frac{1}{3} \mathcal{P}\right), \\
    \mathcal{R}_{22} &=& -C_R \frac{\varepsilon}{TKE} \left(TKY - \frac{1}{3} TKE \right) - C_2 \left( \mathcal{P}_{22} - \frac{1}{3} \mathcal{P}\right).
\end{subeqnarray}
The $C_R$ term is responsible for the return to isotropy and is active for the unstrained and strained cases, whilst the $C_2$ term is for the isotropisation of production and will only be active for the strain cases. Whilst \cite{Launder_1996_IntroductionSinglePointClosure} suggests a value of $C_R=1.8$, RMI-induced mixing layers remain anisotropic, so the value was modified to improve the agreement with the unstrained simulation, using a value of unity here in order to accurately predict the $TKX$ and $TKY$ components. With a value of $C_R=1$, the unstrained system will maintain the initial anisotropic distribution of turbulent kinetic energy across the three directions. To align with the observed relationship between $C_R$ and $C_2$ in \cite{Launder_1996_IntroductionSinglePointClosure}, the value of $C_2$ was set to 0.77 which should allow the model to still describe free shear flows. 

\begin{figure}
    \FigureAndLabel{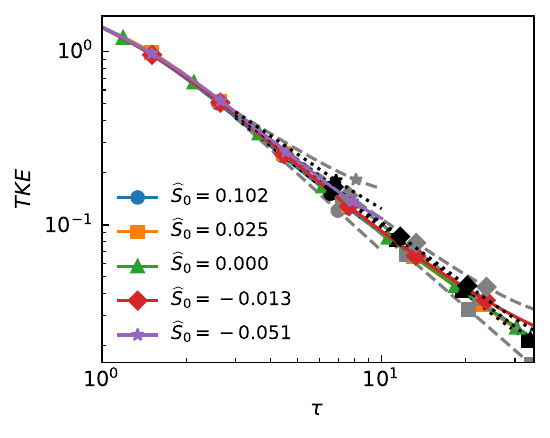}{a}{0.48}{1}
    \hfill
    \FigureAndLabel{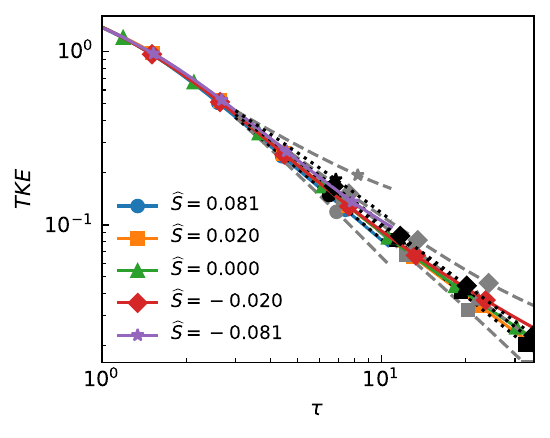}{b}{0.48}{1}
    \caption{Total turbulent kinetic energy for (\textit{a}) constant velocity and (\textit{b}) constant strain rate. Solid lines indicate ILES, gray dashed lines indicate the turbulent kinetic energy model, and black dotted lines indicate the corrected turbulent kinetic energy model.}
    \label{fig:iLES_TKE}
\end{figure}

\begin{figure}
    \FigureAndLabel{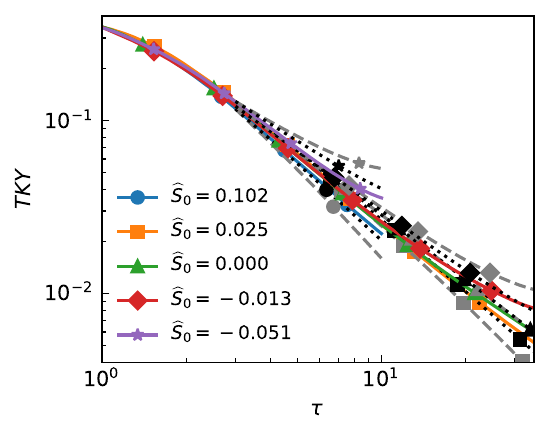}{a}{0.48}{1}
    \hfill
    \FigureAndLabel{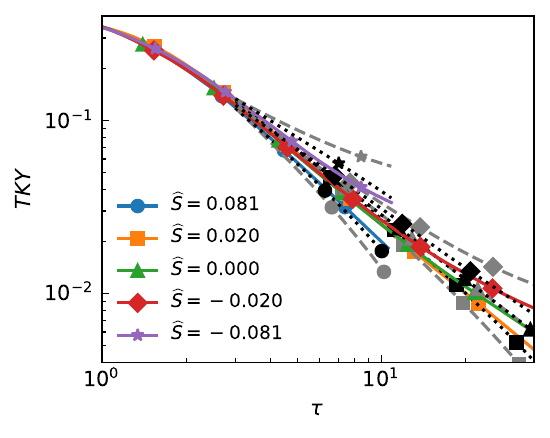}{b}{0.48}{1}
    \caption{Turbulent kinetic energy in the $y$-direction for (\textit{a}) constant velocity and (\textit{b}) constant strain rate. Solid lines indicate ILES, gray dashed lines indicate the turbulent kinetic energy model, and black dotted lines indicate the corrected turbulent kinetic energy model.}
    \label{fig:iLES_TKY}
\end{figure}

The application of the model for $TKE$ is shown in figure \ref{fig:iLES_TKE} alongside the simulation results. The simulation data shows the expected trends of the compression cases having greater $TKE$ due to the positive shear-production contribution. However, the model as shown in dashed grey lines, over-predicts the influence of the strain rate on the total turbulent kinetic energy. The results for the model for the transverse component, $TKY$, is shown in figure \ref{fig:iLES_TKY} and shows the same trend. 

\begin{figure}
    \FigureAndLabel{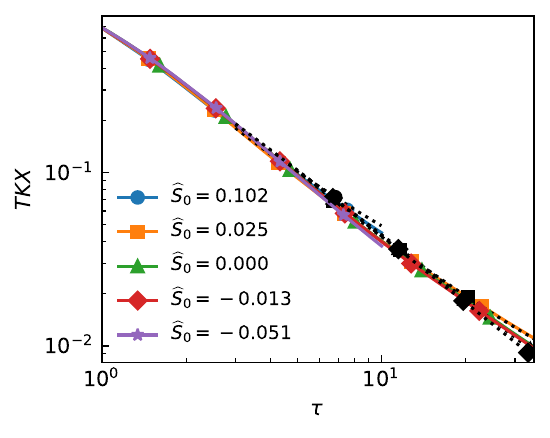}{a}{0.48}{1}
    \hfill
    \FigureAndLabel{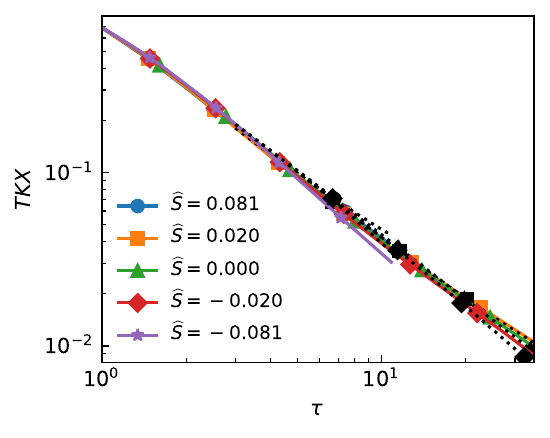}{b}{0.48}{1}
    \caption{Turbulent kinetic energy in the $x$-direction for (\textit{a}) constant velocity and (\textit{b}) constant strain rate. Solid lines indicate ILES, and black dotted lines indicate the corrected turbulent kinetic energy model.}
    \label{fig:iLES_TKX}
\end{figure}

The inaccuracy of the total energy in the domain suggests the model is not accurate and  there is an issue with closure for either the pressure-rate-of-strain tensor or the dissipation rate. For the pressure-rate-of-strain tensor, excessive shear production for the strain cases could be caused  by the model not accurately capturing the redistribution of the energy between the transverse and axial directions. For the compression cases, the alignment with ILES would improve if the transverse energy was redistributed to the axial direction, decreasing the transverse energy and the shear production contribution whilst increasing the $TKX$ component. By the same process, redistribution of energy for the expansion cases should result in lower $TKX$. The results from the ILES in figure \ref{fig:iLES_TKX} show the opposite trend, with instead the expansion cases exhibiting slightly higher $TKX$ than the unstrained case, implying that the energy is not just being redistributed into/from $TKX$ for the strain-cases. Instead, the issue lies with the modelling of the total dissipation rate. It is commonly assumed in turbulence models that bulk compression or expansion will scale the turbulent lengthscale \citep{Dimonte_2006_KLTurbulenceModel,Besnard_1992_TurbulenceTransportEquations}. To account for the transverse compression, the turbulent lengthscale is modified to scale with the geometric mean of the expansion factors:
\begin{align}
    \varepsilon = C_\epsilon \frac{TKE^{3/2}}{W \sqrt{M} \Lambda^{2/3}}
\end{align}
As the strain rates are only applied in two out of the three dimensions, the resulting scale is to the power of 2/3. The results for this corrected model are also plotted in figures \ref{fig:iLES_TKE} and \ref{fig:iLES_TKY} as the black dotted lines, which show an improved agreement with the simulation results. The modified dissipation rate counteracts the effect of shear production, reducing the deviation from the unstrained case. This modification improves the $TKE$ estimates to align with the ILES simulation results. The corrected model and the simulation results for $TKX$ are plotted in figure \ref{fig:iLES_TKX}, with the original model omitted due to the proximity of the data-lines. The combination of the return-to-isotropy, isotropisation of production, and strain-dependent dissipation is able to produce a similar results to the simulations, showing that the influence of the strain rates on $TKX$ amounts to very little variation from the unstrained case. The expansion cases do have slightly higher $TKX$ which can explain the increased integral width for the expansion cases. By having a larger $TKX$, the mixing layer is able to entrain slightly more fluid, resulting in an increased growth rate.

\begin{figure}
    \FigureAndLabel{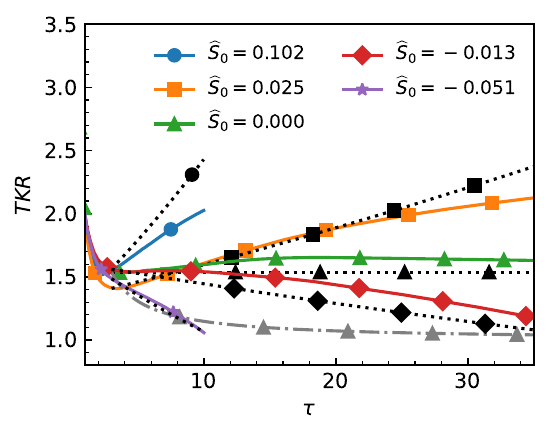}{a}{0.48}{1}
    \hfill
    \FigureAndLabel{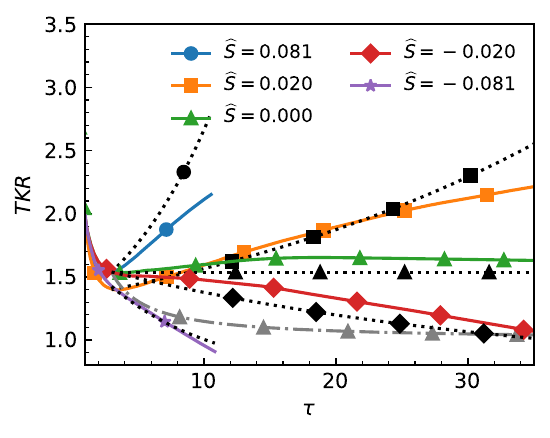}{b}{0.48}{1}
    \caption{Anisotropy of the turbulent kinetic energy for (\textit{a}) constant velocity and (\textit{b}) constant strain rate.  Solid lines indicate ILES, black dotted lines indicate the corrected turbulent kinetic energy model, and grey dot-dash lines indicate the turbulent kinetic energy model with the original LRR-IP coefficients.}
    \label{fig:iLES_Anisotropy}
\end{figure}

The turbulent kinetic energy anisotropy can be calculated from the components using the equation:
\begin{align}
    TKR = \frac{2 TKX}{TKY + TKZ}.
\end{align}
RMI has been shown to be persistently anisotropic, biased in the $x$-direction. The unstrained case achieves an anisotropy value of 1.49 at the final time of $\tau=246$ from \textsf{FLAMENCO} in the $\theta$-group collaboration \citep{ThetaGroup}. The simulation results in figure \ref{fig:iLES_Anisotropy} shows the unstrained case plateauing near this final value. The strained cases diverge from the asymptotic value, and it is observed that the expansion cases obtain a larger values of $TKR$, becoming more anisotropic due to the shear production removing turbulent kinetic energy in the transverse directions. The compression cases head towards isotropy, with the compression amplifying the transverse turbulent kinetic energy. Such behaviour is also observed in spherical simulations such as by \cite{Wang_2023_EvolutionTurbulentMixing} and \cite{ElRafei_2019_ThreedimensionalSimulationsTurbulent} between the first and second re-shock. The anisotropy from the turbulent kinetic energy model is also plotted in figure \ref{fig:iLES_Anisotropy}. The unstrained case maintains a constant anisotropy due to the assignment of $C_R=1.0$, keeping the model and ILES profiles closely aligned, with the main deviation between the two profiles arising from the slight increase in the ILES anisotropy which cannot be reproduced by the model. For the unstrained case, the model which uses the original LRR-IP coefficients is also plotted, which shows a much faster trend to isotropy than is observed from the simulation. Using a value of $C_R=1.8$ causes the anisotropy to exponentially decay towards the isotropic value of $TKR=1$. The half-life of the anisotropy, that is the time required for the anisotropic deviation from unity to reduce by a factor of one-half, is around 2.5$\tau$ for the original coefficients of the LRR-IP model. The corrected turbulent kinetic energy model's anisotropy predictions for the strain cases shows reasonable agreement, with the anisotropy being driven by the shear production but is slightly counteracted by the isotropisation of production term in the model. The largest deviations between the model and ILES results occurs for the expansion cases, with the ILES showing reduced anisotropic growth, suggesting some tendency towards either isotropy or the original anisotropic value. This behaviour may be accounted for with improved modelling of the pressure-rate-of-strain tensor. Whilst the turbulent Mach number remains small throughout the simulation, using a compressible turbulence model for the pressure-rate-of-strain tensor would have the benefit of not being restricted to zero net contribution to the total turbulent kinetic energy.

\subsubsection{Turbulent mass flux}
\begin{figure}
    \FigureAndLabel{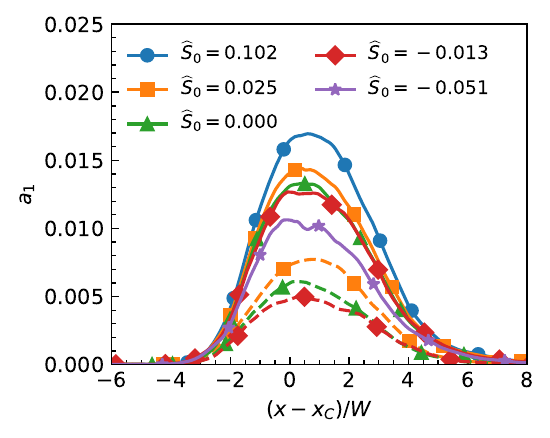}{a}{0.48}{1}
    \hfill
    \FigureAndLabel{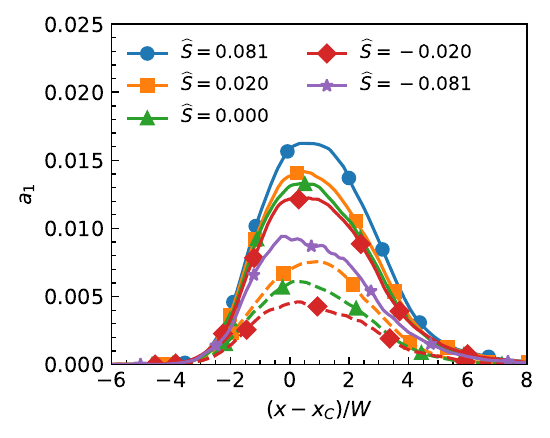}{b}{0.48}{1}
    \caption{Turbulent mass flux for (\textit{a}) constant velocity, and (\textit{b}) constant strain rate.  Solid lines indicate results at $\tau=9.843$, dashed lines indicate results at $\tau=34.451$.}
    \label{fig:iLES_a}
\end{figure}

In compressible flows, the turbulent mass flux, $a_i=\tilde{u_i}-\bar{u}_i = -\overline{u_i''}$, represents the difference between the mean and density weighted velocities. The turbulent mass flux is commonly used in closure of compressible Reynolds-averaged Navier-Stokes models, as the generation of turbulent kinetic energy (or the production of the Reynolds stresses) from potential energy is proportional to the turbulent mass flux. With the transverse velocity-gradients removed, the homogeneity of the $y$-$z$ plane means that $a_2$ and $a_3$ should statistically be zero for the plane. The axial component, $a_1$ is not zero, and is plotted in figure \ref{fig:iLES_a} for all cases at $\tau=9.843$ (solid lines), and for the unstrained and low-magnitude strain rate cases at $\tau=34.451$ (dashed lines). The results show the compression cases have a decreased amount of axial turbulent mass flux compared to the unstrained case. The turbulent mass flux can be generated by a term equivalent to shear production,
\begin{align}
    \frac{ \partial \bar{\rho} a_i}{\partial t} \propto -\bar{\rho} a_j \frac{\partial \bar{u}_i}{\partial x_j},
\end{align}
and with the applied strain rates in the transverse direction, this term will act on components $a_2$ and $a_3$, not $a_1$. The components $a_2$ and $a_3$ will remain statistically zero, however the influence of the strain rate will affect the standard deviation/second moment of the $u_i''$ distribution, which is the turbulent kinetic energy and investigated in \S \ref{sec:iles_tke}. Under axial strain rates, $a_1$ is amplified under compression as observed in \cite{Pascoe2024}, the opposite of the trend observed in figure \ref{fig:iLES_a}. This decrease in $a_1$ matches the behaviour observed for $TKX$ which decreased under compression due to the change in dissipation rate, showing that the effect of the modified turbulent lengthscale and dissipation rate is observable with the turbulent mass flux as well.

\subsubsection{Enstrophy}
\begin{figure}
    \FigureAndLabel{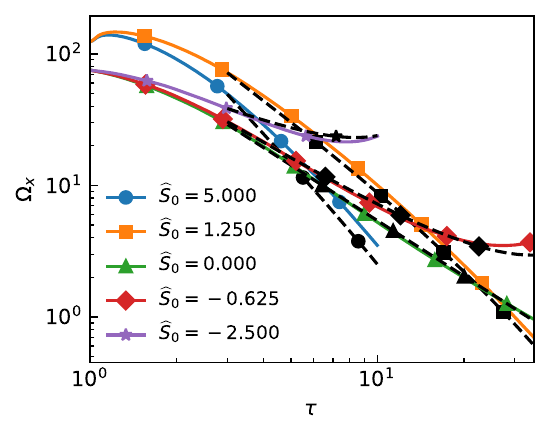}{a}{0.48}{1}
    \hfill
    \FigureAndLabel{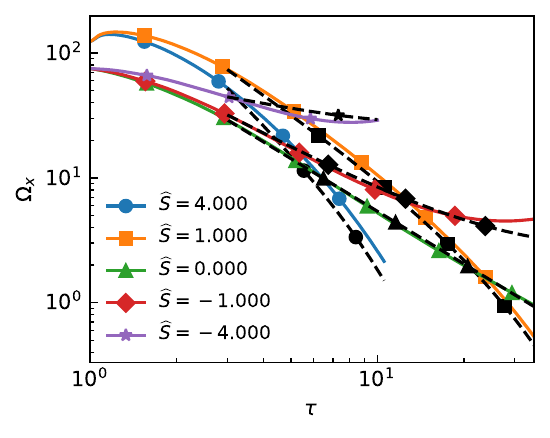}{b}{0.48}{1}
    \caption{Enstrophy in the $y-z$ plane for (\textit{a}) constant velocity and (\textit{b}) constant strain rate. Solid lines indicate ILES, dashed lines indicate the enstrophy model.}
    \label{fig:iLES_OmegaX}
\end{figure}

\begin{figure}
    \FigureAndLabel{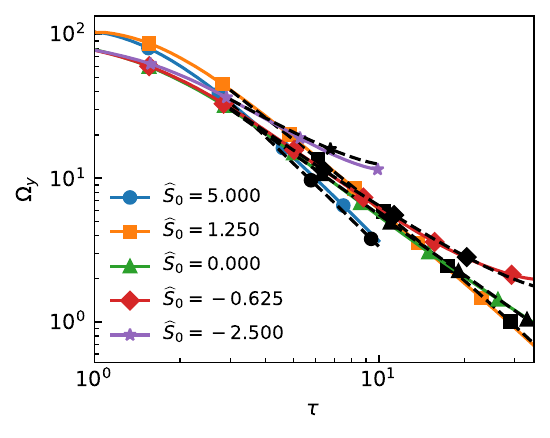}{a}{0.48}{1}
    \hfill
    \FigureAndLabel{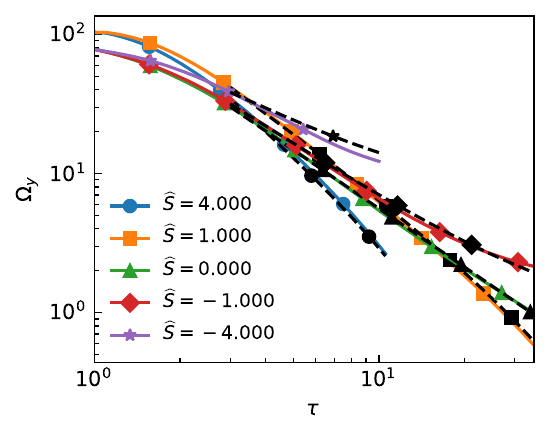}{b}{0.48}{1}
    \caption{Enstrophy in the $x-z$ plane for (\textit{a}) constant velocity and (\textit{b}) constant strain rate. Solid lines indicate ILES, dashed lines indicate the enstrophy model.}
    \label{fig:iLES_OmegaY}
\end{figure}

The modelling of the pressure-dilatation tensor can be avoided by analysing the vorticity or enstrophy of the flow. The vorticity is defined by the curl of the velocity field,
\begin{align}
    \omega_i = \epsilon_{ijk} \frac{\partial u_k}{\partial x_j}.
\end{align} 
For a compressible, viscous fluid, the transport equation for the vorticity components is given by \cite{Kida_1990_EnstrophyBudgetDecaying,Porter_2015_VORTICITYSHOCKSMAGNETIC} as
\begin{align}
    \frac{\partial \omega_i}{\partial t} + \frac{\partial}{\partial x_j} \left(u_j \omega_i\right) = \omega_j \frac{\partial u_i}{\partial x_j} &+ \frac{\epsilon_{ijk}}{\rho^2} \frac{\partial \rho}{\partial x_j} \frac{\partial p}{\partial x_k} \\
    \notag &+\frac{1}{Re_0} \left[ \frac{1}{\rho} \frac{\partial^2 \omega_i}{\partial x_j \partial x_j} - \frac{\epsilon_{ijk}}{\rho^2} \frac{\partial \rho}{\partial x_j}\left( \frac{\partial^2 u_k}{\partial x_l x_l} +\frac{1}{3} \frac{\partial^2 u_l}{\partial x_k \partial x_l}\right)\right].
\end{align}
The domain integrated enstrophy components will be defined by integrating the density-weighted enstrophy over the domain,
\begin{align}
    \Omega_{ij} = \iiint \rho \omega_i \omega_j \,\mathrm{d}V.
\end{align}
For simplicity the notation of $\Omega_x = \Omega_{11}$ will be used for directional components, and $\Omega=\Omega_{ii}$ will be used to denote the total enstrophy respectively. The transport equation for the enstrophy components can be obtained by integrating the equation over the time-varying domain. It is also assumed that after $\tau=1$, the pressure in the simulation is uniform in space and the baroclinic source term can be neglected. As the simulations were performed as ILES, the viscous effects were represented implicitly through the numerical scheme used to solve the flow \citep{Grinstein_2007_ImplicitLargeEddy}. As a result, the final term which is inversely proportional to the Reynolds number is replaced with a symbolic dissipation rate:
\begin{align}
    \frac{\mathrm{d}}{\mathrm{d}t}\begin{pmatrix}
        \Omega_x \\ \Omega_y \\ \Omega_z
    \end{pmatrix}
    = 
    \begin{pmatrix}
        \Omega_x \frac{\partial \bar{u}_1}{\partial x_1}\\ \Omega_y \frac{\partial \bar{u}_2}{\partial x_2} \\ \Omega_z \frac{\partial \bar{u}_3}{\partial x_3}
    \end{pmatrix} -
    \begin{pmatrix}
        \Omega_x \\ \Omega_y \\ \Omega_z
    \end{pmatrix} \frac{\partial \bar{u}_i}{\partial x_i}
    -\mathbf{\varepsilon}_\Omega
\end{align}
The source terms on the right-hand side correspond to the vortex stretching, vortex compression, and dissipation. The vortex stretching amplifies the vorticity in the direction of the strain rate, causing vortices to tilt towards strained axis for expansion or away in the case of compression. The vortex compression term scales all vorticity components according to the mean compression rate. In the rapid distortion limit, the dissipation is neglected as it occurs on a larger timescale than the strain \citep{Hunt_1990_RapidDistortionTheory,Blaisdell_1996_RapidDistortionTheory}. The present simulations are not in the rapid distortion limit, with the dissipation still playing a dominant role. In order to model the dissipation, the self-similar solution of the vorticity will be utilised. With a late-time power scaling of $\Omega \propto \tau^{-1.4}$ \citep{Zhou_2020_DependenceEnstrophyTransport}, then for the unstrained case the enstrophy and dissipation rate will be modelled according to
\begin{align}
    \Omega &= \Omega_0 \left(\frac{\tau}{\tau_0}\right)^{-n},\\
    \varepsilon_\Omega &= - \frac{d\Omega}{d\tau} = \frac{n}{\Omega_0^{1/n} \tau_0} \Omega^{(n+1)/n}. \label{eqn:enstrophy_dissipation}
\end{align}
The dissipation rate has been expressed in equation (\ref{eqn:enstrophy_dissipation}) as function of the current value of the enstrophy as opposed to a function of time, allowing the dissipation rate to maintain dependence or proportionality to the simultaneous enstrophy value. The dissipation rate also depends on a prescribed power law decay exponent and some initial conditions to scale it accordingly. Each enstrophy component is assumed to maintain this dissipation scaling, using the respective enstrophy component to calculate the corresponding dissipation rate. The model equation for the $x$- and $y$-component can be re-written, now using the dissipation model and substituting in the transverse strain rate to take the form
\begin{align}
    \frac{d}{d\tau}\begin{pmatrix}
        \Omega_x \\ \Omega_y
    \end{pmatrix}
    = 
    \begin{pmatrix}
        0 \\ \Omega_y \hat{S}
    \end{pmatrix} -2 \hat{S}
    \begin{pmatrix}
        \Omega_x \\ \Omega_y
    \end{pmatrix}  
    - \frac{n}{\tau_0}
    \begin{pmatrix}
        \Omega_{x0}^{-1/n} \Omega_x^{(n+1)/n} \\ \Omega_{y0}^{-1/n} \Omega_{y}^{(n+1)/n}
    \end{pmatrix}.
\end{align}
The results for this model are plotted in figure \ref{fig:iLES_OmegaX} for $\Omega_X$ and in figure \ref{fig:iLES_OmegaY} for $\Omega_Y$. Whilst the unstrained and compression cases start from the same initial enstrophy values, the expansion cases have a higher initial value due to the interpolation on to the finer mesh. The interpolation process conserves the velocity components, ensuring the domain-integrated velocity and velocity derivatives are unchanged. The interpolation process however can steepen local gradient calculations, and as the enstrophy depends upon these gradients, the domain-integrated enstrophy is not conserved and will increase. For DNS, the flow field should be adequately resolved and smooth meaning that such effects are small under refinement, however for ILES the effect is observable in the flow-field following bulk refinement. The difference in the calculated enstrophy is larger for $\Omega_x$ which represents the enstrophy in the homogeneous $y$-$z$ plane which was interpolated in both directions, whilst $\Omega_y$ represents the out-of-plane direction which includes the unchanged axial direction. Despite this, the trends of the enstrophy components is clear. The combination of the vortex stretching and vortex compression terms create a source term that linearly scales with the negative of the strain rate. It is observed in the plots that the expansion cases show a stronger decrease in enstrophy than the unstrained case, whilst the compression cases show a relative increase in the enstrophy components. For $\Omega_x$, the compression cases even achieve a net production at the final simulation times. The model in the plots starts at $\tau=3$ as after this time the unstrained case is matches the self-similar decay profile, as show by the agreement with the model. The model accurately predicts the trends for the strain-simulation and doesn't require additional scaling for the dissipation rate, unlike the turbulent kinetic energy. The changes in $\Omega_x$ agree with the observation of increased mixedness within the mixing layer; higher vorticity in the $y$-$z$ plane should correspond to greater mixing to achieve more in-plane homogeneity. The out-of-plane vorticity which corresponds to $\Omega_y$ would be expected to help move the two fluids through the mixing layer, however it does not appear to have a noticeable effect of the mean volume fraction profile of $\bar{f}_1$, and instead may just aid in the total mixedness. The change in enstrophy may serve as an indicator of the change in mixedness for the mixing layer. In contrast, the turbulent kinetic energy serves as a better indicator of the effects of how the strain rate affects the growth rate of the mixing layer.

\subsubsection{Buoyancy-drag model}
\label{sec:buoyancydrag}
The buoyancy-drag mixing model is a simple mode based on ordinary differential equations (ODEs) to calculate the growth of the mixing layer width $W$ and the growth velocity $V$. This methodology was inspired by the modelling of bubble penetration in the $At=1$ case for RTI by \cite{Layzer_1955_InstabilitySuperposedFluids}, however there have been many works trying to derive and calibrate the buoyancy-drag model to accurately represent the RMI and RTI for all Atwood numbers \citep{Baker_1981_HeuristicModelNonlinear,Dimonte_2000_SpanwiseHomogeneousBuoyancydrag,Hansom_1990_RadiationDrivenPlanar,Oron_2001_DimensionalityDependenceRayleigh,Ramshaw_1998_SimpleModelLinear}. The simplicity of the buoyancy-drag model has also inspired other models such as the K-L turbulence model \citep{Dimonte_2006_KLTurbulenceModel} which collapses to the buoyancy-drag model under a self-similar analysis. The most relevant buoyancy-drag model to the cases investigated here are the models calibrated to the quarter-scale $\theta$-group case by \citet{Youngs_2020_BuoyancyDragModelling,Youngs_2020_EarlyTimeModifications}. The integral width, bubble height, and spike height are each governed by a pair of ODEs:
\begin{subeqnarray}
    \frac{\mathrm{d}W}{\mathrm{d}t} = V, &\quad& \frac{\mathrm{d}V}{\mathrm{d}t} = - \frac{V^2}{l^{\text{eff}}(\bar{\lambda},W)}, \label{eqn:IW_BD} \\
    \frac{\mathrm{d}h_b}{\mathrm{d}t} = V_b, &\quad& \frac{\mathrm{d}V_b}{\mathrm{d}t} = - \frac{V_b^2}{l^\text{eff}_b(\bar{\lambda},h_b)},\\
    \frac{\mathrm{d}h_s}{\mathrm{d}t} = V_s, &\quad& \frac{\mathrm{d}V_s}{\mathrm{d}t} = - \frac{V_s^2}{l^\text{eff}_s(\bar{\lambda},h_b)}.
\end{subeqnarray}
The growth of the heights is equal to the respective velocity component in the unstrained buoyancy-drag model. As this model is calibrated for the RMI-induced mixing layer, the velocity ODE only contains a drag term, however for RTI flows a buoyancy term is also included. \cite{Youngs_2020_EarlyTimeModifications} calibrated the effective drag lengthscale used in the drag term for the integral width to accurately predict the early-time growth. This analysis was extended to the bubble and spike heights in \cite{Youngs_2020_BuoyancyDragModelling}. The drag lengthscales for each height are given by
\begin{subeqnarray}
    \frac{l^\text{eff}}{\bar{\lambda}} &=& \max\left\{a-b\left(1-e^{-cW/\bar{\lambda}}\right), \frac{\theta}{1-\theta} \left(\frac{W}{\bar{\lambda}}-d\right)  \right\}\label{eqn:W_lengthscale}, \\
    \frac{l^\text{eff}_b}{\bar{\lambda}} &=& \max\left\{a_b-b_s\left(1-e^{-c_b h_b/\bar{\lambda}}\right), \frac{\theta}{1-\theta} \left(\frac{h_b}{\bar{\lambda}}-d_b\right)  \right\} ,\label{eqn:hB_lengthscale}\\
    \frac{l^\text{eff}_s}{\bar{\lambda}} &=& \max\left\{a_s-b_s\left(1-e^{-c_s h_b/\bar{\lambda}}\right), \frac{\theta}{1-\theta} R \left(\frac{h_b}{\bar{\lambda}}-d_s\right)  .\right\}\label{eqn:hS_lengthscale}
\end{subeqnarray}

The effective lengthscale for each equation is a piece-wise function that transitions between a drag lengthscale dependent upon the spectrum perturbation to a drag lengthscale that scales linearly with the outer lengthscale. For the late-time growth a theoretical power-law value of $\theta=1/3$ was used, as suggested in the work of \cite{Elbaz_2018_ModalModelMean}. It is important to note that the bubble and spike are expected to grow self-similarly, and so the drag terms for the spike height is calculated using the bubble height and a calibrated fit value of $R=1.1$ to describe the asymptotic proportionality. The remaining coefficients of the drag lengthscale equations are listed in table \ref{tab:BD_OrigCoeff}, based off the values in the original works \citep{Youngs_2020_BuoyancyDragModelling,Youngs_2020_EarlyTimeModifications}. As the bubble and spike heights were calibrated to a slightly different case, the values of $c_S$ and $d_S$ are updated to provide a more accurate fit to the present unstrained model.

\begin{table}
  \begin{center}
\def~{\hphantom{0}}
  \begin{tabular}{lcccc}
      Lengthscale     & $a$     &  $b$  & $c$   & $d$ \\[3pt]
      Integral width  & 0.3     & 0.176 & 8.35  & 0.237 \\
      Bubble          & 0.7     & 0.297 & 6.0~  & 0.283 \\
      Spike           & 1.4     & 1.19~ & 0.8~  & 0.70~
  \end{tabular}
      \caption{Buoyancy-drag coefficients for $\textit{At}=0.5$, narrowband RMI \citep{Youngs_2020_BuoyancyDragModelling,Youngs_2020_EarlyTimeModifications}}
  \label{tab:BD_OrigCoeff}
  \end{center}
\end{table}

\begin{figure}
    \begin{subfigure}[h]{0.48\textwidth}
        \centering
        \begin{tikzpicture}
            \node[anchor=north west] (image) at (0,0) { \includegraphics[width=\textwidth,trim=0.2cm 0.2cm 0.2cm 0.2cm, clip]{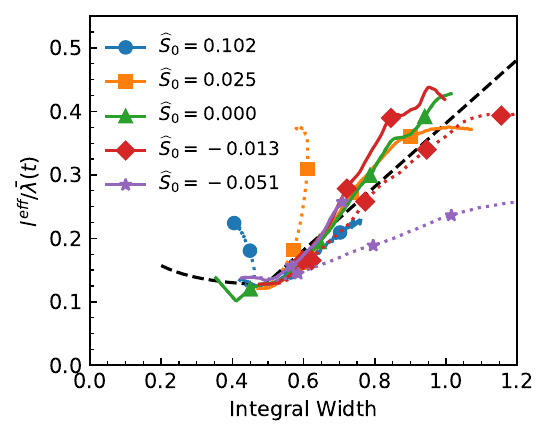}};
            \node[anchor=north west] at (0,-0) {(\textit{a})};
        \end{tikzpicture}
    \end{subfigure}
    \hfill
    \begin{subfigure}[h]{0.48\textwidth}
        \centering
        \begin{tikzpicture}
            \node[anchor=north west] (image) at (0,0) { \includegraphics[width=\textwidth,trim=0.2cm 0.2cm 0.2cm 0.2cm, clip]{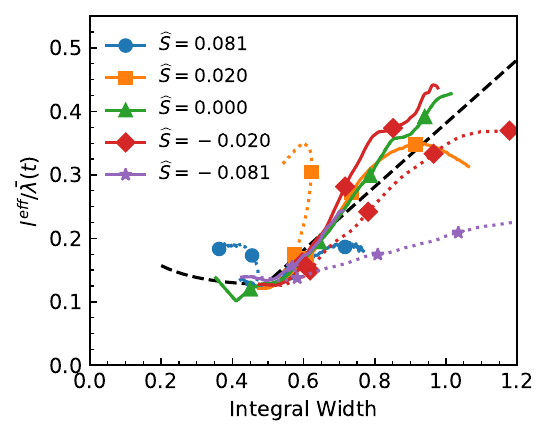}};
            \node[anchor=north west] at (0,-0) {(\textit{b})};
        \end{tikzpicture}
    \end{subfigure}
    \caption{Effective drag lengthscale as a function of non-dimensionalised integral width for (\textit{a}) constant velocity and (\textit{b}) constant strain rate. Solid lines indicate integral width non-dimensionalised by $\bar{\lambda}_0$, dotted lines indicate integral width non-dimensionalised by $\bar{\lambda}(t)$, and dashed lines indicate planar narrowband effective lengthscale model.}
    \label{fig:BD_lengthscale_iw}
\end{figure}

In the planar configuration, there is no variation in the mean wavelength used to calculate the effective drag lengthscale. \cite{Miles_2004_BubbleMergerModel,Miles_2009_BlastWaveDrivenInstabilityVehicle} utilised the time-varying wavelength as the drag lengthscale, whilst \cite{ElRafei_2020_NumericalStudyBuoyancy} calibrated a buoyancy-drag model by fitting the effective drag lengthscale as a function of the time-varying wavelength, $l^\text{eff}/\bar{\lambda}(t) = f(\bar{\lambda}(t),W/\bar{\lambda}(t))$. The effective drag lengthscale can be calculated from equation (\ref{eqn:IW_BD}) using the form
\begin{align}
    l^\text{eff} = - \frac{\dot{W}^2}{\ddot{W}}.
\end{align}
As this equation relies on the second derivative of the integral width measurements, it is inherently noisy. The noise in the profile was reduced through a down-sampling of the data. In figure \ref{fig:BD_lengthscale_iw}, there are two variations of the effective drag lengthscale for the integral width plotted. The first is as a function of $W/\bar{\lambda}(t)$, denoted by the dotted lines, which corresponds to the approach used by \cite{ElRafei_2020_NumericalStudyBuoyancy}. For the strain-cases presented, the profiles do not collapse upon the unstrained case. The low-magnitude compression cases do show alignment with the unstrained model, and also possess a similar gradient to the unstrained case as was observed by \cite{ElRafei_2020_NumericalStudyBuoyancy}. The effective drag lengthscale variation as a function of $W/\bar{\lambda}_0$ is denoted by the solid lines, which all fall within the vicinity of the unstrained model.

The collapse of the effective drag lengthscale profiles suggests the model to be used for the transverse strain cases should be function of $W/\bar{\lambda}_0$,
\begin{align}
    \frac{l^\text{eff}}{\bar{\lambda}(t)} = \max\left\{a-b\left(1-e^{-cW/\bar{\lambda}_0}\right), \frac{\theta}{1-\theta} \left(\frac{W}{\bar{\lambda}_0}-d\right)  \right\}.
\end{align}
In this form, the original effective drag lengthscale calculation is used, however the effective drag lengthscale is proportionally scaled to the time-varying wavelength. 

\begin{figure}
    \FigureAndLabel{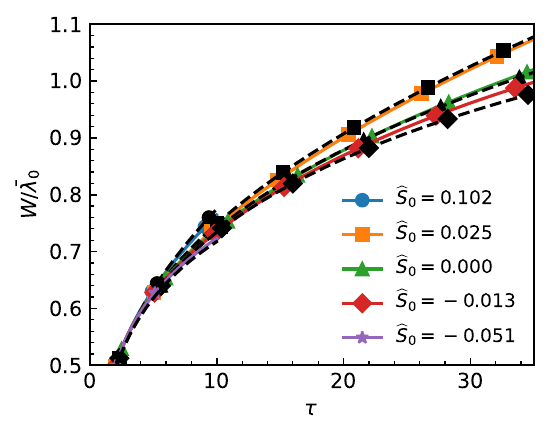}{a}{0.48}{1}
    \hfill
    \FigureAndLabel{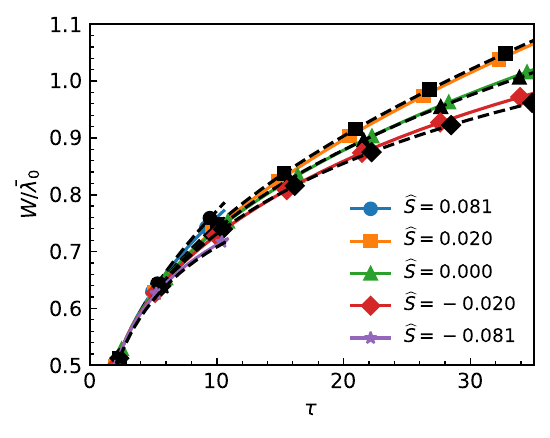}{b}{0.48}{1}
    \caption{Buoyancy-drag model for integral width: (\textit{a}) constant velocity and (\textit{b}) constant strain rate. Solid lines indicate ILES results, dashed lines indicate the buoyancy-drag model.}
    \label{fig:BD_IW}
\end{figure}

\begin{figure}
    \FigureAndLabel{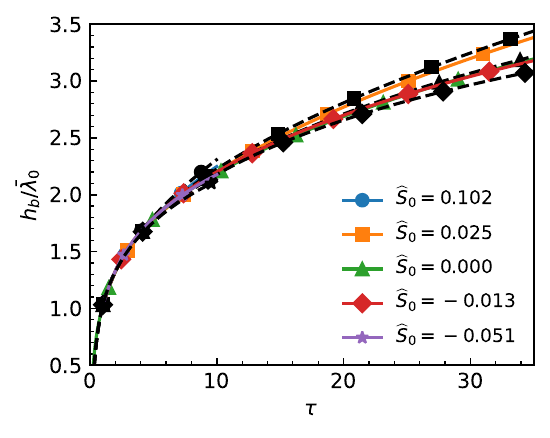}{a}{0.48}{1}
    \hfill
    \FigureAndLabel{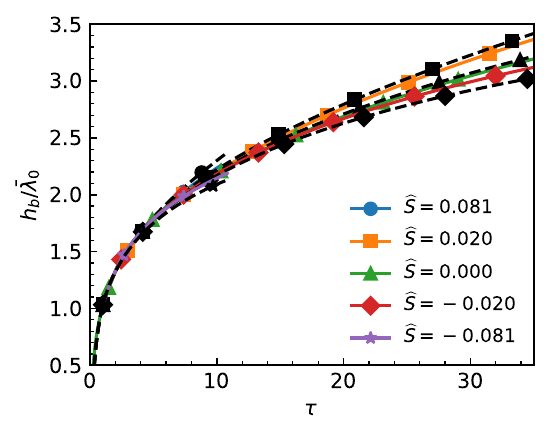}{b}{0.48}{1}
    \caption{Buoyancy-drag model for bubble height: (\textit{a}) constant velocity and (\textit{b}) constant strain rate. Solid lines indicate ILES results, dashed lines indicate the buoyancy-drag model.}
    \label{fig:BD_Bubble}
\end{figure}

\begin{figure}
    \FigureAndLabel{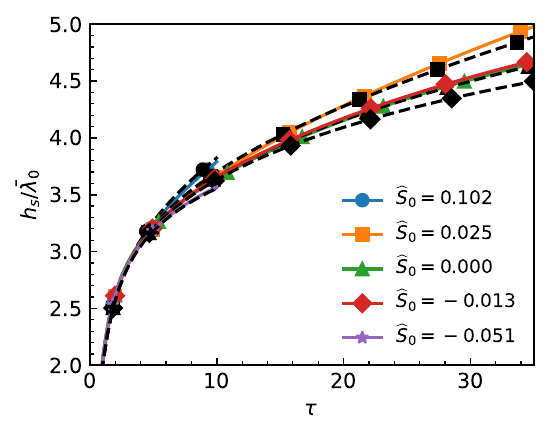}{a}{0.48}{1}
    \hfill
    \FigureAndLabel{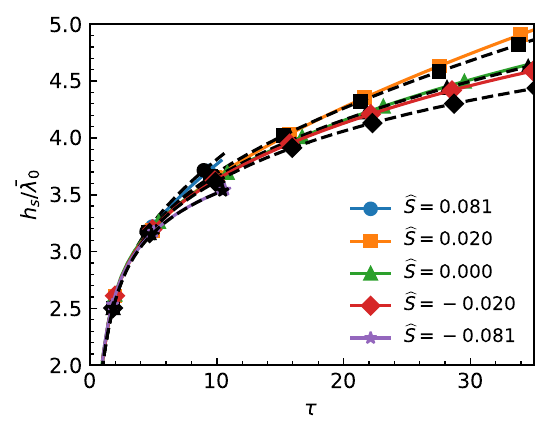}{b}{0.48}{1}
    \caption{Buoyancy-drag model for spike height. (\textit{a}) constant velocity and (\textit{b}) constant strain rate. Solid lines indicate ILES results, dashed lines indicate the buoyancy-drag model.}
    \label{fig:BD_Spike}
\end{figure}

To validate this model, the equations are integrated in time using the initial conditions provided in \cite{Youngs_2020_BuoyancyDragModelling}. An offset of $\tau=0.08$ is used to align the buoyancy-drag model with the simulations, as the buoyancy-drag model is fitted to the post-shock behaviour and does not describe the shock transition. The initial heights and velocities are given by
\begin{subeqnarray}
    W_0 &=& 0.5642 C \sigma_0,\\
    {h_s}_0 &=& {h_b}_0 = 1.1 \times 2.0 C \sigma_0,\\
    V_0 &=& 0.5642 C \bar{k} \sigma_0 \Delta u \textit{At} \times F_W^{nl},\\
    {V_b}_0 &=& 1.1 \times 2.0 C \bar{k} \sigma_0 \Delta u \textit{At} \times F_b^{nl},\\
    {V_s}_0 &=& 1.1 \times 2.0 C \bar{k} \sigma_0 \Delta u \textit{At} \times F_s^{nl},
\end{subeqnarray}
with compression factor $C=0.576$, and non-linearity factors: $F_W^{nl} = 0.85$, $F_b^{nl} = 0.60$, and $F_s^{nl}=1.0$. The solution for all cases are identical until $\tau=1$, after which the transverse strain rates will begin to change $\bar{\lambda}(t)$ and the effective drag lengthscale. The results for integral width are plotted in figure \ref{fig:BD_IW}, and the buoyancy-drag model shows good alignment for all cases except for the weak compression cases. This trend is also observed for bubble heights in figure \ref{fig:BD_Bubble} and the spike heights in figure \ref{fig:BD_Spike}. The increased growth for the weak compression cases could be a result of the redistribution of the turbulent kinetic energy from the transverse direction to the axial direction which is causing some additional growth for these components. Such behaviour could be accounted for by including additional terms in the drag lengthscale that would correspond to the shear production terms, as was done in \cite{Pascoe2024}.

\section{Conclusion}
\label{sec:conclusion}
The influence of transverse strain rates on the Richtmyer--Meshkov instability, or more generally for anisotropic, inhomogeneous mixing-layers, has been investigated by applying transverse strain rates to simulations in planar geometry. Within the linear regime, a linearised potential flow model was derived to predict how the application of the transverse strain rate would affect the initial RMI perturbation growth. Using the strain rate framework, the solution obtained was found to be equivalent to the Bell--Plesset results, meaning that strained planar simulations can replicate the behaviour of convergent simulations in the linear regime. Resolved two-dimensional numerical simulations were conducted in planar geometry, applying expansive or compressive strain rates. The simulations and model showed agreement while the amplitude was smaller than the time-varying wavelength, $a < 0.1\lambda(t)$, with compressive strain rates amplifying the instability growth-rate whilst expansive strain rates inhibit the growth. \\
To investigate the effects of transverse strain rates in the transitional and self-similar regime, strained simulations were conducted using the quarter-scale $\theta$-group case as the initial conditions. These implicit large eddy simulations were initialised at $\tau=1$ from the original case, a multi-mode narrowband RMI-induced mixing layer. Unlike the linear regime, the ILES results show that the compressive strain rates cause the mixing layer to have a slightly decreased growth rate, whilst the expansion cases grow slightly faster. This change in growth rate is explained by the modification of the turbulent length scale with the applied strain rates. Whilst shear-production from the mean velocity gradients will generate turbulent kinetic energy in the transverse direction under compressive strain rates, the turbulent lengthscale will decrease, increasing the dissipation rate inside the mixing layer. As a result the axial turbulent kinetic energy will slightly decrease under compression as the increased dissipation rate will counteract the energy redistribution, and the mixing layer will attain a slightly decreased growth rate. This was further investigated by comparing the effective drag-lengthscale for a Buoyancy-Drag model, showing that the effective drag-lengthscale is best captured by linearly scaling the effective drag-lengthscale with the time-varying mean wavelength, $l^{eff}=\bar{\lambda}(t) f(\bar{\lambda}_0,W)$.\\
The shear-production effect from the strain rate was observed to increase the transverse turbulent kinetic energy, albeit by a reduced amount due to the counteracting dissipation rate modification. This increase in transverse turbulent kinetic energy caused the mixing layer to no longer attain the same asymptotic self-similar state as the unstrained case. The anisotropy of the turbulent kinetic energy increased for expansion cases. The compression cases instead head towards isotropy, with some cases achieving greater turbulent kinetic energy in the transverse direction. The mixedness of the mixing layer was also affected, with the compression cases attaining higher levels of mixedness whilst the expansion cases became less mixed. These changes in mixedness are correlated with the changes in the turbulent kinetic energy and enstrophy. 
\backsection[Acknowledgements]{The authors acknowledge the Sydney Informatics Hub and the use of the University of Sydney’s high performance computing cluster, Artemis. This research was supported by the Australian Government's National Collaborative
Research Infrastructure Strategy (NCRIS), with access to computational resources
provided by the Setonix supercomputer (https://doi.org/10.48569/18sb-8s43) through the National Computational Merit Allocation Scheme. The authors would like to thank EPSRC for the computational time made available on the UK supercomputing facility ARCHER/ARCHER2 via the UK Turbulence Consortium (EP/R029326/1).}

\backsection[Funding]{This research received no specific grant from any funding agency, commercial or not-for-profit sectors.}

\backsection[Declaration of interests]{The authors report no conflict of interest.}

\backsection[Author ORCIDs]{B. Pascoe, https://orcid.org/0000-0002-2286-8463;\\
M. Groom, https://orcid.org/0000-0003-2473-7229;\\
D.L. Youngs, https://orcid.org/0000-0003-2853-5341;\\
B. Thornber, https://orcid.org/0000-0002-7665-089X}

\appendix

\section{Convergence}
\label{sec:appendix}
As the analysis of the effects of strain rates on the transitional-to-self-similar regime in section \ref{sec:SSML} are based upon the quarter-scale $\theta$-group case \citep{ThetaGroup}, it is necessary to assure that the simulations are adequately resolved under the applied strain rates. As the simulations are ILES, they do not attempt to resolve to the Kolmogorov scale, and instead simulate the larger scales whilst relying on the inherent numerical dissipation of the solver to model the smallest scales. The original quarter-scale $\theta$-group case utilised 512 cells in each transverse direction and the same mesh was used for the compression strain rate cases. The expansion cases used a mesh with 1024 cells across, interpolating the original solution onto the finer mesh, in order to maintain the resolution at maximum expansion at the end of the simulations. To show that the integral properties of the simulations are converged, simulations are performed from $\tau=1$ with various meshes, interpolating the original solution onto the new meshes. The three meshes used are listed in table \ref{tab:convergence}, which consist of the original mesh, the transversely refined mesh used for expansion, and a moderately refined mesh which was refined in the axial and transverse directions.
\begin{table}
  \begin{center}
\def~{\hphantom{0}}
    \begin{tabular}{cccc}
         Cells & $\widehat{S}=-0.081$ &$\widehat{S}=0.0$&$\widehat{S}=0.081$  \\
         $~720\times~512^2$ & $\checkmark$ & $\checkmark$ & $\circ$ \\
         $1080\times~768^2$ & $\circ$ & $\circ$ & $\circ$ \\
         $~720\times1024^2$ &  & $\circ$ & $\checkmark$ \\
    \end{tabular}
    \caption{Test cases employed for the convergence study. Check-marks indicate the mesh used for results in the paper, circles indicate meshes used for convergence study.}
    \label{tab:convergence}
  \end{center}
\end{table}

\begin{figure}
    \centering
    \label{fig:enter-label}
    \FigureAndLabel{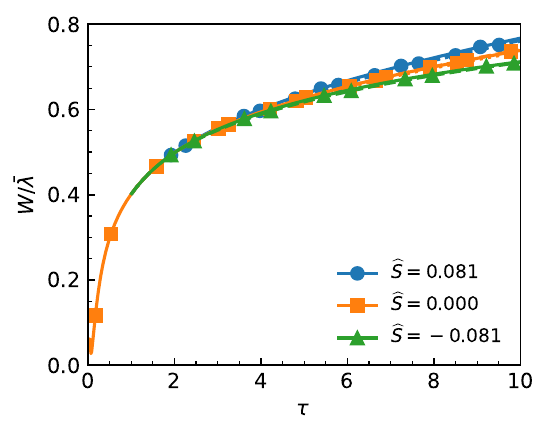}{a}{0.48}{1}
    \hfill
    \FigureAndLabel{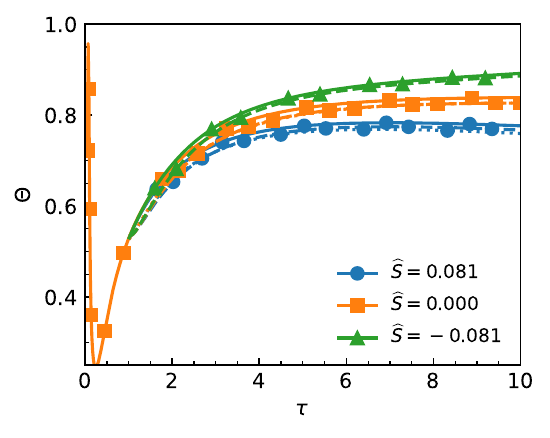}{b}{0.48}{1}
    \caption{Convergence of constant strain rate simulations under transverse compression for (a) integral width and (b) molecular mixing fraction. Solid lines indicate results for 512 cells across, dashed lines for 768 cells, dotted lines for 1024 cells.}
    \label{fig:convergence}
\end{figure}

The simulations are conducted for the for the unstrained case and the two high-magnitude constant strain rate cases, $\hat{S}=\pm0.081$, until a time of $\tau = 10$. The results for the integral width and molecular mixing fraction are plotted in figure \ref{fig:convergence}. The results for the integral width show a very small difference between the 512 cells cases and the higher resolution cases, with the higher resolution cases showing a slightly smaller integral width, but following the same identifiable trends dependent upon the strain rate. The same observation can be made for the the molecular mixing fraction where the mixedness is slightly decreased with the increased mesh resolution, but the same trends are discerned.

\bibliographystyle{jfm}
\bibliography{MyLibrary}

\end{document}